\documentclass[traditabstract]{aa}  

\usepackage{graphicx}
\usepackage{rotating}
\usepackage{txfonts}

\usepackage{amssymb}
\usepackage[fleqn]{amsmath}

\usepackage{url}

\usepackage[pdfencoding=auto,psdextra]{hyperref}
\hypersetup{
    colorlinks=true,
    linkcolor=blue,
    filecolor=magenta,      
    urlcolor=blue,
    citecolor=blue
}
\usepackage{xcolor}

\begin{document}

   \title{CHEX-MATE: Dynamical masses for a sample of 101 {\it Planck} Sunyaev-Zeldovich-selected galaxy clusters
   \thanks{Based in part on observations collected at the European Southern Observatory under ESO programmes 0110.A-4192 and 0111.1-0186.}
   \thanks{Table~\ref{tab_masses} is only available in electronic form at the CDS via anonymous ftp to \url{cdsarc.u-strasbg.fr} (130.79.128.5) or via \url{http://cdsweb.u-strasbg.fr/cgi-bin/qcat?J/A+A/.}}
   }


   \author{
    Mauro Sereno \inst{\ref{inaf-oas},\ref{infn-bo}\thanks{\email{mauro.sereno@inaf.it}}}
        \and
        Sophie Maurogordato \inst{\ref{oca}}
        \and 
        Alberto Cappi \inst{\ref{inaf-oas},\ref{oca}}
        \and
            Rafael Barrena \inst{\ref{iac},\ref{ull}}
            \and
            Christophe Benoist \inst{\ref{oca}}
            \and
            Christopher~P. Haines \inst{\ref{uda}}
            \and
            Mario Radovich \inst{\ref{inaf-oapd}}
            \and
            Mario Nonino\thanks{Deceased} \inst{\ref{inaf-oats}}
                \and
                Stefano Ettori  \inst{\ref{inaf-oas},\ref{infn-bo}}
                \and
                Antonio Ferragamo \inst{\ref{unina}}
                \and
                Rapha\"el Gavazzi \inst{\ref{lam},\ref{iap}}
                \and
                Sophie Huot \inst{\ref{iap}}
                \and
                Lorenzo Pizzuti \inst{\ref{unimib}}
                \and
                Gabriel~W. Pratt \inst{\ref{cea}}
               \and 
                Alina Streblyanska \inst{\ref{iac},\ref{ull}}                
                \and
                Stefano Zarattini \inst{\ref{iac},\ref{ull}} 
                    \and
                    Gianluca Castignani \inst{\ref{inaf-oas}}
                    \and
                    Dominique Eckert \inst{\ref{unige}}
                    \and
                    Fabio Gastaldello \inst{\ref{inaf-iasf}}
                    \and
                    Scott~T. Kay \inst{\ref{unimanch}}
                    \and
                    Lorenzo Lovisari \inst{\ref{inaf-iasf},~\ref{cfa}}
                    \and
                    Ben~J. Maughan \inst{\ref{ubristol}}
                    \and
                    Etienne Pointecouteau \inst{\ref{irap}}
                    \and
                    Elena Rasia \inst{\ref{inaf-oats},~\ref{ifpu},~\ref{unimichigan}}
                    \and
                    Mariachiara Rossetti \inst{\ref{inaf-iasf}}
                    \and
                    Jack Sayers \inst{\ref{caltech}}
          }              
\institute{
INAF - Osservatorio di Astrofisica e Scienza dello Spazio di Bologna, via Piero Gobetti 93/3, I-40129 Bologna, Italy
\label{inaf-oas}
\and
INFN, Sezione di Bologna, viale Berti Pichat 6/2, I-40127 Bologna, Italy \label{infn-bo}
\and
Universit\'e C\^ote d'Azur, Observatoire de la C\^ote d'Azur, CNRS, Laboratoire Lagrange, Bd de l'Observatoire, CS 34229, 06304 Nice Cedex 4, France \label{oca}
\and
Instituto de Astrof\'{\i}sica de Canarias, C/ V\'{\i}a L\'{a}ctea s/n, E-38205 La Laguna, Tenerife, Spain \label{iac}
\and
Universidad de La Laguna, Departamento de Astrof\'{i}sica, E-38206 La Laguna, Tenerife, Spain \label{ull}
\and
Instituto de Astronom\'{i}a y Ciencias Planetarias de Atacama (INCT), Universidad de Atacama, Copayapu 485, Copiap\'{o}, Chile \label{uda}
\and
INAF - Osservatorio Astronomico di Padova, Via dell'Osservatorio 5, 35122 Padova, Italy \label{inaf-oapd}
\and
INAF - Trieste Astronomical Observatory, via Tiepolo 11, 35143, Trieste, Italy \label{inaf-oats}
\and
Dipartimento di Fisica `E. Pancini', Universit\`a degli Studi di Napoli Federico II, Via Cinthia, 21, I-80126 Napoli, Italy\label{unina}
\and
Laboratoire d'Astrophysique de Marseille, CNRS, Aix-Marseille Universit\'e, CNES, Marseille, France\label{lam}
\and
Institut d'Astrophysique de Paris, UMR 7095, CNRS, and Sorbonne Universit\'e, 98 bis boulevard Arago, 75014 Paris, France\label{iap}
\and
Dipartimento di Fisica G. Occhialini, Universit\`a degli Studi di Milano Bicocca, Piazza della Scienza 3, I-20126 Milano, Italy\label{unimib}
\and
Universit\'{e} Paris-Saclay, Universit\'{e} Paris Cit\'{e}, CEA, CNRS, AIM, 91191, Gif-sur-Yvette, France \label{cea}
\and
Department of Astronomy, University of Geneva, ch. d'Ecogia 16, 1290 Versoix, Switzerland   \label{unige}
\and
INAF - IASF Milano, via A. Corti 12, I-20133 Milano, Italy \label{inaf-iasf}
\and
Jodrell Bank Centre for Astrophysics, Department of Physics and Astronomy, The University of Manchester, Oxford Road, Manchester M13 9PL, UK \label{unimanch}
\and
Center for Astrophysics $|$ Harvard $\&$ Smithsonian, 60 Garden Street, Cambridge, MA 02138, USA \label{cfa}
\and
HH Wills Physics Laboratory, University of Bristol, Bristol, UK \label{ubristol}
\and
IRAP, CNRS, Université de Toulouse, CNES, UT3-UPS, (Toulouse), France \label{irap}
\and
IFPU, Institute for Fundamental Physics of the Universe, Via Beirut 2, 34014 Trieste, Italy \label{ifpu}
\and
Department of Physics; University of Michigan, Ann Arbor, MI 48109, USA \label{unimichigan}
\and
California Institute of Technology, 1200 E. California Blvd., MC 367-17, Pasadena, CA 91125, USA \label{caltech}
}


 
  \abstract{
   The Cluster HEritage project with {\it XMM-Newton} – Mass Assembly and Thermodynamics at the Endpoint of structure formation (CHEX-MATE) is a programme to study a minimally biased sample of 118 galaxy clusters detected by {\it Planck} through the Sunyaev–Zeldovich effect.
   Accurate and precise mass measurements are required to exploit CHEX-MATE as an astrophysical laboratory and a calibration sample for cosmological probes in the era of large surveys. We measured masses based on the galaxy dynamics, which are highly complementary to weak-lensing or X-ray estimates.
   We analysed the sample with a uniform pipeline that is stable both for poorly sampled or rich clusters ---using spectroscopic redshifts from public (NED, SDSS, and DESI) or private archives--- and dedicated observational programmes. We modelled the halo mass density and the anisotropy profile. Membership is confirmed with a cleaning procedure in phase space. We derived masses from measured velocity dispersions under the assumed model.
   We measured dynamical masses for 101 CHEX-MATE clusters with at least ten confirmed members within the virial radius $r_\text{200c}$. Estimated redshifts and velocity dispersions agree with literature values when available. Validation with weak-lensing masses shows agreement within $8 \pm 16~\text{(stat.)} \pm 5~\text{(sys.)} \,\%$, and confirms dynamical masses as an unbiased proxy.
   Comparison with {\it Planck} masses shows them to be biased low by $34 \pm 3~\text{(stat.)} \pm 5~\text{(sys.)} \,\%$. A follow-up spectroscopic campaign is underway to cover the full CHEX-MATE sample.
}

    \authorrunning{M. Sereno et al.}
    \titlerunning{CHEX-MATE: Dynamical masses}
    
    \maketitle
%

\section{Introduction}

Cluster number counts as a function of mass and redshift, being sensitive both to the growth factor of structures and to the geometry of the Universe, are a well-established cosmological probe. The huge statistical power of surveys such as Euclid \citep{euclid_pre_sca+al22,euclid_I_24}, eROSITA \citep{erosita_bul+al24,erosita_ghi+al24}, and LSST (Legacy Survey of Space and Time) of the \textit{Vera C. Rubin }Observatory \citep{lsst_ive+al19} is pushing the boundaries of the precision and accuracy regimes. Cosmology with galaxy clusters requires a complete understanding of the selection function and a well-calibrated mass--observable relation, as cluster mass is generally not a direct observable \citep{sar+al16,euclid_ada+19}. Related systematic uncertainties are intertwined and are particularly challenging to account for. Cluster mass estimation is a long-standing challenge, which has been addressed with different approaches based, for example, on galaxy dynamics, the assumption of hydrostatic equilibrium (HE) using gas thermodynamics measured in X-rays or Sunyaev-Zeldovich (SZ), and weak lensing (WL) \citep{pra+al19}. Any bias in the mass calibration impacts the cosmological constraints.

The dynamical approach to estimate cluster masses has the great advantage of being independent of the gas properties. 
The velocity dispersion of cluster galaxies has been commonly used as a valuable proxy for the total mass \citep{biv+al06}. 

The dark matter (DM) halo velocity dispersion is regarded as a low-scatter mass proxy \citep{evr+al08, se+et15_comalit_I}, and masses derived in this way are highly complementary to WL or HE masses. Numerical simulations show that the 1D velocity dispersion, $\sigma_\text{1D}$, of DM halos is strongly correlated with cluster mass, with a low ($\sim 5\%$) dispersion \citep{evr+al08}. However, observations provide estimates of the line-of-sight projected velocity dispersions of galaxies, $\sigma_\text{los}$, and the scaling and bias with respect to the deprojected 1D velocity dispersion of the DM halo has to be estimated \citep{sar+al13,mun+al13,fer+al21}. 

Biases from orientation, halo triaxiality, and noise from large-scale structure are less severe than for WL, whereas deviations from equilibrium are less impactful than for HE masses \citep{gav05}. Moreover, dynamical masses are used to anchor intracluster-medium (ICM)-determined masses in X-ray or SZ cluster surveys \citep{sif+al16,spt_boc+al19}. Together with lensing, galaxy dynamics can
better constrain density profiles and break the mass--concentration degeneracy \citep{biv+al13}.


The velocity dispersion of galaxies may be biased with respect to DM particles (with tidal stripping being more active on large halos and dynamical friction disproportionately affecting the velocities of brighter galaxies, resulting in a velocity bias of the order of $5$-$10\,\%$ \citep{mun+al13, sar+al13,arm+al18,xxl_XXIII_far+al18}. This bias can be significantly reduced with pure and complete cluster member selection \citep{sar+al13,sif+al16,fer+al20}.
Numerical simulations can provide key guidance for the bias correction by assessing the role played by halo triaxiality, large-scale structure, selection processes (colour--luminosity selection, number of galaxies used, and radial window), and the presence of interlopers \citep{sar+al13,sif+al16,fer+al21}.



\begin{figure}
\resizebox{\hsize}{!}{\includegraphics{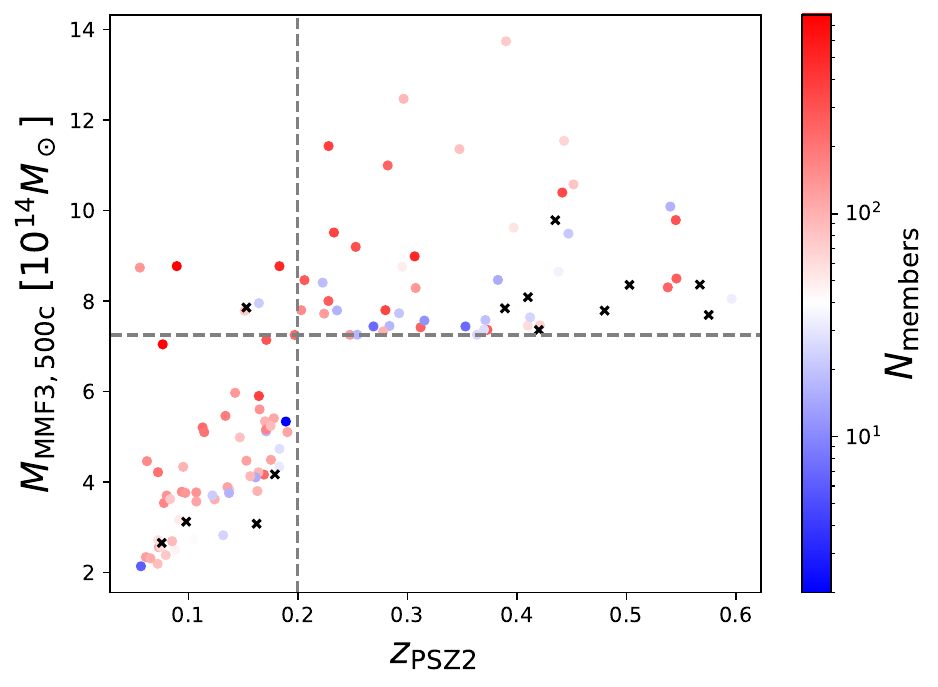}}
\caption{Distribution of the CHEX-MATE clusters in the redshift--mass plane. Points are colour coded according to the number of confirmed spectroscopic members. Black crosses mark the clusters without confirmed members.}
\label{fig_z_M500_members}
\end{figure}

In this paper, we present cluster masses of the CHEX-MATE \citep[Cluster HEritage project with {\it XMM-Newton}: Mass Assembly and Thermodynamics at the Endpoint of structure formation;][]{chexmate+al21} clusters estimated with a dynamical analysis of the velocity distribution of galaxies. The aim of CHEX-MATE is to obtain a deep understanding of the baryonic properties and the interplay of gravitational and non-gravitational processes, and to constrain the biases in cluster mass determination. CHEX-MATE serves as a high-quality, multi-wavelength laboratory for astrophysics in the cluster regime, and as a calibration sample and anchor for large surveys. CHEX-MATE was allocated 3Ms of observation time as a Multi-Year Heritage Programme with {\it XMM-Newton}, and targets a homogeneous and complete sample of 118 galaxy clusters selected by {\it Planck} in SZ (very near to mass selection) with a high signal-to-noise ratio $(\text{S/N})_\text{MMF3} > 6.5$ (see Fig.~\ref{fig_z_M500_members}). Here and in the following, the subscript MMF3 denotes quantities based on or derived from the multi-frequency matched filter MMF3 catalogues \citep{planck_2013_XXIX}.
CHEX-MATE includes a census of the population at the most recent time (Tier 1: {\it Planck} clusters selected in the northern hemisphere with $(\text{S/N})_\text{MMF3} > 6.5$, and $0.05<z<0.2$) and the most massive systems to have formed (Tier 2: $(\text{S/N})_\text{MMF3} > 6.5$, $z<0.6, M_\text{MMF3, 500c} > 7.25\times10^{14} M_\odot$). The Tier 2 subsample contains massive clusters by design, whereas due to the small local volume, clusters in Tier 1 are mostly of low mass, spanning a range of $2\times10^{14} M_\odot \la M_\text{MMF3, 500c} \la  9\times10^{14}  M_\odot$. There are four clusters common to the two tiers.

Investigating baryonic physics from various angles will allow us to probe the mass scale on a well-defined and minimally biased sample. Deep X-ray observations will provide HE mass estimates precise to $15$-$20\,\%$. To realise the full potential of the high-fidelity Heritage data, a multi-wavelength campaign is underway. Mass measurements accurate to $\sim30\%$ per cluster (including the intrinsic scatter from projection effects) can be achieved thanks to WL, or galaxy dynamics \citep{ume+al16b,euclid_pre_XLII_ser+al24}. When the full set of multi-wavelength data is in hand, it will be possible to compare a wide range of observables and derived proxies ---for example, from X-ray luminosity, the SZ effect, or optical richness with WL, HE, or dynamical mass estimates--- and to assess the bias and
scatter in these different probes. Combining various mass proxies with different dependencies will be key to accounting for systematic uncertainties.

In Sect.~\ref{sec_halo}, we introduce the halo model, followed by Sect.~\ref{sec_obs} where we describe the data. In Sect.~\ref{sec_memb}, we detail the method used to select members and measure velocity dispersion. Estimates of redshift and velocity dispersion are compared to values from the literature in Sect.~\ref{sec_reds}. In Sect.~\ref{sec_mass}, we discuss and validate the measurements of the CHEX-MATE dynamical masses, followed by Sect.~\ref{sec_plan}, where we quantify the bias of the {\it Planck} masses. In Sect.~\ref{sec_syst}, we review possible systematic errors. In Section~\ref{sec_conc}, we outline some final considerations. In Appendix~\ref{sec_nfw}, we detail the halo model. Estimated redshifts, masses, and velocity dispersions are listed in Appendix~\ref{sec_tab_masses}. In Appendix~\ref{sec_efosc2} we report the results of the first run of our observational campaign.

As noted above, MMF3 denotes masses and quantities based on the multi-frequency matched filter MMF3 \citep{planck_2013_XXIX}. The notation SZ or PSZ2 indicates results extracted from the PSZ2 union catalogue \citep{planck_2015_XXVII}.

If needed, we adopt a flat $\Lambda$CDM model with (total) present day matter density parameter $\Omega_\text{m}=0.30$ and Hubble constant $H_0=70\,\text{km~s~Mpc}^{-1}$. As usual, $H(z)$ is the redshift dependent Hubble parameter, $E_z\equiv H(z)/H_0$, and $h = H_0 / (100\,\text{km~s~Mpc}^{-1})$.

We use $O_{\Delta\text{i}}$, with $\text{i}= \text{c}$ or $\text{i}= \text{m}$, to denote a cluster property, $O$, measured within the radius $r_{\Delta\text{i}}$, which encloses a mean over-density of $\Delta$ times the critical ($\text{i}= \text{c}$) or the matter ($\text{i}= \text{m}$) density at the cluster redshift, with $\rho_\mathrm{c}(z)\equiv3H^2(z)/(8\pi G)$. In the following, unless stated otherwise, we consider quantities within the overdensity radius $r_\text{200c}$. 

The notation `$\log$' is the logarithm in base 10, and `$\ln$' is the natural logarithm. 
Unless stated otherwise, the central location and scale of distributions are computed as $C_\text{BI}$ (Centre BIweight) and $S_\text{BI}$ \citep[Scale BIweight,][]{bee+al90}. Probabilities are computed considering the marginalised posterior distributions.

Updated CHEX-MATE products and catalogues can be recovered from \url{http://xmm-heritage.oas.inaf.it/}. Products related to the dynamical analysis, including redshifts and masses, are also stored at \url{http://pico.oabo.inaf.it/\textasciitilde sereno/CHEX-MATE/sigma/}.




\section{Halo model}
\label{sec_halo}

Gravity is the main driver of cluster formation and evolution. All cluster properties, to some degree, can be inferred from the cluster mass $M_\Delta$ \citep{kai84,voi05}. The typical concentration of the halo is related to the mass \citep{di+jo19}, and the scale radius of the anisotropy velocity profile, $\beta_\sigma$, is related to the scale radius of the mass density profile \citep{mam+al10}. 

Here, we adopt as reference a simple, one-parameter model for the halo, where all properties are derived from the mass, $M_{200\text{c}}$. 
We assume that the tracers, that is, the galaxies, follow the mass distribution. The inference of the spatial distribution of the tracers in our cluster sample is hampered by the heterogeneous nature of the datasets, collected from a variety of sources, whose completeness and homogeneity is difficult to assess working at the catalogue level. On the other hand, our working hypothesis is well tested \citep{mam+al13}.

We model the cluster density profile as a Navarro-Frenk-White profile \citep[NFW,][]{nfw96}. Observational results and theoretical predictions agree that  concentrations $c_{200\text{c}}$ are higher for lower mass haloes, and were smaller at early times \citep{ser+al17_psz2lens}, with a possible upturn at very high masses and redshifts. Here, we adopt the mass-concentration relation proposed in \citet{di+jo19}.

We consider orbits of cluster galaxies which are isotropic near the centre and more radial outside \citep{biv+al21}, and we model the velocity anisotropy profile, $\beta_\sigma$, as in \citet[ML,][]{ma+lo05}, with the anisotropy radius fixed to the scale radius of the mass density profile.

The velocity dispersion within an aperture, $\sigma_\text{ap}(R)$, is fully determined by the mass and anisotropy profiles \citep{lo+ma01},
\begin{equation}
\label{eq_kin_mass_1}
\sigma_\text{ap}(R) = \sigma_\text{ap}(R | M(r), \beta_\sigma(r)),
\end{equation}
where $r$ is the three-dimensional radius and $R$ is the projected radius.
Given a measured velocity dispersion within a given aperture radius, as in the left-hand side in Eq.~(\ref{eq_kin_mass_1}), and a one-parameter model for mass and anisotropy profiles (the right-hand side), Eq.~(\ref{eq_kin_mass_1}) can be inverted to estimate the halo mass. The model is uniquely determined by the mass, and we can, in turn, estimate, for example, the halo concentration, or the 1D velocity dispersion. The model is detailed in Appendix~\ref{sec_nfw}.


\section{Archival data and observations}
\label{sec_obs}

\begin{table}
\caption{
Clusters with a minimum number of candidate member galaxies.}
\label{tab_archives}
\centering
\resizebox{\hsize}{!} {
\begin{tabular}[c]{l  r r r r r}
        \hline
        \noalign{\smallskip}  
        $N_\text{candidates}$  & All &   NED+SDSS+DESI &    SDSS  & DESI &  NED-     \\
                        \noalign{\smallskip}  
        \hline
        \noalign{\smallskip}  
        $\ge ~~1$  & 111  & 106 & 68 & 5 & 100   \\
        $\ge ~~5$  & 106  & 95 & 67 & 3 & 76   \\
        $\ge 10$   & 103 & 90 & 66 & 3 & 68   \\
        $\ge 25$   & 89 & 85 & 55 & 3 & 67   \\
        $\ge 50$   & 79 & 76 & 35 & 3 & 61   \\
        $\ge 100$   & 68 & 66 & 19 & 0 & 53   \\
        \end{tabular}
       }
\tablefoot{
We report the total number of CHEX-MATE clusters with at least $N_\text{candidates}$ (col. 1) per archive in the query area within $3\,\theta_\text{MMF3,500c}$ from the X-ray peak, and within a search window of $\pm 12000~\text{km~s}^{-1}$ in velocity rest-frame from the {\it Planck} estimate. The columns All, NED+SDSS+DESI, SDSS, DESI, NED- report the number of unique candidates from all the datasets, from either NED or SDSS or DESI, from SDSS only, from DESI only, or from NED (without entries in common with SDSS or DESI) only, respectively.
}
\end{table}

Some CHEX-MATE clusters, mostly the more massive Tier 2 clusters, already have one or more measurements of their galaxy velocity dispersion published in the literature. However, those are generally estimated with heterogeneous methodologies or member selections. This may add supplementary scatter or systematic uncertainties to the measurements. Furthermore, a few other clusters, as, for example, newly discovered, low-mass, local clusters from Tier 1, may lack previous analyses.

We reprocess the full sample with a homogeneous analysis. We start by compiling the galaxy redshift catalogue, and apply the same methodology to the whole dataset. For this purpose, we compiled spectroscopic catalogues for each cluster from archival data when available, and started an observational campaign to map the remaining poorly covered clusters.

We query existing spectroscopic redshifts in a search area centred on the location of the X-ray peak of the cluster \citep{chexmate_bar+al23}. The search radius of the cylinder is chosen to encompass a radius of $ 3 r_\text{MMF3, 500c}$ (corresponding to nearly $2r_\text{200c}$) from the cluster centre, where $r_\text{MMF3, 500c}$ is derived from $M_\text{MMF3, 500c}$. 

\subsection{Archival data}

We use public resources to retrieve available spectroscopic redshifts for each CHEX-MATE cluster. The Sloan Digital Sky Surveys (SDSS) is now in stage V \citep{sdss_V_kol+19}. We query the eighteenth data release (DR18) using SkyServer.\footnote{\url{https://SkyServer.sdss.org/dr18/}} We select galaxies from the \texttt{SpecPhoto} catalogue, which is a pre-computed join of spectroscopy and photometry.

The Dark Energy Spectroscopic Instrument (DESI) completed its five-month Survey Validation (SV) in May 2021 \citep{desi_edr+24}. The Early Data Release includes spectral information from 3207569 objects.\footnote{\url{https://data.desi.lbl.gov/doc/releases/edr/vac/zcat/}} We select objects classified as galaxies with high quality redshifts (\texttt{ZWARN}=0).

The NASA Extragalactic Database (NED) comprises large survey releases or targeted observations and allows us to combine data from various campaigns performed at different times. Since we query the SDSS archive separately, SDSS sources in NED (which includes DR6) are filtered out. From NED queries, we keep galaxies with redshifts flagged as spectroscopic, whereas we exclude either sources other than galaxies or galaxies whose redshift is flagged as photometric. For galaxies with redshifts flagged as unknown, we keep only objects with measured uncertainties on redshifts, and select only those with a redshift error smaller than $\delta z = 0.001$, to prevent including photometric redshifts. 

In the following, we denote as `SDSS' the sample of candidates from SDSS DR18, as `DESI' the sample from DESI-SV, as `NED-' the sample from NED excised of SDSS or DESI sources, and as `NED+SDSS+DESI' the matched samples of unique sources.  Duplicates are matched in a searching radius of 1\arcsec and eliminated, giving priority to objects from the homogeneous SDSS, or DESI archives.
In summary, 106 clusters are spectroscopically covered by either SDSS, DESI, or NED (see Table~\ref{tab_archives}).

\subsection{Other datasets}
\label{sec_obs_other}

We complement the public, large samples with other catalogues based on previous studies.
The Arizona Cluster Redshift Survey \citep[ACReS,][]{haines+al13,haines+al15} is a spectroscopic survey of 30 massive clusters at $0.15{<}z{<}0.30$  carried out using the Hectospec multi-fibre spectrograph on the 6.5m MMT (Multiple Mirror Telescope). Targets were selected from wide-field ($52^{\prime}{\times}52^{\prime}$) $J$- and $K$-band imaging acquired using the WFCAM instrument on the 3.8m United Kingdom Infrared Telescope, as those having $J-K$ colours consistent with being at the cluster redshift \citep{haines+al09}. This simple approach permits cluster galaxies to be selected by stellar mass and without bias with respect to their star-formation history. The capacity of the Hectospec instrument to place galaxies on 300 fibres anywhere over its 1-degree diameter field-of-view permitted cluster galaxies to be targeted efficiently out to $\sim 2$-$3\,r_\text{200c}$.  
The ACReS data-base covers 12 CHEX-MATE clusters (PSZ2~G049.22+30.87, PSZ2~G067.17+67.46, PSZ2~G124.20-36.48, PSZ2~G107.10+65.32, PSZ2~G340.36+60.58, PSZ2~G072.62+41.46, PSZ2~G159.91-73.50, PSZ2~G313.33+61.13, PSZ2~G186.37+37.26, PSZ2~G073.97-27.82, PSZ2~G092.71+73.46, PSZ2~G187.53+21.92) for a total of 4026 galaxies in the defined search space.

{\it Planck} clusters were targeted with two multi-year observational programmes at Roque de los Muchachos Observatory in La Palma, Spain, \citep{bar+al20,fer+al21,agu+al22} to optically validate and characterise the first \citep[PSZ1,][]{planck_2013_XXIX}, or the second (PSZ2) {\it Planck} cluster sample. The programmes carried out multi-object spectroscopy using the DoLoRes spectrograph at the 3.5m TNG (Telescopio Nazionale Galileo), or OSIRIS at the 10.4m GTC (Gran Telescopio Canarias), in order to obtain radial velocities of a large set of cluster members. In this work, we consider 4 clusters, also included in the CHEX-MATE sample (PSZ2~G283.91+73.87, PSZ2~G204.10+16.51, PSZ2~G087.03-57.37, PSZ2~G206.45+13.89), for a total of 221 candidate galaxy members with spectroscopic redshifts. 

\citet{noo+al20} studied the environmental dependence of X-ray AGN activity at $z \sim 0.4$. Follow-up VIMOS (VIsible Multi-Object Spectrograph) optical spectroscopy was used to determine  cluster members. The sample covers three CHEX-MATE clusters (PSZ2~G324.04+48.79, PSZ2~G205.93-39.46, and PSZ2~G057.25-45.34) for a total of 278 candidates in the search area.

PSZ2~G004.45-19.55 is a strong-lensing cluster hosting a radio relic \citep{sif+al14,alb+al17}. The cluster was targeted for spectroscopic follow-up \citep{sif+al14,alb+al17}. The data consist of 35 redshifts and were made available to the community by the authors.\footnote{\url{https://github.com/cristobal-sifon/data}}

\subsection{Observational campaign}

The mass calibration goal can be achieved by combining archival data, ongoing surveys, and dedicated proposals. A multi-wavelength campaign is ongoing to cover the full CHEX-MATE sample. The Heritage consortium was awarded $\sim 39h$ with OmegaCam at VST (VLT Survey Telescope) from semester S19B to S21B, $\sim 21h$ at MegaCam at CFHT (Canadian-France-Hawaii Telescope) in semesters S19B and S20A, and $\sim 25h$ at HSC (Hyper Suprime-Cam) at the Subaru telescope in semesters S19B and S20A for multi-band photometry.  

In order to obtain measurements of velocity dispersions on the whole CHEX-MATE sample, an observational campaign has been led at ESO (PIs S. Maurogordato and M. Sereno) and at Roque de los Muchachos Observatory (PI R. Barrena) facilities to cover the clusters with no or few spectroscopically confirmed members. 

Multi-object spectroscopy of scarcely covered clusters was performed with ESO facilities in 2022-2023 (runs 0110.A-4192 and 0111.1-0186) with 19.15 nights allocated with EFOSC2 at NTT (New Technology Telescope), and 13.3 hours with FORS2 at VLT-UT1 (Very Large Telescope Unit 1) to observe nearby or distant clusters, respectively. 
Eight clusters were observed with EFOSC2 at NTT  
(PSZ2~G263.68-22.55, PSZ2~G346.61+35.06, PSZ2~G266.83+25.08, PSZ2~G340.94+35.07, PSZ2~G325.70+17.34, PSZ2~G062.46-21.35, PSZ2~G106.87-83.23, PSZ2~G259.98-63.43)
and 3 clusters with FORS2 at VLT (PSZ2~G243.15-73.84, PSZ2~G004.45-19.55, and PSZ2~G155.27-68.42).

We present hereafter the results of the ESO campaign with EFOSC2 at NTT, whose observations and data reduction were recently completed. Due to a EFOSC2 punching machine failure, the runs originally scheduled in multi-object mode were performed in long slit mode, resulting in a lower efficiency than that anticipated. Spectra are extracted and calibrated using a pipeline based on tasks available in \texttt{PyRAF}.\footnote{\url{https://iraf-community.github.io/pyraf.html}}. Redshifts are measured by cross-correlating galaxy spectra with templates using the \texttt{xcsao} task in the \texttt{rvsao} package \citep{kur+al92}.
The number of measured redshifts is between 2 and 26 per cluster with a median of 20 candidates per cluster, for a total of 164 redshifts measured in the cluster fields of view. The redshifts are reported in Appendix~\ref{sec_efosc2}


\section{Membership}
\label{sec_memb}

Galaxy clusters lie at the highest density nodes of the cosmic web. One of the main difficulties in estimating the velocity dispersion is the identification of galaxies which are real members of the cluster.
The only available component of the 3D velocity field is its projection along the line-of-sight. The observed phase-space distribution differs from the real one due to projection effects, and galaxies at large proper distances from the cluster centre that are physically located outside the virialised halo may appear to lie at small projected radii.

The fraction of contaminants, called interlopers, can be as large as $\sim40\, \%$ \citep{mam+al10}, and it can significantly bias the velocity dispersion to higher values \citep{woj+al07}. Numerous approaches have been developed to reduce the fraction of interlopers \citep{old+al14}, from the pioneering 3-sigma clipping method \citep{ya+vi77}, to methods using the projected phase-space distribution, such as the shifting gapper \citep{fad+al96, gi+me01, bi+gi03, bay+al16} or the CLEAN method \citep{mam+al13}. We eliminate interlopers with an adaptation of the CLEAN method, which we detail in the following subsections.

\subsection{Estimators}

Robust summary estimators of the velocity distribution have been proposed \citep{bee+al90}. 
Here, to measure location and scale of the velocity distribution, we use the bi-weighted estimators \citep{bee+al90}, which are very robust even for the very low number of galaxy redshifts per cluster of some CHEX-MATE clusters. Uncertainties are obtained by bootstrap resampling with replacement of the velocity distribution for each cluster. 

Our collection of data is heterogeneous, and the spatial coverage of the candidate member galaxies is not uniform, or, in some occasions, it can be limited to the inner regions. We rely on some working hypotheses. Firstly, we assume that the galaxies follow the mass distribution.
Secondly, we derive an effective aperture radius, $R_\text{ap}$, as two times the mean radial distance of the candidates from the centre, as for galaxies following a near isothermal density profile. When we measure the velocity dispersion of the confirmed galaxies, we interpret it as an aperture velocity distribution measured within $R_\text{ap}$. In this way, we correct for not uniform, or incomplete coverage.

\subsection{Initial candidates}

Candidate member galaxies are selected in phase space. The radius and velocity search windows are chosen to properly map the velocity distribution. We select galaxies  in a cylinder centred on the location of the X-ray peak of the cluster \citep{chexmate_bar+al23}.

The redshift selection is performed using a search window of $\pm12000~\text{km~s}^{-1}$ in rest-frame velocity, corresponding to $\Delta z \sim 0.04~(1+z)$ around the redshift of the cluster. For this first step, we take the redshifts from the {\it Planck} PSZ2 catalogue as revised in \citet{chexmate+al21}.

The rest-frame velocity window is by design poorly constraining and allows the algorithm to readjust in case of an inaccurate initial redshift estimate, as might be the case, for example, for a {\it Planck} cluster validated with a few photometric redshifts, or for multi-modal velocity distributions. Since {\it Planck} redshifts are in general accurate, we verify that a smaller search window in rest-frame velocity (e.g., $\pm6000~\text{km~s}^{-1}$) would only have impacted three clusters, which are discussed in the following as bimodal or misplaced clusters.

The first-step selection process ensures a high membership completeness at the detriment of purity, which is increased by the second-step interloper rejection (see Sect.~\ref{sec_clea}). Some summary statistics for the initial candidates are reported in Table~\ref{tab_archives}. There are 106 (103) clusters with at least 5 (10) initial candidates, for a total number of 23292 candidate members.

\subsection{Cleaning}
\label{sec_clea}

\begin{figure}
\resizebox{\hsize}{!}{\includegraphics{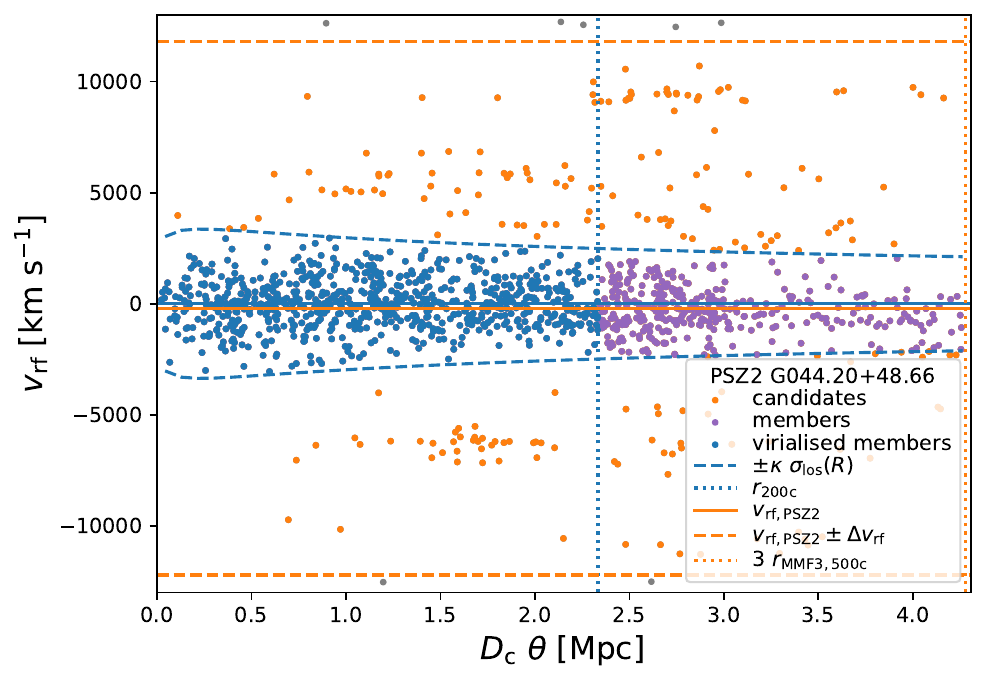}} 
\caption{Projected phase-space diagram of the galaxies in the field of PSZ2~G044.20+48.66 (Abell~2142) and member selection. The distance $D_\text{c}$ is the angular diamater distance to the cluster.  This is the CHEX-MATE cluster with the greatest number of confirmed redshifts in the virialised region. The initial candidates within a radial distance of $3\theta_\text{MMF3, 500c}$ from the X-ray peak, and within a window in velocity of $\pm \Delta_\text{vrf, PSZ2}$ with 
$\Delta_\text{vrf, PSZ2}=12000~{\rm km~s^{-1}}$
from the nominal PSZ2 estimate, are shown in orange. The virialised members (in blue) are selected within a radial distance of $r_\text{200c}$ and a rest-frame velocity window of $\pm\kappa~\sigma_\text{los}(R)$, with $\kappa=2.7$, where $\sigma_\text{los}$ is the velocity dispersion along the line-of-sight at projected radius $R$. 
Radial distances are shown assuming the final cluster redshift.
}
\label{fig_phase_space_PSZ2G044.20+48.66_r200c_rap20_Deltavrf12000}
\end{figure}

CLEAN selects cluster members based on their location in projected phase-space, $R$--$v_\text{rf}$, where $R$ is the projected radial distance from the cluster centre and $v_\text{rf} = c (z - z_\text{loc})/(1 + z_\text{loc})$ is the rest-frame velocity, with the cluster redshift $z_\text{loc}$ estimated on the identified cluster members \citep{mam+al13}. For the cluster centre, we keep the position of the X-ray peak. Galaxies whose peculiar rest-frame velocities lie within $\pm \kappa~\sigma_\text{los}(R)$ from the central velocity are identified as members with an iterative method. Here, we describe our adaptation of the method.

As a first estimate, we measure the redshift location and the velocity dispersion of the candidate member galaxies within $\theta_\text{MMF3, 500}$ from the X-ray peak. The relatively small aperture is meant to favour selection of galaxies in the main clump and filter out substructures at larger radii which could significantly bias high the initial estimate of the velocity dispersion. However, we verify that our procedure is very stable with respect to different choices.

Given the measured velocity dispersion within the estimated aperture radius, we use the halo model described in Sect.~\ref{sec_halo} and Appendix~\ref{sec_nfw} to invert Eq.~(\ref{eq_kin_mass_1}) and solve for the halo mass $M_\text{200c}$
under the hypothesis that the tracers follow the mass distribution.
From the mass, we can compute the virial radius and the expected line-of-sight velocity distribution. The membership is assigned by selecting galaxies with rest-frame velocities smaller (in absolute value) than $\kappa \times \sigma_\text{los} (R)$. We adopt $\kappa = 2.7$, which preserves the LOS velocity dispersion profile \citep{mam+al10,mam+al13}. The radius $r_\text{200c}$, as well as the radial distances from the centre in proper length, are recomputed at each step of the iterative process.

We compute the velocity centre and dispersion of the filtered sample. Only members within $r_\text{200c}$ from the centre are considered for this step. We iterate until convergence, checking that the velocity dispersion had changed by no more than 0.05 per cent. The result of the cleaning procedure is shown in Fig.~\ref{fig_phase_space_PSZ2G044.20+48.66_r200c_rap20_Deltavrf12000} for PSZ2~G044.20+48.66, aka Abell 2142, the CHEX-MATE cluster with the most confirmed members in the virialised region (i.e., 778 members within $r_\text{200c}$). Even though radial distances are recomputed at each iteration step with the newly determined redshift location, in the plot, for better visualisation, we consider the final redshift at all steps.

If we find the velocity distribution of the selected members to be multi-modal after convergence, we split the distribution in clumps and re-run the cleaning on each clump. We detail the procedure below.



\subsection{Substructures}
\label{sec_subs}

\begin{figure}
\resizebox{\hsize}{!}{\includegraphics{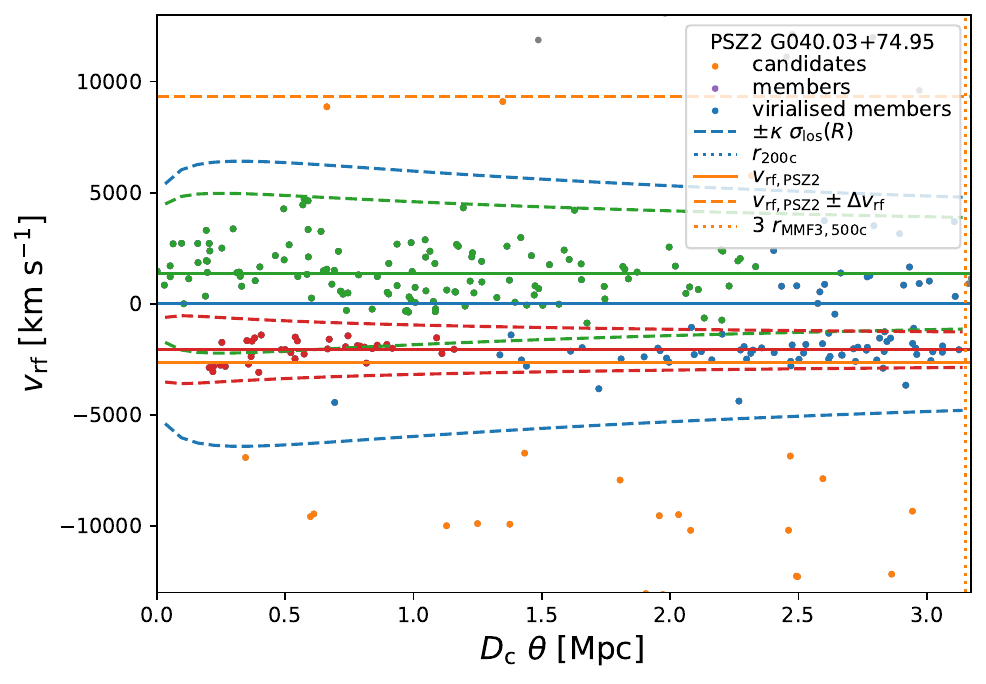}} 
\caption{Projected phase-space diagram of the galaxies in the field of PSZ2~G040.03+74.95 (Abell 1831) and member selection. The distribution is clearly bimodal. The initial candidates within a radial distance of $3\theta_\text{MMF3, 500c}$ from the X-ray peak, and within a window in velocity of $\pm \Delta_\text{vrf, PSZ2}$ with 
$\Delta_\text{vrf, PSZ2}=12000~{\rm km~s^{-1}}$
from the nominal PSZ2 velocity, are shown in orange. The candidate virialised members that passed the first cleaning procedure (in blue) are selected within a radial distance of $r_\text{200c}$ and a rest-frame velocity window of $\pm\kappa~\sigma_\text{los}(R)$, with $\kappa=2.7$. In green and red, we show the result of the cleaning procedure on each sub-clump. For visualisation purposes, radial distances are here shown assuming the same cluster redshift (the result of the first cleaning procedure) for all galaxies.
For this cluster, all members selected at the first step (violet) are also virialised members (blue, which is superimposed on violet).
}
\label{fig_phase_space_PSZ2G040.03+74.95_r200c_rap20_Deltavrf12000}
\end{figure}

The cleaning procedure in Sect.~\ref{sec_clea} is robust against irregular or bimodal clusters, and it usually identifies the main halo, but some substructures can still persist. The original CLEAN method selects the main clump before the iterative process \citep{mam+al13}, but here we prefer to check for substructures in a second step due to the uneven coverage of candidate members.

The regularity of the velocity distribution of the filtered members is checked with the dip test \citep{ha+ha85}, which measures multimodality by the maximum difference between the empirical distribution function and the unimodal distribution function that minimises that maximum difference. If the null hypothesis of unimodality is rejected at 3-$\sigma$, that is, the $p$-value of the dip statistics on the filtered redshift distribution is smaller than $(1-0.9973)$, we analyse the two larger clumps.

Galaxies are associated with clumps with the relative velocity gap technique \citep{wa+th76}, with gapper coefficient $C = 4$ \citep{gir+al93,mam+al13}. If the original data distribution of the redshifts was Gaussian with 100 sampled items, we expected to find $\sim 99.7\,\%$ of the rescaled weighted gaps to be less than $C = 4$.

For each clump, we repeat the iterative cleaning (see Sect.~\ref{sec_clea}), now considering as first guess the distribution of all galaxies in each clump. We identify the {\it Planck} cluster as the most massive clump. The cleaning procedure for a cluster that is bimodal in phase space is illustrated in Fig.~\ref{fig_phase_space_PSZ2G040.03+74.95_r200c_rap20_Deltavrf12000}.

We find only two cases with significant substructures after the first cleaning procedure. The cluster PSZ2~G040.03+74.95 at $z = 0.076$, aka Abell 1831 or RXC J1359.2+2758, shows two peaks in the velocity distribution \citep[see our Fig.~\ref{fig_phase_space_PSZ2G040.03+74.95_r200c_rap20_Deltavrf12000};][]{ko+ko10,mau+al97}. It is part of a supercluster with high multiplicity \citep{cho+al14}. Abell 1831 is a visual superposition of the two clusters Abell 1831A and Abell 1831B, nearly aligned along the line of sight, but at different redshifts, $z \sim 0.063$ and $z \sim 0.075$, respectively \citep{ko+ko10,lak+al14}. Abell 1831B is a much richer cluster than A1831A, the latter being a poor foreground cluster. The X-ray emission is mostly associated with Abell~1831B, and its peak, as derived in \citet{chexmate_bar+al23}, coincides with the galaxy CGCG~162-041 at $z \sim 0.076$ identified in \citet{mau+al97}. According to our results, the main clump, $M_\text{200c} = (20 \pm 4) \times 10^{14} M_\odot$, is much more massive than the foreground cluster, $M_\text{200c} = (1.4 \pm 0.5) \times 10^{14} M_\odot$, in line with previous results \citep{ko+ko10}. The subsystem Abell 1831A was the first to be covered with spectroscopic measurements, and its redshift was usually adopted as the redshift of the entire A1831 cluster \citep{ko+ko10}, as also done for the {\it Planck} catalogue.

The other bimodal cluster is PSZ2~G218.81+35.51, aka MS 0906.5+1110 or RXC (MCXC) J0909.1+1059, at $z=0.178$. The cluster is coupled with a large region of diffuse extended emission $\sim 2.3\arcmin$ away, which corresponds to cluster Abell 750 at $z=0.164$ \citep{log+al22}. The masses of the two systems are comparable, $M_\text{200c} = (5.1 \pm 0.9) \times 10^{14} M_\odot$ for the CHEX-MATE cluster versus $M_\text{200c} = (3.4 \pm 0.6) \times 10^{14} M_\odot$ for the companion.

\subsection{Confirmed members}

\begin{figure}
\resizebox{\hsize}{!}{\includegraphics{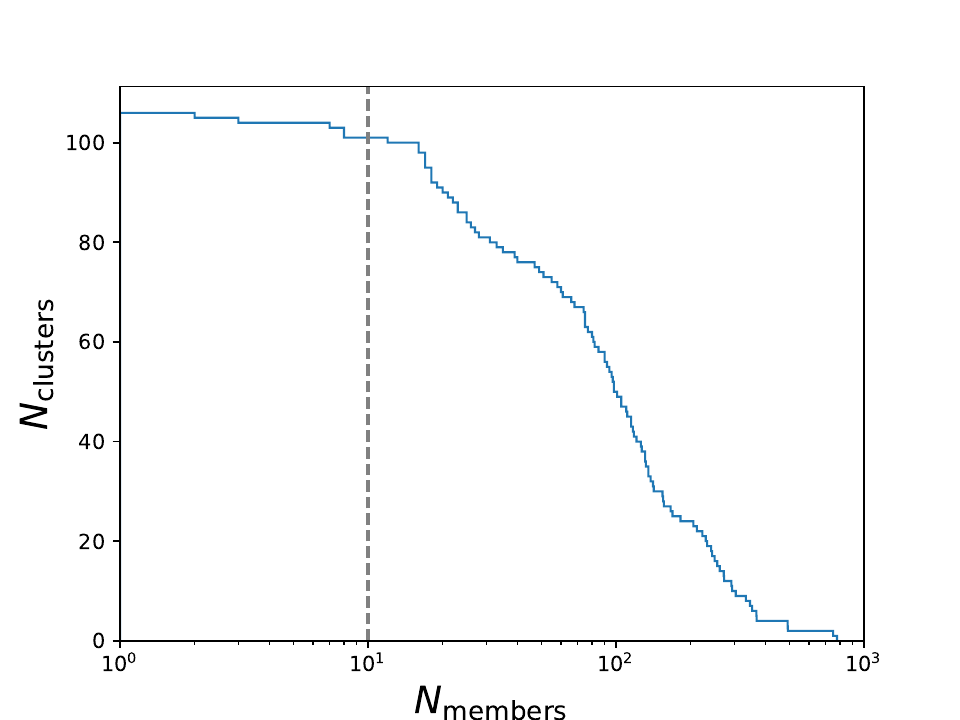}}
\caption{Reverse cumulative distribution of the number of confirmed member galaxies with spectroscopic redshifts within $r_\text{200c}$ per cluster.}
\label{fig_N_members}
\end{figure}

\begin{figure}
\resizebox{\hsize}{!}{\includegraphics{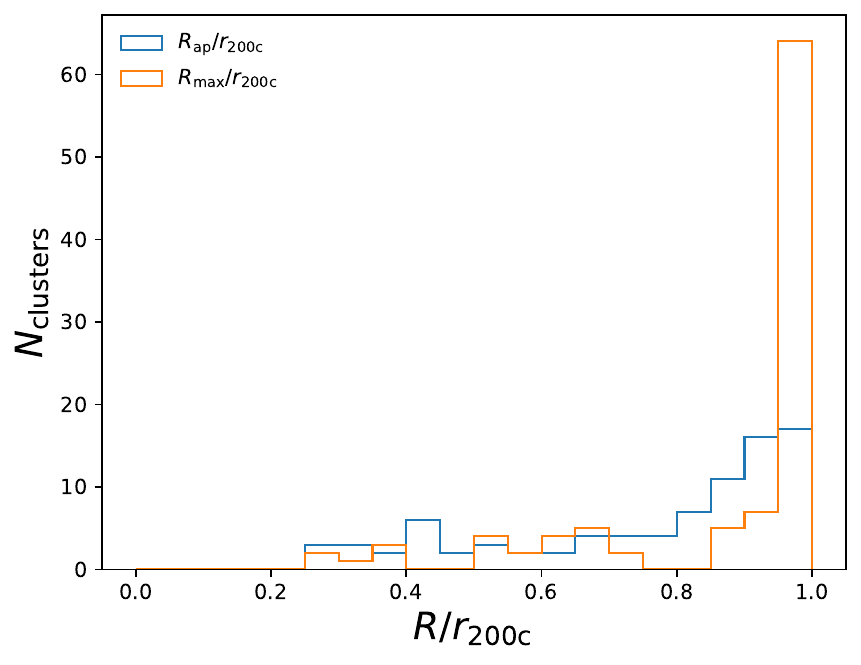}}
\caption{Distribution of the maximum radius, $R_\text{max}$, and of the effective aperture radius, $R_\text{ap}$, of the confirmed member galaxies within $r_\text{200c}$. Radii are in units of $r_\text{200c}$.}
\label{fig_hist_R_ap_R_max}
\end{figure}

\begin{figure}
\resizebox{\hsize}{!}{\includegraphics{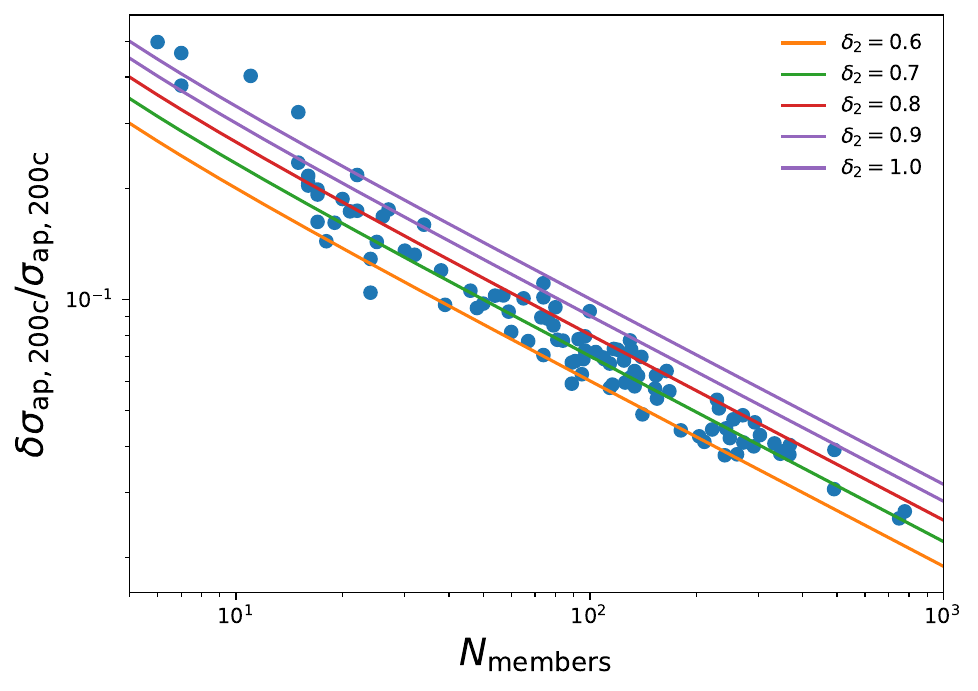}}
\caption{Precision in the estimate of the aperture velocity dispersion within $r_\text{200c}$ as a function of the confirmed members within the aperture. The precision scales approximately as $\delta_2 / \sqrt{N_\text{members}-1}$.}
\label{fig_delta_sigma}
\end{figure}

The number of confirmed members per cluster varies significantly (see Fig.~\ref{fig_N_members}). We find at least two confirmed member galaxies within $r_\text{200c}$ for 105 out of 118 CHEX-MATE clusters. For 7 clusters, there are no candidate galaxies in the initial search space. For four clusters, there is just one candidate. There is one cluster with 5 candidates but no confirmed members, and one cluster with 9 candidates but just one confirmed member.

The median number of confirmed members within $r_\text{200c}$ is 82.5 for the full CHEX-MATE sample, 96 for clusters with at least two confirmed members, and 97 for clusters with at least 10 confirmed members. The cluster with the most confirmed members is PSZ2~G044.20+48.66, aka Abell 2142 or RXC J1558.3+2713, with 778 (see Fig.~\ref{fig_phase_space_PSZ2G044.20+48.66_r200c_rap20_Deltavrf12000}).

The virial region is usually well sampled (see Fig.~\ref{fig_hist_R_ap_R_max}). For most of the clusters, the radial distribution of confirmed members reaches or extends beyond $r_\text{200c}$. For 77 out of 105 clusters with measured velocity dispersion, $R_\text{max} / r_\text{200c} \ge 0.9$, where $R_\text{max}$ is the maximum radial distance. The distributions are also homogeneous, with $R_\text{ap} / R_\text{max} \ge 0.9$ for 64 clusters out of 105 clusters.

For a reliable estimate of the velocity dispersion, we require at least 10 members with spectroscopic redshifts within $r_\text{200c}$ \citep{bee+al90,gir+al93}, a cut passed by 101 clusters. Even though a reliable estimate could be based on just 5 members if randomly drawn from the parent sample \citep{bee+al90}, selection biases,
for example in luminosity \citep{biv+al92,old+al14}, could affect our heterogeneous dataset, and a threshold of 10 members is more conservative \citep{gir+al93}. 

The velocity dispersions are well recovered. For clusters with at least 10 members, we estimate the aperture velocity dispersion within $r_\text{200c}$ with a precision of $69\pm36~\text{km~s}^{-1}$ ($7\pm 4\,\%$) for the sample. The precision scales with the number of members galaxies, with $\delta \sigma_\text{ap} \propto 1 / \sqrt{N_\text{members} -1}$ (see Fig.~\ref{fig_delta_sigma}). 

The clusters with measured $\sigma$ are highly representative of the parent sample (see Fig.~\ref{fig_z_M500_members}). The Kolmogorov-Smirnov test gives $p$-values in excess of $0.999$ for both the mass or redshift distributions of the full sample. The sample includes 55 (50) out of 61 Tier 1 (2) clusters.


\section{Redshift and velocity dispersion}
\label{sec_reds}

The measurements of location and scale of the velocity distribution are performed on the clean sample, that is, the confirmed member galaxies with spectroscopic redshift within $r_\text{200c}$. The results of our analysis are presented in Table~\ref{tab_masses}.

\subsection{Cluster redshifts}

\begin{figure}
\resizebox{\hsize}{!}{\includegraphics{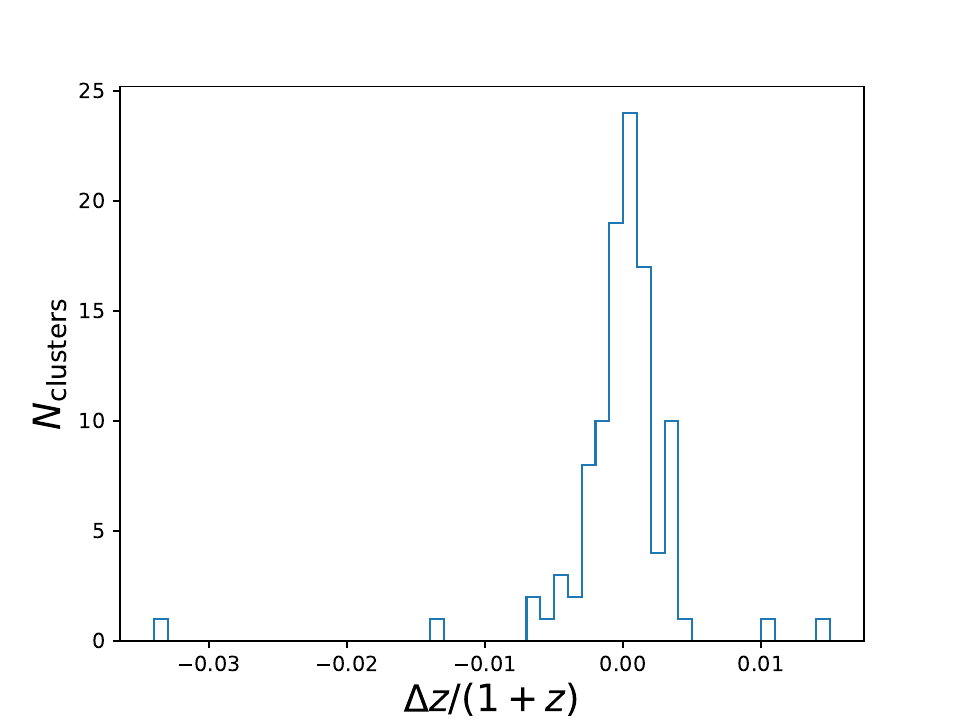}} \\
\caption{Normalised redshift difference between the cluster redshift from the PSZ2 catalogue and our estimates.}
\label{fig_Delta_z}
\end{figure}

Here, we compare the estimated cluster redshifts with the estimates from the PSZ2 catalogue. We limit the comparison to the 105 CHEX-MATE clusters with at least two confirmed members. Our analysis confirms the reliability of the PSZ2 redshifts (see Fig.~\ref{fig_Delta_z}), with only four PSZ2 redshifts differing from our estimates by more than $0.01 \times (1+z)$. We discuss the exceptions in the following.

The cluster PSZ2~G062.46-21.35, aka RXC J2104.9+1401 or ACT-CL J2104.8+1401, was assigned a redshift of $z_\text{PSZ2}=0.1615$ in the PSZ2 catalogue. It is one of 4195 optically confirmed, SZ selected galaxy clusters detected with signal-to-noise ratio larger than 4 by the Atacama Cosmology Telescope \citep{act_hil+al21}. The {\it Planck} redshift estimate is based on only one bright galaxy, likely assumed to be the BCG. Our EFOSC2 observations confirm the galaxy redshift, but show that it is a foreground galaxy. We update the cluster redshift to $z=0.202\pm0.001$ based on 15 confirmed members within $r_\text{200c}$.

The cluster PSZ2~G004.45-19.55 is a newly confirmed {\it Planck} cluster at $z_\text{PSZ2}=0.54$. 
\citet{sif+al14} and \citet{alb+al17} updated the redshift to $z = 0.52$ with spectroscopic data from dedicated follow-up programmes. For our determination, we use the same data (see Sect.~\ref{sec_obs}) and obtain a consistent result ($z = 0.519\pm0.002$) based on 16 confirmed members.

The cluster PSZ2~G143.26+65.24, aka RM~J115914.9+494748.4, at $z_\text{PSZ2}=0.36321$ was validated in the {\it Planck} catalogue with redMaPPer, whose catalogue reports a photometric redshift of $z_\lambda = 0.36321 \pm 0.01366$ and a spectroscopic redshift of $z= 0.350117$. Our estimate of $z=0.349\pm0.001$ is based on 25 confirmed members. It is in full agreement with the spectroscopic redMaPPer redshift and consistent within $1$-$\sigma$ with the photometric estimate used for the ${\it Planck}$ catalogue.

The cluster PSZ2~G040.03+74.95, aka Abell 1831, is a bimodal system already discussed in Sect.~\ref{sec_subs}. In the {\it Planck} catalogue, the cluster is associated with the foreground clump A1831A, whereas we identify A1831B as the main clump at $z=0.0756\pm 0.0004$ based on 125 confirmed members.

\subsection{Velocity dispersion}
\label{sec_sc}

\begin{figure}
\resizebox{\hsize}{!}{\includegraphics{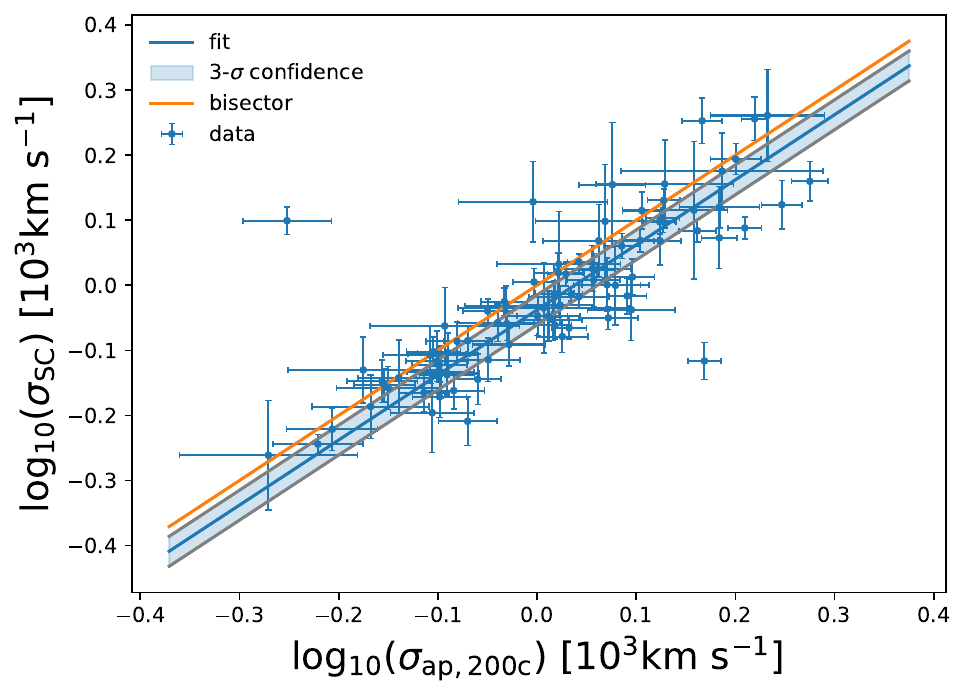}} \\
\caption{Aperture velocity dispersions from our analysis compared to results from the literature. The bisector line tracks $y=x$.}
\label{fig_chexmate_sigma_sc}
\end{figure}


A number of CHEX-MATE clusters were the subjects of dedicated studies and spectroscopic campaigns, and their velocity dispersions are available in previous studies. To compare our results with the literature, we consider the version 2 of the Sigma Catalogs of velocity dispersions \citep[SC,][]{se+et15_comalit_IV}, two meta-catalogues collecting and homogenising data for 4544 clusters (3476 unique items) from 29 sources.\footnote{The catalogues are available at \url{http://pico.oabo.inaf.it/\textasciitilde sereno/CoMaLit/sigma/}.}

\texttt{SC-single} is the catalogue of unique entries. We identify counterparts by matching with clusters from the CHEX-MATE sample whose redshifts differ by less than $\Delta z = 0.05\,(1+z)$ and whose projected distance in the sky does not exceed 10\arcmin. We find 89 matches with the full CHEX-MATE sample. \texttt{SC-all} comprises the full body of information and it can contain multiple entries per cluster.

Out of the 89 clusters in the matched SC sample, there are 86 clusters which we could measure a velocity dispersion for. For the subsample of 84 clusters with reported info on the members in \texttt{SC-single}, the median number of members is 115 (89) for our sample (SC). Whereas we limit the analysis to members within $r_\text{200c}$, this is not necessarily true for the literature sample.

We compare aperture velocity dispersions. A direct comparison is not straightforward as estimates in the literature can use a variety of data, aperture radii, velocity dispersion estimators, or methods for interloper removal. 
Notwithstanding these caveats, the agreement is good (see Fig.~\ref{fig_chexmate_sigma_sc}). There are only two clusters for which
\begin{equation}
\frac{\sigma_\text{SC}-\sigma_\text{ap, 200c}}{ (\delta^2_{\sigma_\text{SC}} + \delta^2_{\sigma_\text{ap, 200c}})^{1/2}} \ge 5 \, ,
\end{equation} 
where $\sigma_\text{SC}$ is the aperture velocity reported in \texttt{SC-single}, $\sigma_\text{ap, 200c}$ is our estimate of the aperture velocity dispersion within $r_\text{200c}$, and the $\delta$'s are the associated uncertainties.

The entry in \texttt{SC-single} for PSZ2 G031.93+78.71, that is, Abell~1775, is from a preliminary analysis ---later updated by \citet{agu+al22}--- that labelled the cluster as clearly substructured, and they deemed the velocity dispersion estimate as not trustable. Our estimate of $\sigma_\text{ap,200c} = 560 \pm 57~\text{km~s}^{-1}$ is in better agreement with other estimates from the literature reported in \texttt{SC-all}. \citet{gir+al98} found $\sigma_\text{ap} = 356 \pm 164~\text{km~s}^{-1}$; \citet{soh+al20} found $\sigma_\text{ap} = 597 \pm 58~\text{km~s}^{-1}$; \citet{rin+al16} found $\sigma_\text{ap} = 577 \pm 62~\text{km~s}^{-1}$.

The cluster PSZ2~G107.10+65.32, aka Abell 1758, is a complex system with a Northern and a Southern component \citep{mon+al17b}. \citet{mon+al17b} performed a detailed multi-probe analysis. The Northern component, associated with the main X-ray peak, is clearly bimodal as seen in WL or X-ray observations. The east–west bimodality in the Northern component cannot be observed in the projected, one-dimensional velocity distribution of the member galaxies, from which \citet{mon+al17b} estimated $\sigma_\text{ap} \sim 1442~\text{km~s}^{-1}$, very close to our estimate of $\sigma_\text{ap,200c} = 1475 \pm 56~\text{km~s}^{-1}$. However, the two-dimensional analysis of the galaxy distribution reveals two peaks in the Northern component with velocity dispersions of $\sigma_\text{ap} \sim 1296~\text{km~s}^{-1}$ and $\sigma_\text{ap} \sim 1075~\text{km~s}^{-1}$ \citep{mon+al17b}. Other analyses of the system from the literature reported in \texttt{SC-all} favour lower values, see \citet{soh+al20}, who found $\sigma_\text{ap}= 765\pm 50~\text{km~s}^{-1}$ (the entry for \texttt{SC-single}), \citet[$\sigma_\text{ap}= 744\pm 107~\text{km~s}^{-1}$]{sif+al15}, or \citet[$\sigma_\text{ap}= 704\pm 84~\text{km~s}^{-1}$]{rin+al16}.

A quantitative comparison can be performed with a linear regression,
\begin{equation}
\label{eq_scaling_1}
\log (\sigma_\text{SC} / \sigma_\text{pivot}) = \alpha + \beta \log (\sigma_\text{ap, 200c} / \sigma_\text{pivot}) + \gamma \log F_z    \, ,
\end{equation}
where $F_z = H(z) / H (z_\text{ref})$ is the redshift dependent Hubble parameter normalised to the reference redshift of the sample, $z_\text{ref} = 0.2$, and $\sigma_\text{pivot} = 1000~{\rm km~s^{-1}}$ is the pivot velocity dispersion. The bias at the reference velocity dispersion and redshift,

\begin{equation}
\label{eq_scaling_1b}
 b = \frac{\sigma_\text{SC}}{\sigma_\text{ap, 200c}}(\sigma_\text{ap, 200c} = \sigma_\text{pivot}, z = z_\text{ref} ) - 1 \, ,
\end{equation}
can be computed as $b = 10^\alpha -1$.

The regression follows the CoMaLit (Comparing Masses from Literature) Bayesian scheme described in \citet{se+et15_comalit_I, ser+al15_comalit_II, se+et15_comalit_IV, se+et17_comalit_V}, which we refer to for details, and is performed with the \textsc{R}-package \texttt{LIRA} \citep{ser16_lira}.\footnote{The package \texttt{LIRA} (LInear Regression in Astronomy) is publicly available from the Comprehensive R Archive Network at \url{https://cran.r-project.org/web/packages/lira/index.html}.}

For our analysis, we assume that both $X = \log (\sigma_\text{ap, 200c} / \sigma_\text{pivot})$, that is, the covariate in Eq.~(\ref{eq_scaling_1}), and $ Y = \log (\sigma_\text{SC} / \sigma_\text{pivot})$, that is, the response, are scattered, and possibly biased, proxies of the true aperture velocity dispersion, $Z$, whose distribution we model as a Gaussian with redshift dependent mean and variance \citep{se+et15_comalit_IV}. We assume that, apart from the intrinsic scatter, the covariate is unbiased with respect to the true aperture velocity dispersion.

We consider non-informative priors \citep{se+et15_comalit_IV, se+et17_comalit_V}: uniform distribution for the normalisation $\alpha$ and the mean of the distribution of $Z$; the Student's $t_1$ distribution with one degree of freedom for the slopes $\beta$ and $\gamma$; the Gamma distribution for the inverse of either variances or intrinsic scatters.

We find marginal agreement with the sample from the literature, with $\alpha = -0.038\pm 0.008$, that is a bias of $-8.3\pm 1.7\,\%$, with no evidence for evolution with velocity dispersion, $\beta =0.99 \pm 0.16$, or time, $\gamma = -0.1\pm0.4$. 
We repeat the regression assuming no evolution ($\beta = 1$, and $\gamma = 0$), and we find $\alpha = -0.038\pm 0.008$, that is, a bias of $-8.4\pm 1.6\,\%$.

A significant number of clusters in the matched sample (47 out of 86) come from \citet{soh+al20}, who constructed the HeCS-omnibus cluster sample exploiting Hectospec at MMT (Multiple Mirror Telescope) or SDSS observations. This subsample drives the comparison and leads to the bias. If we consider only this subsample, we find a bias of $-11\pm 2\,\%$, whereas, if we consider only the remaining clusters, we find a bias of $-3\pm 3\,\%$, which is compatible with no bias.

The bias might be due to the different selection of member candidates. \citet{soh+al20} used the caustic technique for cluster membership and mass determination. Caustic-based methods can efficiently remove the high-velocity interlopers, but require a large number of spectroscopic members, and they might be less effective at rejecting low-velocity galaxies near the turnaround radius \citep[and references therein]{sar+al13}. As a result, velocity dispersions computed after rejection of interlopers based upon caustic techniques might be lower than results based on other methods \citep{sar+al13}.

\section{Dynamical masses}
\label{sec_mass}

\begin{figure}
\begin{tabular}{c}
\resizebox{\hsize}{!}{\includegraphics{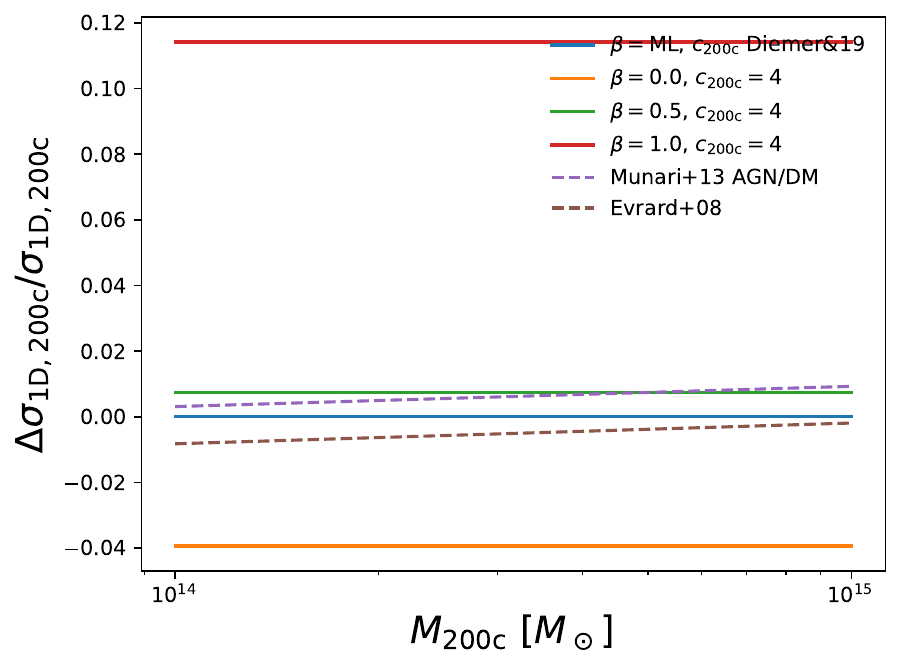}} \\
\resizebox{\hsize}{!}{\includegraphics{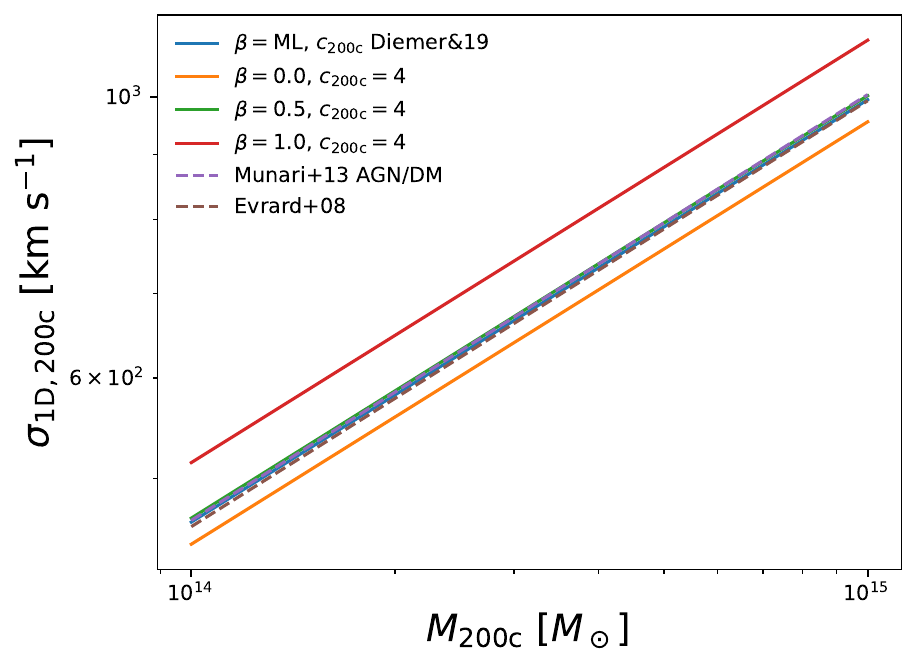}} \\
\end{tabular}
\caption{Relation between mass and 1D velocity dispersion, both for analytical models (full lines) and for results from numerical simulations (dashed lines). \emph{Top}: Relative difference with respect to the reference model as a function of mass. \emph{Bottom}: Mass--velocity dispersion relation.}
\label{fig_M_sigma_theory}
\end{figure}

We present how we derive dynamical masses and how we validate them with comparison to WL estimates.

\subsection{Derivation}

The relation between mass and velocity dispersion is well established on theoretical grounds. Numerical simulations and analytical models agree on a nearly self-similar scenario \citep{kai86,voi05}, where $\sigma$ is a reliable proxy of the mass.

The internal kinematics of member galaxies in a spherical halo can be conveniently derived assuming a shape for the total mass profile, $M(r)$ (or equivalently the gravitational potential), a velocity anisotropy profile, $\beta_\sigma (r)$, and a shape for the 3D velocity distribution \citep{mam+al13}. To derive the main, integrated kinematic properties, we do not need to model the full velocity distribution (see Appendix~\ref{sec_nfw}). The radial velocity dispersion, $\sigma_\text{r}(r)$, can be obtained by solving the spherical Jeans equation under the hypothesis that the tracers follow the mass distribution.
The line-of-sight, $\sigma_\text{los}(R)$, and the aperture, $\sigma_\text{ap}(R)$, velocity dispersion can then be derived by integration of the (weighted) radial velocity \citep{lo+ma01}.

In Fig.~\ref{fig_M_sigma_theory}, we compare results from simple analytical models with numerical simulations \citep{evr+al08,mun+al13}. As representative results from studies based on numerical simulations, we consider the DM-only analysis from \citet{evr+al08}, and the results for DM particles in a hydrodynamical simulation set with cooling and feedback from active galactic nuclei (AGN) or supernovae (SN) from \citet{mun+al13}. Realistic models and simulations are in good agreement, with differences at the subpercent level. Only unrealistic models, as, for example, the strongly anisotropic model with only radial motions ($\beta_\sigma=1$), show large deviations.

In this work, we compute dynamical masses at the same time and consistently with the member selection. We adopt the same one-parameter halo model used for cleaning. The model is uniquely determined by the mass. Given the measured velocity dispersion within a given aperture radius, we invert Eq.~(\ref{eq_kin_mass_1}) to estimate the halo mass. The model can be used to estimate, for example, the halo concentration and the aperture or 1D velocity dispersion within the virial radius. Dynamical masses, $M_{\sigma,r_{\Delta\text{c}}}$, are reported in Table~\ref{tab_masses}.


\subsection{Validation with WL masses}
\label{sec_LC2}

\begin{figure}
\resizebox{\hsize}{!}{\includegraphics{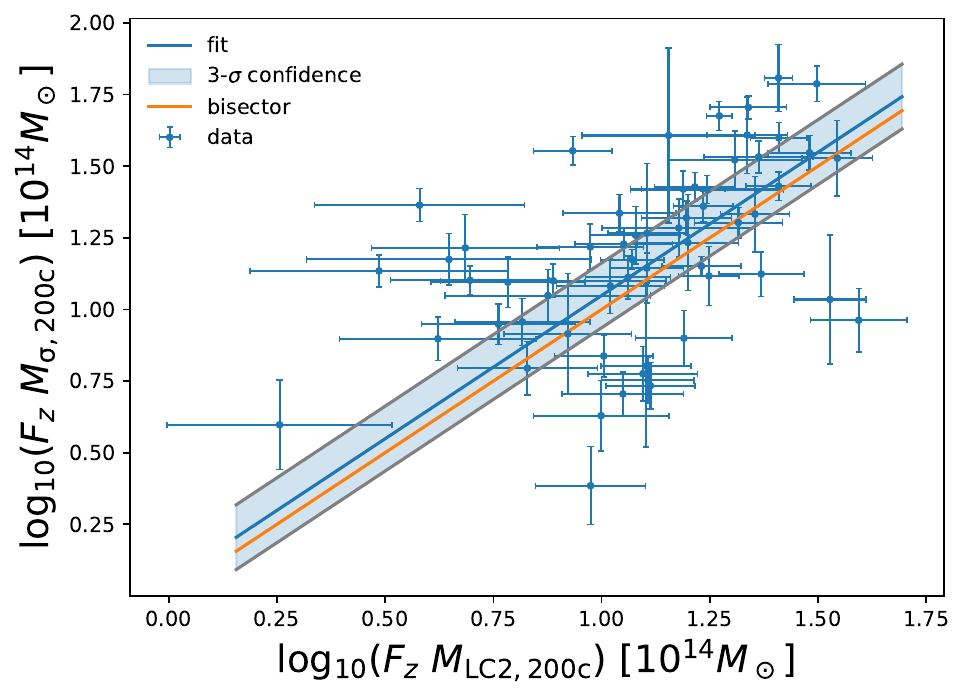}} \\
\caption{WL masses from the literature vs dynamical masses. The bisector line tracks $y=x$.}
\label{fig_chexmate_MLC2_Msigma}
\end{figure}

Dynamical masses can be validated by comparison with WL masses. WL by galaxy clusters is well understood \citep{ba+sc01,ume20} and offers a reliable tool to accurately measure cluster masses, either from targeted observations \citep[see, e.g.,][]{hoe+al12, wtg_I_14, wtg_III_14, hoe+al15, ume+al14, ume+al16b,ok+sm16,die+al19} or from survey data \citep[see, e.g., ][]{ser+al17_psz2lens,ume+al20,mel+al15,hsc_med+al18b,ser+al17_psz2lens,ser+al18_psz2lens,euclid_pre_XLII_ser+al24}.

For comparison with the literature, we consider the Literature Catalogue of weak Lensing Clusters of galaxies \citep[LC$^2$ or LC2,][]{ser15_comalit_III}. The LC2s are meta-catalogues of WL masses retrieved from the literature and  homogenised.\footnote{The catalogues are available at \url{http://pico.oabo.inaf.it/\textasciitilde sereno/CoMaLit/LC2/}.} The latest compilation (v3.9) contains 1501 clusters and groups (806 unique) from 119 bibliographic sources. 
We consider the \texttt{LC2-single} catalogue of unique clusters, where a single mass measurement per cluster is reported.

The WL meta-catalogue is matched to the CHEX-MATE sample by finding clusters whose redshifts differ for less than $\Delta z = 0.05\,(1+z)$ and whose projected distance in the sky does not exceed 10\arcmin, as also done in Sect.~\ref{sec_sc}. We find 61 matches, 59 of them with measured velocity dispersion and dynamical mass.

Agreement is quantified similarly to Sect.~\ref{sec_sc} by investigating the scaling relation,
\begin{equation}
\label{eq_scaling_2}
\log \left( F_z \frac{M_{\sigma, \text{200c}}}{M_\text{pivot}} \right ) = \alpha + \beta \log \left( F_z \frac{M_\text{LC2, 200c}}{M_\text{pivot}} \right ) + \gamma \log F_z   \, , 
\end{equation}
where $M_\text{pivot} = 1 \times 10^{15}M_\odot$. 

We find no evidence for bias ($\alpha = 0.03\pm 0.06$), dependence on mass ($\beta = 1.1\pm 0.4$), or evolution with redshift ($\gamma = 0.2 \pm 1.3$). Assuming no evolution ($\beta = 1$, and $\gamma = 0$), we measure $\alpha = 0.05\pm 0.04$ (see Fig.~\ref{fig_chexmate_MLC2_Msigma}). 


\section{{\it Planck} mass bias}
\label{sec_plan}

\begin{figure}
\resizebox{\hsize}{!}{\includegraphics{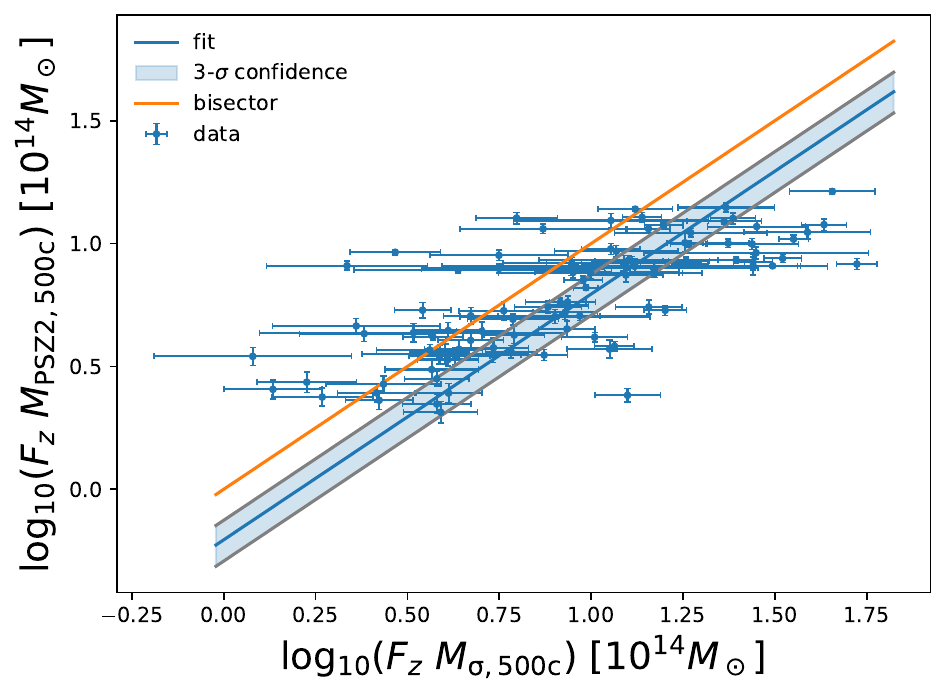}} \\
\caption{CHEX-MATE dynamical masses vs {\it Planck} masses. The bisector line tracks $y=x$.}
\label{fig_chexmate_MVD500c_MPSZ500c_self}
\end{figure}

The era of Stage-IV surveys \citep{euclid_I_24} now includes the successful launch of the {\it Euclid} satellite, which is operating and scanning the sky. This opens a new chapter for cosmological probes based on cluster number counts when results from Stage-III surveys  or multi-wavelength precursors are still far from conclusive. A number of analyses \citep{planck_2013_XX, planck_2015_XXIV, des_cos+al21} found results in tension with multiple cosmological probes, for example, SN, baryon acoustic oscillations, cosmic shear, galaxy clustering, or CMB (cosmic microwave background) anisotropies. This has been seen either as a possible sign of physics beyond the $\Lambda$CDM model or, much less glamorously, as the result of unaccounted for systematic uncertainties. In fact, a recent analysis of the KiDS survey, with a complete and pure cluster sample and well controlled mass assessment through WL showed results consistent with other probes \citep{les+al22}.

The {\it Planck} sample is still a primary source for cosmology and astrophysics, and there is still a need for unbiased mass measurements. {\it Planck} masses, that is the masses reported in the PSZ2 catalogue, were calibrated with a local subsample of relaxed clusters with precise HE mass measurements \citep{planck_2013_XX}. It is well understood that a bias can exist in HE or X-ray calibrated mass measurements \citep{se+et15_comalit_I}, 
\begin{equation}
b_\text{PSZ2} = \frac{M_\text{PSZ2,500c}}{M_\text{500c}} -1 \, .
\end{equation}
In our notation, the bias is negative for under-estimated masses.\footnote{In the alternative notation $ M_\text{PSZ2,500c} = (1 - b^{(1)}_\text{PSZ2}) M_\text{500c}$, the bias is positive for underestimated masses. The bias could be alternatively defined as $b^{(2)}_\text{PSZ2} = \ln (M_\text{PSZ2,500c} / M_\text{500c})$ \citep{se+et17_comalit_V}.} Very large bias values of $b_\text{PSZ2} \sim -0.4$ are needed to fully reconcile the official {\it Planck} number count analysis with other probes \citep{planck_2013_XX, planck_2015_XXIV}.

Based on a suite of numerical simulations \citep{bat+al12,kay+al12}, the {\it Planck} team estimated $b_\text{PSZ2}=-0.2^{+0.2}_{-0.1}$. Observational constraints mostly agree with this estimate. {\it Planck} masses are biased low with respect to WL calibrated masses \citep{lin+al14,se+et17_comalit_V,ser+al17_psz2lens}. Values of $b_\text{PSZ2}\sim -0.3$-$0.4$ were found for some high-mass, intermediate redshift subsamples \citep{lin+al14}, even though other results from cluster WL \citep{smi+al16} or CMB lensing \citep{planck_2015_XXIV} showed a nearly null bias. However, most WL results agrees on values of $b_\text{PSZ2}\sim -0.2$ when selection effects are accounted for (see, for example, \citealt{se+et17_comalit_V} for a homogeneous reanalysis of WL cluster samples). When accounting for redshift evolution too, there is some evidence for a bias that is more pronounced for high redshift clusters \citep{se+et17_comalit_V,aym+al24}, which could alleviate the cosmological tension even for biases that are less extreme than $\sim -0.4$.

The previous results are mostly based on WL masses, which can also be biased \citep{se+et15_comalit_I}. It is useful to asses the bias with independent methods.  \citet{les+al23} investigated the  clustering of the {\it Planck} clusters, focusing on the redshift-space two-point correlation function, to find $b_\text{PSZ2} \sim -0.4$.
Previous assessments of the {\it Planck} mass bias based on dynamical masses are of particular interest to our analysis. \citet{amo+al17} measured a mass bias of $b_\text{PSZ2} = -0.36\pm0.11$ using dynamical mass measurements based on velocity dispersion of 17 {\it Planck} clusters. \citet{fer+al21} estimated a mass bias of $b_\text{PSZ1} = -0.17\pm0.07$ from a sample including 270 clusters from the PSZ1 sample. Recently, \citet{agu+al22} determined $b_\text{PSZ2} = -0.20\pm0.06$ from a complete subsample of 388 clusters from the PSZ2 catalogue.

We can estimate the mass bias by comparing the CHEX-MATE dynamical masses to the {\it Planck} masses,
\begin{equation}
\label{eq_scaling_3}
\log \left( F_z \frac{M_\text{PSZ2, 500c}}{M_\text{pivot}} \right ) = \alpha + \beta \log\left( F_z\frac{M_{\sigma, \text{500c}}}{M_\text{pivot}} \right ) + \gamma \log F_z   \, , 
\end{equation}
where $M_\text{pivot} = 7.25 \times 10^{14}M_\odot$ (i.e., the mass threshold for the Tier 2 clusters). At the pivot mass, and at the reference redshift, the bias can be derived as $b_\text{PSZ2} = 10 ^ \alpha  - 1$.

Selection effects or Eddington/Malmquist biases may affect the linear regression in Eq.~\eqref{eq_scaling_3}. The PSZ2 sample is selected by SZ properties. CHEX-MATE clusters are selected by $(\text{S/N})_\text{MMF3}$ and, on top of this, Tier-2 clusters pass the condition $M_\text{MMF3, 500c} > 7.25\times10^{14} M_\odot$. We can account for most of the selection effects by modelling the conditional probability of the {\it Planck} mass measurements as a truncated normal distribution \citep{se+et15_comalit_IV,ser16_lira,se+et17_comalit_V},
\begin{equation}
y_i|Y_i \sim {\cal N}(Y_i, \delta_{y,i}^2) {\cal U}(y_i-y_{\mathrm{th}}) \, ,
\end{equation}
where ${\cal N}$ is the Gaussian distribution, ${\cal U}$ is the step function,
$Y$ is the true value of $\log ( F_z M_\text{PSZ2, 500c} / M_\text{pivot} )$, $y$ is the measured value of $Y$, and $\delta_{y,i}$ is the measurement uncertainty. We fit for $M_\text{PSZ2, 500c}$ whereas the cut in mass is in $M_\text{MMF3, 500c}$. We account for this by taking as mass threshold the minimum mass of the Tier-2 clusters and by also considering an uncertainty in the threshold, $y_\text{th}$, as large as the measurement uncertainty.

The data are not sufficient to probe evolution with redshift ($\gamma = 1.8 \pm 0.9$), whereas we find some not conclusive evidence for dependence on mass ($\beta = 0.4\pm 0.1$). Some systematic uncertainties could be mass dependent (see Sect.~\ref{sec_syst}), and we should better understand them before drawing any conclusion. In the following, we do not discuss the mass dependence but we focus on the normalisation after marginalising over the other parameters. When we consider a model where $\beta$ and $\gamma$ are free, we find that {\it Planck} masses are under-estimated with respect to dynamical masses, $\alpha =-0.18\pm0.02$. At the reference mass and redshift, we find a mass bias of $b_\text{PSZ2} = -0.34 \pm 0.03$. Assuming no evolution ($\beta = 1$, and $\gamma = 0$), we find $\alpha = -0.21\pm0.03$, and $b_\text{PSZ2} = -0.38\pm 0.04$ (see Fig.~\ref{fig_chexmate_MVD500c_MPSZ500c_self}). 
Our analysis cannot disprove the very high mass bias required to reconcile the {\it Planck} number count analysis with other probes.

 
\section{Systematic errors}
\label{sec_syst}

\begin{figure}
\resizebox{\hsize}{!}{\includegraphics{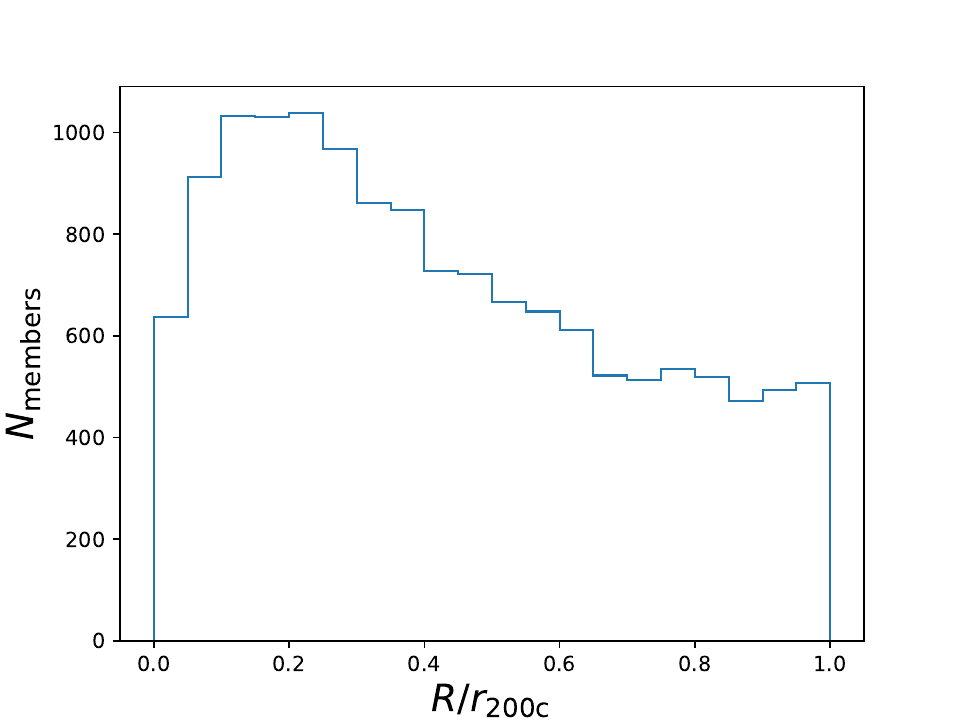}} \\
\caption{Distribution of radial distances, in units of $r_\text{200c}$ of the confirmed member galaxies with spectroscopic redshift within $r_\text{200c}$.}
\label{fig_R_members}
\end{figure}

\begin{table}
\caption{Systematic uncertainties in dynamical mass estimate.}
\label{tab_systematics}
\centering
\begin{tabular}[c]{l  r}
        \hline
        \noalign{\smallskip}  
        effect  & mass error    \\
        \noalign{\smallskip}  
        \hline
        \noalign{\smallskip}  
        velocity bias   & $\sim 5\,\%$   \\
        selection bias  & $\la 1\,\%$   \\
        interlopers   & $\la 3\,\%$   \\
        halo model  & $\la 1\,\%$   \\
        radial coverage and aperture & $\la 1\,\%$   \\
        physical region  & $\la 1\,\%$   \\
        redshift uncertainties & $\la 1\,\%$   \\
        total & $\sim 5\,\%$
        \end{tabular}
\end{table}

We review some systematic errors that could affect our results. For our model and a self-similar scenario, $M_\Delta \sim \sigma_\Delta^3$, and the relative uncertainty in the mass is three times as large as the relative uncertainty in velocity dispersion.


\subsection{Velocity bias}

Theoretical results are mostly based on DM whereas we measure the velocity dispersion of galaxies. Results from numerical simulations vary according to the tracer of the potential field, such as DM particles, subhaloes, red or blue galaxies, and to the assumed gas and baryonic physics, for example, radiative cooling or feedback from AGN, or SN and stars \citep{mun+al13}. 

Galaxy velocity dispersion can be biased with respect to DM velocity dispersion. Competing effects, such as dynamical friction or mergers, coupled with selection effects in colour, magnitude, or aperture, can bias the velocity dispersion low or high. For the CHEX-MATE mass range, the mass bias is predicted to be of the order of $\sim 5$ percent \citep{sar+al13,xxl_XXIII_far+al18,amo+al17,arm+al18}, comparable to the statistical uncertainty.

\subsection{Selection bias}

Galaxy velocity dispersion estimates depend on the selected galaxies. Member galaxies at different evolution stages or with different properties can experience a different dynamical status. Velocity dispersion can change based on galaxy luminosity or type, or due to segregation effects. 
A sample of a few dozens of confirmed luminous member galaxies can be representative of the full population. By selecting the subset of the 100 (30) most luminous red-sequence galaxies within a cluster, the velocity dispersion bias is $\sim 0\,\%$ ($\sim -4 \,\%$) \citep{sar+al13}. The median number of spectroscopic redshifts per cluster of our sample is 97, and the bias is expected to be negligible.

The bias is significantly smaller if the galaxies are randomly selected \citep{sar+al13}. Due to the heterogeneous nature of the data that we collected, it is difficult to establish whether galaxies were actually selected by luminosity for some CHEX-MATE clusters, even though in a few cases we know that candidate member galaxies were selected with no bias with respect to their formation history (see e.g. Sect.~\ref{sec_obs_other}).

We find no significant difference in our results by considering subsamples with a larger number of confirmed members. If we limit our analysis to clusters with at least 25 confirmed members within $r_\text{200c}$, we find that the normalisation of the $\sigma_\text{SC}$-$\sigma_\text{ap,200c}$, $M_{\sigma}$-$M_\text{LC}$, or $M_\text{PSZ2}$-$M_{\sigma}$ relation change by $0.4\pm2\,\%$, $5\pm9\,\%$, or $2\pm4\,\%$, respectively. These variations are compatible with zero, that is, a null effect.


\subsection{Interlopers}

Interlopers can significantly bias the mass estimate \citep{cen97,dia99,biv+al06}. However, we perform a conservative interloper removal (see Sect.~\ref{sec_memb}). The CLEAN method has proved to be very effective, and the mass can be recovered at the percent level for different dynamical settings \citep{mam+al13}. 

The interloper velocity bias is minimum if the velocity dispersion is computed within an aperture near the virial radius \citep{sar+al13}. The bias for well sampled clusters ($\sim 100$ members within $r_\text{200c}$) can be at the percent level in the redshift and mass range covered by the CHEX-MATE clusters \citep{sar+al13}. Even though the velocity bias is small, interlopers can increase the scatter \citep{sar+al13}.

The systematic uncertainty due to interlopers can be measured by comparing the results of different rejection methods. Our comparison with literature values in Sect.~\ref{sec_sc} hints at systematic uncertainties as large as $10\,\%$, even though differences are usually smaller.

The member selection is very stable with respect to the settings. At the ensemble level, we find no significant differences, for example, by considering a search window of $\pm6000~\text{km~s}^{-1}$ in rest-frame velocity for the initial cut or by selecting members with rest-frame velocities smaller than $\pm3\,\sigma_\text{los}(R)$.

\subsection{Halo model}

We derive dynamical masses based on the ML anisotropy profile and the mass-concentration relation from \citet{di+jo19}. These are known to be a good description of real cluster profiles on average \citep{biv+al21}. 

Adopting different models would not make a large difference in the membership selection, since they would predict similar velocity dispersion profiles \citep{mam+al13,biv+al21}. 

However, if the adopted profiles are not a good description of the individual cluster profile, dynamical masses might be biased. The scatter of the proxy depends on how close the chosen model is to reality. As we verify by comparison with WL masses, we find no sign of statistically significant bias.

Results based on alternative models with $c_\text{200c} \sim 4$ and $\beta_\sigma \sim 0.5$, or from numerical simulations differ from our results by less than $1\, \%$ in the CHEX-MATE cluster mass range (see Fig.~\ref{fig_M_sigma_theory}), a systematic difference which is not significant in comparison to the statistical precision for our sample.

\subsection{Radial coverage and aperture}

The ratio of kinetic to potential energy depends on the mass accretion history \citep{ser+al21}, and it can be overestimated in the inner regions with respect to larger radii \citep{pow+al12}. Aperture effects should not bias results in the redshift and mass range covered by CHEX-MATE \citep{sar+al13}. We select galaxies over a representative radial range (see Figs.~\ref{fig_hist_R_ap_R_max} and \ref{fig_R_members}, with  $R_\text{ap} / r_\text{200c} =0.9 \pm 0.2)$, and our estimates should not be significantly impacted due to incomplete coverage.

To estimate the effective aperture radius, we assume a nearly isothermal density distribution for the galaxies, which produces a nearly uniform distribution of radial distances after weighting for the larger area at larger radii. This assumption can be checked by considering the distribution of normalised radial distances of the confirmed members (see Fig.~\ref{fig_R_members}). Inner regions are on average better covered by observations, even though most CHEX-MATE clusters are usually covered up to $r_\text{200c}$ (see Fig.~\ref{fig_hist_R_ap_R_max}). If we assume a constant, flat surface density (a quite extreme scenario) rather than the isothermal density distribution, the aperture radius can be estimated as $1.5$ times the mean radius. With this conversion factor to estimate the aperture radius in the iterative cleaning procedure, we estimate that $\sigma_\text{ap,200c}$ for the ensemble is larger by $2.4 \pm0.1 \,\%$, with a scatter of $\la 1 \,\%$. Alternatively, assuming a more concentrated galaxy distribution and a conversion factor of 2.5 in the iterative cleaning procedure (a quite extreme scenario), we estimate that $\sigma_\text{ap,200c}$ for the ensemble is smaller by $-2.0 \pm0.1 \,\%$, with a scatter of $\la 1 \,\%$.

\subsection{Physical region}

We conservatively measure velocity dispersions based on members within $r_\text{200c}$, that is, a region where the cluster is mostly virialised. Alternatively, we can consider members up to larger radii. The most recently accreted particles that have passed through the pericentre of their orbit once since their infall pile up near the apocentre of their first orbit, thus creating a sharp density enhancement or caustic in the halo outskirts at the splashback radius \citep{di+kr14}. If we consider members within $r_\text{200m}$, which is close to the splashback radius, the estimated $\sigma_\text{ap,200c}$ for the ensemble is larger by $0.6 \pm0.3 \,\%$, with a scatter of $\sim 3 \,\%$.

The number of members for the velocity dispersion estimate can be maximised by considering members within a fixed, large aperture. If we consider members within $3\,r_\text{MMF3,500c}$ ($3~\text{Mpc}$), the estimated $\sigma_\text{ap,200c}$ for the ensemble is larger by $0.5 \pm0.3 \,\%$ ($0.3 \pm0.3 \,\%$), with a scatter of $\sim 4 \,\%$ ($\sim 3 \,\%$). Due to the larger number of considered members, the typical statistical uncertainty is $\sim 10\,\%$ ($\sim 7\,\%$) smaller for a maximum aperture of $3\,r_\text{MMF3,500c}$ ($3~\text{Mpc}$). Thus, we find that our results do not depend strongly on the exact choice of physical region.

\subsection{Redshift uncertainties}

Uncertainties in the redshift measurements inflate the velocity dispersion estimate \citep{dan+al80}. Correction formulae have been proposed for the standard deviation as estimator for the velocity dispersion \citep{dan+al80}. Even though we use the $S_\text{BI}$, we still consider that correction to quantify the associated systematic error.

We rely on spectroscopic estimation of redshifts, and, if applicable, we select galaxies with high quality observations and high reliability measurements (see Sect.~\ref{sec_obs}). In most cases, uncertainties are small, $\delta v \la 50~\text{km~s}^{-1}$ (see Appendix~\ref{sec_efosc2}).

Let us consider a cluster with $\sigma_\text{ap,200c} \sim 1000~\text{km~s}^{-1}$ at $z\sim 0.2$, nearly at the median values of the CHEX-MATE sample. For $\langle \delta v ^2 \rangle^{1/2} \sim 50~\text{km~s}^{-1}$, we expect the velocity dispersion to be over-estimated by $\la 0.1\,\%$. The error is small ($\sim 1\,\%$) even in more extreme cases, for example, small clusters with $\sigma_\text{ap,200c} \sim 500~\text{km~s}^{-1}$ and large uncertainties, $\langle \delta v ^2 \rangle^{1/2} \sim 100~\text{km~s}^{-1}$.

\subsection{Summary}

Different systematic uncertainties could balance out. For example, the effect of a large intrinsic velocity bias at small radii can counter the member contamination at large radii \citep{sar+al13}. In summary, we estimate the total systematic error on our dynamical masses to be of order of $\sim 5\, \%$ (see Table~\ref{tab_systematics}). 

The systematic error could be reduced to the percent level if a sufficient number of member galaxies is secured ($\ga 50$ members), and if the radial coverage extends to the virial radius  ($R_\text{ap} / r_\text{200c} \sim 1$) \citep{sif+al16}. These conditions are met, on average by our sample. However, a better understanding of the velocity bias, that is, the main source of systematic uncertainty, is still needed.

Cross comparison of different methods is also needed to validate the assessment of systematic uncertainties, as testing of a single method, for example, on numerical simulations or under specific settings, could underestimate some biases. \citet{old+al14} compared different cluster dynamical mass estimation techniques that utilise the positions, velocities, and colours of galaxies. 
They found that masses determined with different methods can differ by $\sim 10\,\%$, 
even though the difference is usually smaller than or comparable to the scatter. Differences in mass estimates can also depend on the calibration factors used for some methods \citep{gif+al13}.

The agreement of our dynamical masses with WL masses, with differences at the $7 \pm 16\,\%$ level (see Sect.~\ref{sec_LC2}), suggests a small level of systematic uncertainties. Consistency should be further tested on larger and homogeneous WL samples. On the other hand, observations and comparison of caustic masses with WL, X-ray, or SZ masses showed some evidence for masses being biased low by $10$-$40\,\%$ \citep{ser+al15_comalit_II,xcop_ett+al19,lov+al20,log+al22}.


\section{Conclusions}
\label{sec_conc}

Mass measurements are anchors for reliable studies of cluster astrophysics and cosmology. Each mass proxy is affected by its own biases \citep{se+et15_comalit_I,ser+al15_comalit_II,se+et17_comalit_V}. The best way to mitigate the overall bias is via multi-wavelength multi-probe analyses \citep{fo+pe02,lim+al13,ser+al17_CLUMP_M1206,ser+al18_CLUMP_I,kim+al24}.

The aims of the CHEX-MATE programme are to provide and study a minimally biased sample of clusters that are not significantly impacted by selection biases and whose masses are only affected by systematic uncertainties that are fully accounted for. We computed dynamical masses based on positions and velocities of member galaxies. Thanks to analyses in phase space and theoretically and observationally motivated cluster models, we can select cluster members with minimal contamination from interlopers. We can then derive the mass based on the velocity dispersion of the selected members.

Adequate spectroscopic data are available for the CHEX-MATE sample. We mainly exploited archival data, either public or kindly shared. These data were complemented with the first results from our follow-up spectroscopic campaign. Full spectroscopic coverage of the sample is still needed to fully exploit its potential.

Dynamical masses as derived in this paper build on a strong theoretical foundation. Members were selected with a cleaning procedure whose parameters are fixed by our understanding of the properties of cluster-sized halos extracted from numerical simulations \citep{mam+al10,mam+al13,old+al14}. Cluster mass and members were consistently derived within the same framework thanks to theoretically motivated models, which ensure stable and accurate results.

We estimated the possible bias in our dynamical masses due to a range of factors. We find that the predicted bias is of the order of $5\,\%$, most of which is bias related to the measured velocity dispersion of galaxies versus the underlying velocity dispersion of the total matter distribution. This low level of bias is further confirmed via a comparison to WL masses, which we found to be consistent with our dynamical masses at $7 \pm 16\,\%$.

Even though the data at hand cover a substantial fraction of the CHEX-MATE clusters and we measured the dynamical mass for 101 clusters with at least ten confirmed members within $r_\text{200c}$ out of 118 clusters, full spectroscopic coverage is crucial to achieving the project goals. Thanks to the extended mass and redshift baseline expected from our completed follow-up campaign, the precision on the scaling relation parameters should improve by $\sim 20\,\%$ with respect to the analysis exploiting only archive data. Accuracy will strongly benefit from complete, wide, and homogeneous sky coverage for the following reasons: (i) archive clusters were often originally targeted for specific objectives, which can make them a biased sample and disrupt the selection function; (ii) complete observations of the Southern clusters nearly doubles the survey area and minimises cosmic variance; (iii) comparison of data from ESO and other facilities is needed to further check for systematic uncertainties that might affect the meta-catalogue of redshifts collected from different sources. The CHEX-MATE collaboration has been working to complete the coverage. 


\begin{acknowledgements}

Based on observations collected at the European Southern Observatory under ESO programmes 0110.A-4192, and 0111.1-0186.

The authors thank Crist\'obal Sif\'on for kindly sharing data.

This research was supported by the International Space Science Institute (ISSI) in Bern, through ISSI International Team project \#565 ({\it Multi-Wavelength Studies of the Culmination of Structure Formation in the Universe}).

This research has made use of NASA's Astrophysics Data System (ADS) and of the NASA/IPAC Extragalactic Database (NED), which is operated by the Jet Propulsion Laboratory, California Institute of Technology, under contract with the National Aeronautics and Space Administration.

SE, MR, and MS acknowledge the financial contribution from the contract Prin-MUR 2022 supported by Next Generation EU (M4.C2.1.1, n.20227RNLY3 {\it The concordance cosmological model: stress-tests with galaxy clusters}).

MS acknowledges the financial contribution from INAF Theory Grant 2023: Gravitational lensing detection of matter distribution at galaxy cluster boundaries and beyond (1.05.23.06.17).

SE acknowledges the financial contributions ASI-INAF Athena 2019-27-HH.0, ``Attivit\`a di Studio per la comunit\`a scientifica di Astrofisica delle Alte Energie e Fisica Astroparticellare'' (Accordo Attuativo ASI-INAF n. 2017-14-H.0),
and from the European Union’s Horizon 2020 Programme under the AHEAD2020 project (grant agreement n. 871158).
 
CPH acknowledges support from ANID through FONDECYT Regular 2021 project no. 1211909.

GC acknowledges the support from the Next Generation EU funds within the National Recovery and Resilience Plan (PNRR), Mission 4 - Education and Research, Component 2 - From Research to Business (M4C2), Investment Line 3.1 - Strengthening and creation of Research Infrastructures, Project IR0000012 – 'CTA+ - Cherenkov Telescope Array Plus'.

DE acknowledges support from the Swiss National Science Foundation (SNSF) through grant agreement 200021\_212576.

LL acknowledges the financial contribution from the INAF grant 1.05.12.04.01.

AF acknowledges the project "Strengthening the Italian Leadership in ELT and SKA (STILES)", proposal nr. IR0000034, admitted and eligible for funding from the funds referred to in the D.D. prot. no. 245 of August 10, 2022 and D.D. 326 of August 30, 2022, funded under the program 'Next Generation EU' of the European Union, 'Piano Nazionale di Ripresa e Resilienza' (PNRR) of the Italian Ministry of University and Research (MUR), 'Fund for the creation of an integrated system of research and innovation infrastructures', Action 3.1.1 'Creation of new IR or strengthening of existing IR involved in the Horizon Europe Scientific Excellence objectives and the establishment of networks'.

BJM acknowledges support from STFC grant ST/Y002008/1.

EP acknowledges support from the French Agence Nationale de la Recherche (ANR), under grant ANR-22-CE31-0010 (project BATMAN).

MR acknowledges the financial contribution from the European Union's Horizon 2020 Programme under the AHEAD2020 project (grant agreement n. 871158).

JS was supported by NASA Astrophysics Data Analysis Program (ADAP) Grant 80NSSC21K1571.

\

Funding for the Sloan Digital Sky Survey V has been provided by the Alfred P. Sloan Foundation, the Heising-Simons Foundation, the National Science Foundation, and the Participating Institutions. SDSS acknowledges support and resources from the Center for High-Performance Computing at the University of Utah. SDSS telescopes are located at Apache Point Observatory, funded by the Astrophysical Research Consortium and operated by New Mexico State University, and at Las Campanas Observatory, operated by the Carnegie Institution for Science. The SDSS web site is \url{www.sdss.org}.

SDSS is managed by the Astrophysical Research Consortium for the Participating Institutions of the SDSS Collaboration, including Caltech, The Carnegie Institution for Science, Chilean National Time Allocation Committee (CNTAC) ratified researchers, The Flatiron Institute, the Gotham Participation Group, Harvard University, Heidelberg University, The Johns Hopkins University, L’Ecole polytechnique f\'{e}d\'{e}rale de Lausanne (EPFL), Leibniz-Institut f\"{u}r Astrophysik Potsdam (AIP), Max-Planck-Institut f\"{u}r Astronomie (MPIA Heidelberg), Max-Planck-Institut f\"{u}r Extraterrestrische Physik (MPE), Nanjing University, National Astronomical Observatories of China (NAOC), New Mexico State University, The Ohio State University, Pennsylvania State University, Smithsonian Astrophysical Observatory, Space Telescope Science Institute (STScI), the Stellar Astrophysics Participation Group, Universidad Nacional Aut\'{o}noma de M\'{e}xico, University of Arizona, University of Colorado Boulder, University of Illinois at Urbana-Champaign, University of Toronto, University of Utah, University of Virginia, Yale University, and Yunnan University.

This research used data obtained with the Dark Energy Spectroscopic Instrument (DESI). DESI construction and operations is managed by the Lawrence Berkeley National Laboratory. This material is based upon work supported by the U.S. Department of Energy, Office of Science, Office of High-Energy Physics, under Contract No. DE–AC02–05CH11231, and by the National Energy Research Scientific Computing Center, a DOE Office of Science User Facility under the same contract. Additional support for DESI was provided by the U.S. National Science Foundation (NSF), Division of Astronomical Sciences under Contract No. AST-0950945 to the NSF’s National Optical-Infrared Astronomy Research Laboratory; the Science and Technology Facilities Council of the United Kingdom; the Gordon and Betty Moore Foundation; the Heising-Simons Foundation; the French Alternative Energies and Atomic Energy Commission (CEA); the National Council of Science and Technology of Mexico (CONACYT); the Ministry of Science and Innovation of Spain (MICINN), and by the DESI Member Institutions: www.desi.lbl.gov/collaborating-institutions. The DESI collaboration is honored to be permitted to conduct scientific research on Iolkam Du’ag (Kitt Peak), a mountain with particular significance to the Tohono O’odham Nation. Any opinions, findings, and conclusions or recommendations expressed in this material are those of the author(s) and do not necessarily reflect the views of the U.S. National Science Foundation, the U.S. Department of Energy, or any of the listed funding agencies.


\end{acknowledgements}

%
%

\citestyle{aa}
\bibliographystyle{aa}
\bibliography{references}

\begin{thebibliography}{118}
\expandafter\ifx\csname natexlab\endcsname\relax\def\natexlab#1{#1}\fi

\bibitem[{{Aguado-Barahona} {et~al.}(2022){Aguado-Barahona},
  {Rubi{\~n}o-Mart{\'\i}n}, {Ferragamo}, {Barrena}, {Streblyanska}, \&
  {Tramonte}}]{agu+al22}
{Aguado-Barahona}, A., {Rubi{\~n}o-Mart{\'\i}n}, J.~A., {Ferragamo}, A.,
  {et~al.} 2022,
  \href{http://dx.doi.org/10.1051/0004-6361/202039980}{\color{blue}\aap},
  \href{https://ui.adsabs.harvard.edu/abs/2022A&A...659A.126A}{659, A126}

\bibitem[{{Albert} {et~al.}(2017){Albert}, {Sif{\'o}n}, {Stroe}, {Mernier},
  {Intema}, {R{\"o}ttgering}, \& {Brunetti}}]{alb+al17}
{Albert}, J.~G., {Sif{\'o}n}, C., {Stroe}, A., {et~al.} 2017,
  \href{http://dx.doi.org/10.1051/0004-6361/201730496}{\color{blue}\aap},
  \href{https://ui.adsabs.harvard.edu/abs/2017A&A...607A...4A}{607, A4}

\bibitem[{{Amodeo} {et~al.}(2017){Amodeo}, {Mei}, {Stanford}, {Bartlett},
  {Melin}, {Lawrence}, {Chary}, {Shim}, {Marleau}, \& {Stern}}]{amo+al17}
{Amodeo}, S., {Mei}, S., {Stanford}, S.~A., {et~al.} 2017,
  \href{http://dx.doi.org/10.3847/1538-4357/aa7063}{\color{blue}\apj},
  \href{https://ui.adsabs.harvard.edu/abs/2017ApJ...844..101A}{844, 101}

\bibitem[{{Applegate} {et~al.}(2014){Applegate}, {von der Linden}, {Kelly},
  {Allen}, {Allen}, {Burchat}, {Burke}, {Ebeling}, {Mantz}, \&
  {Morris}}]{wtg_III_14}
{Applegate}, D.~E., {von der Linden}, A., {Kelly}, P.~L., {et~al.} 2014,
  \href{http://dx.doi.org/10.1093/mnras/stt2129}{\color{blue}\mnras},
  \href{http://adsabs.harvard.edu/abs/2014MNRAS.439...48A}{439, 48}

\bibitem[{{Armitage} {et~al.}(2018){Armitage}, {Barnes}, {Kay}, {Bah{\'e}},
  {Dalla Vecchia}, {Crain}, \& {Theuns}}]{arm+al18}
{Armitage}, T.~J., {Barnes}, D.~J., {Kay}, S.~T., {et~al.} 2018,
  \href{http://dx.doi.org/10.1093/mnras/stx3020}{\color{blue}\mnras},
  \href{https://ui.adsabs.harvard.edu/abs/2018MNRAS.474.3746A}{474, 3746}

\bibitem[{{Aymerich} {et~al.}(2024){Aymerich}, {Douspis}, {Pratt}, {Salvati},
  {Soubri{\'e}}, {Andrade-Santos}, {Forman}, {Jones}, {Aghanim}, {Kraft}, \&
  {van Weeren}}]{aym+al24}
{Aymerich}, G., {Douspis}, M., {Pratt}, G.~W., {et~al.} 2024,
  \href{http://dx.doi.org/10.1051/0004-6361/202449513}{\color{blue}\aap},
  \href{https://ui.adsabs.harvard.edu/abs/2024A&A...690A.238A}{690, A238}

\bibitem[{{Barrena} {et~al.}(2020){Barrena}, {Ferragamo},
  {Rubi{\~n}o-Mart{\'\i}n}, {Streblyanska}, {Aguado-Barahona}, {Tramonte},
  {G{\'e}nova-Santos}, {Hempel}, {Lietzen}, {Aghanim}, {Arnaud},
  {B{\"o}hringer}, {Chon}, {Dahle}, {Douspis}, {Lasenby}, {Mazzotta}, {Melin},
  {Pointecouteau}, {Pratt}, \& {Rossetti}}]{bar+al20}
{Barrena}, R., {Ferragamo}, A., {Rubi{\~n}o-Mart{\'\i}n}, J.~A., {et~al.} 2020,
  \href{http://dx.doi.org/10.1051/0004-6361/202037552}{\color{blue}\aap},
  \href{https://ui.adsabs.harvard.edu/abs/2020A&A...638A.146B}{638, A146}

\bibitem[{{Bartalucci} {et~al.}(2023){Bartalucci}, {Molendi}, {Rasia}, {Pratt},
  {Arnaud}, {Rossetti}, {Gastaldello}, {Eckert}, {Balboni}, {Borgani},
  {Bourdin}, {Campitiello}, {De Grandi}, {De Petris}, {Duffy}, {Ettori},
  {Ferragamo}, {Gaspari}, {Gavazzi}, {Ghizzardi}, {Iqbal}, {Kay}, {Lovisari},
  {Mazzotta}, {Maughan}, {Pointecouteau}, {Riva}, \&
  {Sereno}}]{chexmate_bar+al23}
{Bartalucci}, I., {Molendi}, S., {Rasia}, E., {et~al.} 2023,
  \href{http://dx.doi.org/10.1051/0004-6361/202346189}{\color{blue}\aap},
  \href{https://ui.adsabs.harvard.edu/abs/2023A&A...674A.179B}{674, A179}

\bibitem[{{Bartelmann} \& {Schneider}(2001)}]{ba+sc01}
{Bartelmann}, M. \& {Schneider}, P. 2001,
  \href{http://dx.doi.org/10.1016/S0370-1573(00)00082-X}{\color{blue}\physrep},
  \href{http://adsabs.harvard.edu/abs/2001PhR...340..291B}{340, 291}

\bibitem[{{Battaglia} {et~al.}(2012){Battaglia}, {Bond}, {Pfrommer}, \&
  {Sievers}}]{bat+al12}
{Battaglia}, N., {Bond}, J.~R., {Pfrommer}, C., \& {Sievers}, J.~L. 2012,
  \href{http://dx.doi.org/10.1088/0004-637X/758/2/74}{\color{blue}\apj},
  \href{http://adsabs.harvard.edu/abs/2012ApJ...758...74B}{758, 74}

\bibitem[{{Bayliss} {et~al.}(2016){Bayliss}, {Ruel}, {Stubbs}, {Allen},
  {Applegate}, {Ashby}, {Bautz}, {Benson}, {Bleem}, {Bocquet}, {Brodwin},
  {Capasso}, {Carlstrom}, {Chang}, {Chiu}, {Cho}, {Clocchiatti}, {Crawford},
  {Crites}, {de Haan}, {Desai}, {Dietrich}, {Dobbs}, {Doucouliagos}, {Foley},
  {Forman}, {Garmire}, {George}, {Gladders}, {Gonzalez}, {Gupta}, {Halverson},
  {Hlavacek-Larrondo}, {Hoekstra}, {Holder}, {Holzapfel}, {Hou}, {Hrubes},
  {Huang}, {Jones}, {Keisler}, {Knox}, {Lee}, {Leitch}, {von der Linden},
  {Luong-Van}, {Mantz}, {Marrone}, {McDonald}, {McMahon}, {Meyer}, {Mocanu},
  {Mohr}, {Murray}, {Padin}, {Pryke}, {Rapetti}, {Reichardt}, {Rest}, {Ruhl},
  {Saliwanchik}, {Saro}, {Sayre}, {Schaffer}, {Schrabback}, {Shirokoff},
  {Song}, {Spieler}, {Stalder}, {Stanford}, {Staniszewski}, {Stark}, {Story},
  {Vanderlinde}, {Vieira}, {Vikhlinin}, {Williamson}, \& {Zenteno}}]{bay+al16}
{Bayliss}, M.~B., {Ruel}, J., {Stubbs}, C.~W., {et~al.} 2016,
  \href{http://dx.doi.org/10.3847/0067-0049/227/1/3}{\color{blue}\apjs},
  \href{https://ui.adsabs.harvard.edu/abs/2016ApJS..227....3B}{227, 3}

\bibitem[{{Beers} {et~al.}(1990){Beers}, {Flynn}, \& {Gebhardt}}]{bee+al90}
{Beers}, T.~C., {Flynn}, K., \& {Gebhardt}, K. 1990,
  \href{http://dx.doi.org/10.1086/115487}{\color{blue}\aj},
  \href{http://ads.astro.puc.cl/abs/1990AJ....100...32B}{100, 32}

\bibitem[{{Biviano} \& {Girardi}(2003)}]{bi+gi03}
{Biviano}, A. \& {Girardi}, M. 2003,
  \href{http://dx.doi.org/10.1086/345893}{\color{blue}\apj},
  \href{https://ui.adsabs.harvard.edu/abs/2003ApJ...585..205B}{585, 205}

\bibitem[{{Biviano} {et~al.}(1992){Biviano}, {Girardi}, {Giuricin},
  {Mardirossian}, \& {Mezzetti}}]{biv+al92}
{Biviano}, A., {Girardi}, M., {Giuricin}, G., {Mardirossian}, F., \&
  {Mezzetti}, M. 1992,
  \href{http://dx.doi.org/10.1086/171695}{\color{blue}\apj},
  \href{https://ui.adsabs.harvard.edu/abs/1992ApJ...396...35B}{396, 35}

\bibitem[{{Biviano} {et~al.}(2006){Biviano}, {Murante}, {Borgani}, {Diaferio},
  {Dolag}, \& {Girardi}}]{biv+al06}
{Biviano}, A., {Murante}, G., {Borgani}, S., {et~al.} 2006,
  \href{http://dx.doi.org/10.1051/0004-6361:20064918}{\color{blue}\aap},
  \href{http://ads.astro.puc.cl/abs/2006A%26A...456...23B}{456, 23}

\bibitem[{{Biviano} {et~al.}(2013){Biviano}, {Rosati}, {Balestra}, {Mercurio},
  {Girardi}, {Nonino}, {Grillo}, {Scodeggio}, {Lemze}, {Kelson}, {Umetsu},
  {Postman}, {Zitrin}, {Czoske}, {Ettori}, {Fritz}, {Lombardi}, {Maier},
  {Medezinski}, {Mei}, {Presotto}, {Strazzullo}, {Tozzi}, {Ziegler},
  {Annunziatella}, {Bartelmann}, {Benitez}, {Bradley}, {Brescia}, {Broadhurst},
  {Coe}, {Demarco}, {Donahue}, {Ford}, {Gobat}, {Graves}, {Koekemoer},
  {Kuchner}, {Melchior}, {Meneghetti}, {Merten}, {Moustakas}, {Munari}, {Reg{\H
  o}s}, {Sartoris}, {Seitz}, \& {Zheng}}]{biv+al13}
{Biviano}, A., {Rosati}, P., {Balestra}, I., {et~al.} 2013,
  \href{http://dx.doi.org/10.1051/0004-6361/201321955}{\color{blue}\aap},
  \href{http://adsabs.harvard.edu/abs/2013A%26A...558A...1B}{558, A1}

\bibitem[{{Biviano} {et~al.}(2021){Biviano}, {van der Burg}, {Balogh},
  {Munari}, {Cooper}, {De Lucia}, {Demarco}, {Jablonka}, {Muzzin}, {Nantais},
  {Old}, {Rudnick}, {Vulcani}, {Wilson}, {Yee}, {Zaritsky}, {Cerulo}, {Chan},
  {Finoguenov}, {Gilbank}, {Lidman}, {Pintos-Castro}, \& {Shipley}}]{biv+al21}
{Biviano}, A., {van der Burg}, R.~F.~J., {Balogh}, M.~L., {et~al.} 2021,
  \href{http://dx.doi.org/10.1051/0004-6361/202140564}{\color{blue}\aap},
  \href{https://ui.adsabs.harvard.edu/abs/2021A&A...650A.105B}{650, A105}

\bibitem[{{Bocquet} {et~al.}(2019){Bocquet}, {Dietrich}, {Schrabback}, {Bleem},
  {Klein}, {Allen}, {Applegate}, {Ashby}, {Bautz}, {Bayliss}, {Benson},
  {Brodwin}, {Bulbul}, {Canning}, {Capasso}, {Carlstrom}, {Chang}, {Chiu},
  {Cho}, {Clocchiatti}, {Crawford}, {Crites}, {de Haan}, {Desai}, {Dobbs},
  {Foley}, {Forman}, {Garmire}, {George}, {Gladders}, {Gonzalez}, {Grandis},
  {Gupta}, {Halverson}, {Hlavacek-Larrondo}, {Hoekstra}, {Holder}, {Holzapfel},
  {Hou}, {Hrubes}, {Huang}, {Jones}, {Khullar}, {Knox}, {Kraft}, {Lee}, {von
  der Linden}, {Luong-Van}, {Mantz}, {Marrone}, {McDonald}, {McMahon}, {Meyer},
  {Mocanu}, {Mohr}, {Morris}, {Padin}, {Patil}, {Pryke}, {Rapetti},
  {Reichardt}, {Rest}, {Ruhl}, {Saliwanchik}, {Saro}, {Sayre}, {Schaffer},
  {Shirokoff}, {Stalder}, {Stanford}, {Staniszewski}, {Stark}, {Story},
  {Strazzullo}, {Stubbs}, {Vanderlinde}, {Vieira}, {Vikhlinin}, {Williamson},
  \& {Zenteno}}]{spt_boc+al19}
{Bocquet}, S., {Dietrich}, J.~P., {Schrabback}, T., {et~al.} 2019,
  \href{http://dx.doi.org/10.3847/1538-4357/ab1f10}{\color{blue}\apj},
  \href{https://ui.adsabs.harvard.edu/abs/2019ApJ...878...55B}{878, 55}

\bibitem[{{Bulbul} {et~al.}(2024){Bulbul}, {Liu}, {Kluge}, {Zhang}, {Sanders},
  {Bahar}, {Ghirardini}, {Artis}, {Seppi}, {Garrel}, {Ramos-Ceja}, {Comparat},
  {Balzer}, {B{\"o}ckmann}, {Br{\"u}ggen}, {Clerc}, {Dennerl}, {Dolag},
  {Freyberg}, {Grandis}, {Gruen}, {Kleinebreil}, {Krippendorf}, {Lamer},
  {Merloni}, {Migkas}, {Nandra}, {Pacaud}, {Predehl}, {Reiprich}, {Schrabback},
  {Veronica}, {Weller}, \& {Zelmer}}]{erosita_bul+al24}
{Bulbul}, E., {Liu}, A., {Kluge}, M., {et~al.} 2024,
  \href{http://dx.doi.org/10.1051/0004-6361/202348264}{\color{blue}\aap},
  \href{https://ui.adsabs.harvard.edu/abs/2024A&A...685A.106B}{685, A106}

\bibitem[{{Cen}(1997)}]{cen97}
{Cen}, R. 1997, \href{http://dx.doi.org/10.1086/304394}{\color{blue}\apj},
  \href{https://ui.adsabs.harvard.edu/abs/1997ApJ...485...39C}{485, 39}

\bibitem[{{CHEX-MATE Collaboration} {et~al.}(2021){CHEX-MATE Collaboration},
  {Arnaud}, {Ettori}, {Pratt}, {Rossetti}, {Eckert}, {Gastaldello}, {Gavazzi},
  {Kay}, {Lovisari}, {Maughan}, {Pointecouteau}, {Sereno}, {Bartalucci},
  {Bonafede}, {Bourdin}, {Cassano}, {Duffy}, {Iqbal}, {Maurogordato}, {Rasia},
  {Sayers}, {Andrade-Santos}, {Aussel}, {Barnes}, {Barrena}, {Borgani},
  {Burkutean}, {Clerc}, {Corasaniti}, {Cuillandre}, {De Grandi}, {De Petris},
  {Dolag}, {Donahue}, {Ferragamo}, {Gaspari}, {Ghizzardi}, {Gitti}, {Haines},
  {Jauzac}, {Johnston-Hollitt}, {Jones}, {K{\'e}ruzor{\'e}}, {Le Brun},
  {Mayet}, {Mazzotta}, {Melin}, {Molendi}, {Nonino}, {Okabe}, {Paltani},
  {Perotto}, {Pires}, {Radovich}, {Rubino-Martin}, {Salvati}, {Saro},
  {Sartoris}, {Schellenberger}, {Streblyanska}, {Tarr{\'\i}o}, {Tozzi},
  {Umetsu}, {van der Burg}, {Vazza}, {Venturi}, {Yepes}, \&
  {Zarattini}}]{chexmate+al21}
{CHEX-MATE Collaboration}, {Arnaud}, M., {Ettori}, S., {et~al.} 2021,
  \href{http://dx.doi.org/10.1051/0004-6361/202039632}{\color{blue}\aap},
  \href{https://ui.adsabs.harvard.edu/abs/2021A&A...650A.104C}{650, A104}

\bibitem[{{Chow-Mart{\'\i}nez} {et~al.}(2014){Chow-Mart{\'\i}nez}, {Andernach},
  {Caretta}, \& {Trejo-Alonso}}]{cho+al14}
{Chow-Mart{\'\i}nez}, M., {Andernach}, H., {Caretta}, C.~A., \& {Trejo-Alonso},
  J.~J. 2014,
  \href{http://dx.doi.org/10.1093/mnras/stu1961}{\color{blue}\mnras},
  \href{https://ui.adsabs.harvard.edu/abs/2014MNRAS.445.4073C}{445, 4073}

\bibitem[{{Costanzi} {et~al.}(2021){Costanzi}, {Saro}, {Bocquet}, {Abbott},
  {Aguena}, {Allam}, {Amara}, {Annis}, {Avila}, {Bacon}, {Benson}, {Bhargava},
  {Brooks}, {Buckley-Geer}, {Burke}, {Carnero Rosell}, {Carrasco Kind},
  {Carretero}, {Choi}, {da Costa}, {Pereira}, {De Vicente}, {Desai}, {Diehl},
  {Dietrich}, {Doel}, {Eifler}, {Everett}, {Ferrero}, {Fert{\'e}}, {Flaugher},
  {Fosalba}, {Frieman}, {Garc{\'\i}a-Bellido}, {Gaztanaga}, {Gerdes},
  {Giannantonio}, {Giles}, {Grandis}, {Gruen}, {Gruendl}, {Gupta}, {Gutierrez},
  {Hartley}, {Hinton}, {Hollowood}, {Honscheid}, {James}, {Jeltema}, {Krause},
  {Kuehn}, {Kuropatkin}, {Lahav}, {Lima}, {MacCrann}, {Maia}, {Marshall},
  {Menanteau}, {Miquel}, {Mohr}, {Morgan}, {Myles}, {Ogando}, {Palmese},
  {Paz-Chinch{\'o}n}, {Plazas}, {Rapetti}, {Reichardt}, {Romer}, {Roodman},
  {Ruppin}, {Salvati}, {Samuroff}, {Sanchez}, {Scarpine}, {Serrano},
  {Sevilla-Noarbe}, {Singh}, {Smith}, {Soares-Santos}, {Stark}, {Suchyta},
  {Swanson}, {Tarle}, {Thomas}, {To}, {Tucker}, {Varga}, {Wechsler}, {Zhang},
  {DES}, \& {SPT Collaborations}}]{des_cos+al21}
{Costanzi}, M., {Saro}, A., {Bocquet}, S., {et~al.} 2021,
  \href{http://dx.doi.org/10.1103/PhysRevD.103.043522}{\color{blue}\prd},
  \href{https://ui.adsabs.harvard.edu/abs/2021PhRvD.103d3522C}{103, 043522}

\bibitem[{{Danese} {et~al.}(1980){Danese}, {de Zotti}, \& {di
  Tullio}}]{dan+al80}
{Danese}, L., {de Zotti}, G., \& {di Tullio}, G. 1980, \aap,
  \href{http://ads.astro.puc.cl/abs/1980A%26A....82..322D}{82, 322}

\bibitem[{{DESI Collaboration} {et~al.}(2024){DESI Collaboration}, {Adame},
  {Aguilar}, {Ahlen}, {Alam}, {Aldering}, {Alexander}, {Alfarsy}, {Allende
  Prieto}, {Alvarez}, {Alves}, {Anand}, {Andrade-Oliveira}, {Armengaud},
  {Asorey}, {Avila}, {Aviles}, {Bailey}, {Balaguera-Antol{\'\i}nez},
  {Ballester}, {Baltay}, {Bault}, {Bautista}, {Behera}, {Beltran}, {BenZvi},
  {Beraldo e Silva}, {Bermejo-Climent}, {Berti}, {Besuner}, {Beutler},
  {Bianchi}, {Blake}, {Blum}, {Bolton}, {Brieden}, {Brodzeller}, {Brooks},
  {Brown}, {Buckley-Geer}, {Burtin}, {Cabayol-Garcia}, {Cai}, {Canning},
  {Cardiel-Sas}, {Carnero Rosell}, {Castander}, {Cervantes-Cota}, {Chabanier},
  {Chaussidon}, {Chaves-Montero}, {Chen}, {Chen}, {Chuang}, {Claybaugh},
  {Cole}, {Cooper}, {Cuceu}, {Davis}, {Dawson}, {de Belsunce}, {de la Cruz},
  {de la Macorra}, {Della Costa}, {de Mattia}, {Demina}, {Demirbozan},
  {DeRose}, {Dey}, {Dey}, {Dhungana}, {Ding}, {Ding}, {Doel}, {Doshi},
  {Douglass}, {Edge}, {Eftekharzadeh}, {Eisenstein}, {Elliott}, {Ereza},
  {Escoffier}, {Fagrelius}, {Fan}, {Fanning}, {Fawcett}, {Ferraro}, {Flaugher},
  {Font-Ribera}, {Forero-Romero}, {Forero-S{\'a}nchez}, {Frenk},
  {G{\"a}nsicke}, {Garc{\'\i}a}, {Garc{\'\i}a-Bellido}, {Garcia-Quintero},
  {Garrison}, {Gil-Mar{\'\i}n}, {Golden-Marx}, {Gontcho A Gontcho},
  {Gonzalez-Morales}, {Gonzalez-Perez}, {Gordon}, {Graur}, {Green}, {Gruen},
  {Guy}, {Hadzhiyska}, {Hahn}, {Han}, {Hanif}, {Herrera-Alcantar}, {Honscheid},
  {Hou}, {Howlett}, {Huterer}, {Ir{\v{s}}i{\v{c}}}, {Ishak}, {Jacques}, {Jana},
  {Jiang}, {Jimenez}, {Jing}, {Joudaki}, {Joyce}, {Jullo}, {Juneau},
  {Kara{\c{c}}ayl{\i}}, {Karim}, {Kehoe}, {Kent}, {Khederlarian}, {Kim},
  {Kirkby}, {Kisner}, {Kitaura}, {Kizhuprakkat}, {Kneib}, {Koposov},
  {Kov{\'a}cs}, {Kremin}, {Krolewski}, {L'Huillier}, {Lahav}, {Lambert},
  {Lamman}, {Lan}, {Landriau}, {Lang}, {Lange}, {Lasker}, {Leauthaud}, {Le
  Guillou}, {Levi}, {Li}, {Linder}, {Lyons}, {Magneville}, {Manera}, {Manser},
  {Margala}, {Martini}, {McDonald}, {Medina}, {Medina-Varela}, {Meisner},
  {Mena-Fern{\'a}ndez}, {Meneses-Rizo}, {Mezcua}, {Miquel}, {Montero-Camacho},
  {Moon}, {Moore}, {Moustakas}, {Mueller}, {Mundet}, {Mu{\~n}oz-Guti{\'e}rrez},
  {Myers}, {Nadathur}, {Napolitano}, {Neveux}, {Newman}, {Nie}, {Nikutta},
  {Niz}, {Norberg}, {Noriega}, {Paillas}, {Palanque-Delabrouille}, {Palmese},
  {Pan}, {Parkinson}, {Penmetsa}, {Percival}, {P{\'e}rez-Fern{\'a}ndez},
  {P{\'e}rez-R{\`a}fols}, {Pieri}, {Poppett}, {Porredon}, {Pothier}, {Prada},
  {Pucha}, {Raichoor}, {Ram{\'\i}rez-P{\'e}rez}, {Ramirez-Solano},
  {Rashkovetskyi}, {Ravoux}, {Rocher}, {Rockosi}, {Ross}, {Rossi}, {Ruggeri},
  {Ruhlmann-Kleider}, {Sabiu}, {Said}, {Saintonge}, {Samushia}, {Sanchez},
  {Saulder}, {Schaan}, {Schlafly}, {Schlegel}, {Scholte}, {Schubnell}, {Seo},
  {Shafieloo}, {Sharples}, {Sheu}, {Silber}, {Sinigaglia}, {Siudek}, {Slepian},
  {Smith}, {Soumagnac}, {Sprayberry}, {Stephey}, {Su{\'a}rez-P{\'e}rez}, {Sun},
  {Tan}, {Tarl{\'e}}, {Tojeiro}, {Ure{\~n}a-L{\'o}pez}, {Vaisakh}, {Valcin},
  {Valdes}, {Valluri}, {Vargas-Maga{\~n}a}, {Variu}, {Verde}, {Walther},
  {Wang}, {Wang}, {Weaver}, {Weaverdyck}, {Wechsler}, {White}, {Xie}, {Yang},
  {Y{\`e}che}, {Yu}, {Yuan}, {Zhang}, {Zhang}, {Zhao}, {Zheng}, {Zhou}, {Zhou},
  {Zou}, {Zou}, \& {Zu}}]{desi_edr+24}
{DESI Collaboration}, {Adame}, A.~G., {Aguilar}, J., {et~al.} 2024,
  \href{http://dx.doi.org/10.3847/1538-3881/ad3217}{\color{blue}\aj},
  \href{https://ui.adsabs.harvard.edu/abs/2024AJ....168...58D}{168, 58}

\bibitem[{{Diaferio}(1999)}]{dia99}
{Diaferio}, A. 1999,
  \href{http://dx.doi.org/10.1046/j.1365-8711.1999.02864.x}{\color{blue}\mnras},
  \href{https://ui.adsabs.harvard.edu/abs/1999MNRAS.309..610D}{309, 610}

\bibitem[{{Diemer} \& {Joyce}(2019)}]{di+jo19}
{Diemer}, B. \& {Joyce}, M. 2019,
  \href{http://dx.doi.org/10.3847/1538-4357/aafad6}{\color{blue}\apj},
  \href{https://ui.adsabs.harvard.edu/abs/2019ApJ...871..168D}{871, 168}

\bibitem[{{Diemer} \& {Kravtsov}(2014)}]{di+kr14}
{Diemer}, B. \& {Kravtsov}, A.~V. 2014,
  \href{http://dx.doi.org/10.1088/0004-637X/789/1/1}{\color{blue}\apj},
  \href{http://adsabs.harvard.edu/abs/2014ApJ...789....1D}{789, 1}

\bibitem[{{Dietrich} {et~al.}(2019){Dietrich}, {Bocquet}, {Schrabback},
  {Applegate}, {Hoekstra}, {Grandis}, {Mohr}, {Allen}, {Bayliss}, {Benson},
  {Bleem}, {Brodwin}, {Bulbul}, {Capasso}, {Chiu}, {Crawford}, {Gonzalez}, {de
  Haan}, {Klein}, {von der Linden}, {Mantz}, {Marrone}, {McDonald},
  {Raghunathan}, {Rapetti}, {Reichardt}, {Saro}, {Stalder}, {Stark}, {Stern},
  \& {Stubbs}}]{die+al19}
{Dietrich}, J.~P., {Bocquet}, S., {Schrabback}, T., {et~al.} 2019,
  \href{http://dx.doi.org/10.1093/mnras/sty3088}{\color{blue}\mnras},
  \href{https://ui.adsabs.harvard.edu/abs/2019MNRAS.483.2871D}{483, 2871}

\bibitem[{{Ettori} {et~al.}(2019){Ettori}, {Ghirardini}, {Eckert},
  {Pointecouteau}, {Gastaldello}, {Sereno}, {Gaspari}, {Ghizzardi},
  {Roncarelli}, \& {Rossetti}}]{xcop_ett+al19}
{Ettori}, S., {Ghirardini}, V., {Eckert}, D., {et~al.} 2019,
  \href{http://dx.doi.org/10.1051/0004-6361/201833323}{\color{blue}\aap},
  \href{https://ui.adsabs.harvard.edu/abs/2019A&A...621A..39E}{621, A39}

\bibitem[{{Euclid Collaboration} {et~al.}(2024){Euclid Collaboration},
  {Sereno}, {Farrens}, {Ingoglia}, {Lesci}, {Baumont}, {Covone}, {Giocoli},
  {Marulli}, {Miranda La Hera}, {Vannier}, {Biviano}, {Maurogordato},
  {Moscardini}, {Aghanim}, {Andreon}, {Auricchio}, {Baldi}, {Bardelli},
  {Bellagamba}, {Bodendorf}, {Bonino}, {Branchini}, {Brescia}, {Brinchmann},
  {Camera}, {Capobianco}, {Carbone}, {Cardone}, {Carretero}, {Casas},
  {Castellano}, {Cavuoti}, {Cimatti}, {Cledassou}, {Congedo}, {Conselice},
  {Conversi}, {Copin}, {Corcione}, {Courbin}, {Courtois}, {Cropper}, {Da
  Silva}, {Degaudenzi}, {Di Giorgio}, {Dinis}, {Dubath}, {Duncan}, {Dupac},
  {Dusini}, {Farina}, {Ferriol}, {Frailis}, {Franceschi}, {Fumana}, {Galeotta},
  {Garilli}, {Gillis}, {Grazian}, {Grupp}, {Haugan}, {Holmes}, {Hook},
  {Hormuth}, {Hornstrup}, {Hudelot}, {Jahnke}, {Joachimi}, {Keih{\"a}nen},
  {Kermiche}, {Kiessling}, {Kubik}, {Kunz}, {Kurki-Suonio}, {Ligori}, {Lilje},
  {Lindholm}, {Lloro}, {Maino}, {Maiorano}, {Mansutti}, {Marggraf}, {Markovic},
  {Martinet}, {Massey}, {Medinaceli}, {Mei}, {Mellier}, {Meneghetti}, {Merlin},
  {Meylan}, {Moresco}, {Munari}, {Niemi}, {Nutma}, {Padilla}, {Paltani},
  {Pasian}, {Pedersen}, {Pettorino}, {Pires}, {Polenta}, {Poncet}, {Popa},
  {Raison}, {Rebolo}, {Renzi}, {Rhodes}, {Riccio}, {Romelli}, {Roncarelli},
  {Rossetti}, {Saglia}, {Sapone}, {Sartoris}, {Schirmer}, {Schneider},
  {Schrabback}, {Secroun}, {Seidel}, {Serrano}, {Sirignano}, {Sirri}, {Stanco},
  {Starck}, {Tallada-Cresp{\'\i}}, {Taylor}, {Tereno}, {Toledo-Moreo},
  {Torradeflot}, {Tutusaus}, {Valentijn}, {Valenziano}, {Vassallo},
  {Veropalumbo}, {Wang}, {Weller}, {Zacchei}, {Zamorani}, {Zoubian}, {Zucca},
  {Boucaud}, {Bozzo}, {Cerna}, {Colodro-Conde}, {Di Ferdinando}, {Farinelli},
  {Israel}, {Mauri}, {Neissner}, {Scottez}, {Tenti}, {Wiesmann}, {Akrami},
  {Allevato}, {Baccigalupi}, {Ballardini}, {Benielli}, {Borgani}, {Borlaff},
  {Burigana}, {Cabanac}, {Cappi}, {Carvalho}, {Castignani}, {Castro},
  {Ca{\~n}as-Herrera}, {Chambers}, {Cooray}, {Coupon}, {Davini}, {De Lucia},
  {Desprez}, {Di Domizio}, {Dole}, {Escartin Vigo}, {Escoffier}, {Ferrero},
  {Gabarra}, {Gaztanaga}, {George}, {Giacomini}, {Gozaliasl}, {Hildebrandt},
  {Kajava}, {Kansal}, {Kirkpatrick}, {Legrand}, {Liebing}, {Loureiro},
  {Macias-Perez}, {Magliocchetti}, {Mainetti}, {Maoli}, {Martinelli},
  {Martins}, {Matthew}, {Maturi}, {Maurin}, {Metcalf}, {Monaco}, {Morgante},
  {Nadathur}, {Nucita}, {Patrizii}, {Peel}, {P{\"o}ntinen}, {Popa}, {Porciani},
  {Potter}, {Reimberg}, {Sakr}, {S{\'a}nchez}, {Schneider}, {Sefusatti},
  {Simon}, {Spurio Mancini}, {Stadel}, {Stanford}, {Steinwagner}, {Teyssier},
  {Valiviita}, \& {Viel}}]{euclid_pre_XLII_ser+al24}
{Euclid Collaboration}, {Sereno}, M., {Farrens}, S., {et~al.} 2024,
  \href{http://dx.doi.org/10.1051/0004-6361/202348680}{\color{blue}\aap},
  \href{https://ui.adsabs.harvard.edu/abs/2024A&A...689A.252E}{689, A252}

\bibitem[{{Euclid Collaboration: Adam} {et~al.}(2019){Euclid Collaboration:
  Adam}, {Vannier}, {Maurogordato}, {Biviano}, {Adami}, {Ascaso}, {Bellagamba},
  {Benoist}, {Cappi}, {D{\'\i}az-S{\'a}nchez}, {Durret}, {Farrens}, {Gonzalez},
  {Iovino}, {Licitra}, {Maturi}, {Mei}, {Merson}, {Munari}, {Pell{\'o}},
  {Ricci}, {Rocci}, {Roncarelli}, {Sarron}, {Amoura}, {Andreon}, {Apostolakos},
  {Arnaud}, {Bardelli}, {Bartlett}, {Baugh}, {Borgani}, {Brodwin}, {Castander},
  {Castignani}, {Cucciati}, {De Lucia}, {Dubath}, {Fosalba}, {Giocoli},
  {Hoekstra}, {Mamon}, {Melin}, {Moscardini}, {Paltani}, {Radovich},
  {Sartoris}, {Schultheis}, {Sereno}, {Weller}, {Burigana}, {Carvalho},
  {Corcione}, {Kurki-Suonio}, {Lilje}, {Sirri}, {Toledo-Moreo}, \&
  {Zamorani}}]{euclid_ada+19}
{Euclid Collaboration: Adam}, R., {Vannier}, M., {Maurogordato}, S., {et~al.}
  2019, \href{http://dx.doi.org/10.1051/0004-6361/201935088}{\color{blue}\aap},
  \href{https://ui.adsabs.harvard.edu/abs/2019A&A...627A..23E}{627, A23}

\bibitem[{{Euclid Collaboration: Mellier} {et~al.}(2024){Euclid Collaboration:
  Mellier}, {Abdurro'uf}, {Acevedo Barroso}, {Ach{\'u}carro}, {Adamek}, {Adam},
  {Addison}, {Aghanim}, {Aguena}, {Ajani}, {Akrami}, {Al-Bahlawan}, {Alavi},
  {Albuquerque}, {Alestas}, {Alguero}, {Allaoui}, {Allen}, {Allevato},
  {Alonso-Tetilla}, {Altieri}, {Alvarez-Candal}, {Amara}, {Amendola}, {Amiaux},
  {Andika}, {Andreon}, {Andrews}, {Angora}, {Angulo}, {Annibali}, {Anselmi},
  {Anselmi}, {Arcari}, {Archidiacono}, {Aric{\`o}}, {Arnaud}, {Arnouts},
  {Asgari}, {Asorey}, {Atayde}, {Atek}, {Atrio-Barandela}, {Aubert}, {Aubourg},
  {Auphan}, {Auricchio}, {Aussel}, {Aussel}, {Avelino}, {Avgoustidis}, {Avila},
  {Awan}, {Azzollini}, {Baccigalupi}, {Bachelet}, {Bacon}, {Baes}, {Bagley},
  {Bahr-Kalus}, {Balaguera-Antolinez}, {Balbinot}, {Balcells}, {Baldi},
  {Baldry}, {Balestra}, {Ballardini}, {Ballester}, {Balogh}, {Ba{\~n}ados},
  {Barbier}, {Bardelli}, {Barreiro}, {Barriere}, {Barros}, {Barthelemy},
  {Bartolo}, {Basset}, {Battaglia}, {Battisti}, {Baugh}, {Baumont},
  {Bazzanini}, {Beaulieu}, {Beckmann}, {Belikov}, {Bel}, {Bellagamba}, {Bella},
  {Bellini}, {Benabed}, {Bender}, {Benevento}, {Bennett}, {Benson},
  {Bergamini}, {Bermejo-Climent}, {Bernardeau}, {Bertacca}, {Berthe},
  {Berthier}, {Bethermin}, {Beutler}, {Bevillon}, {Bhargava}, {Bhatawdekar},
  {Bisigello}, {Biviano}, {Blake}, {Blanchard}, {Blazek}, {Blot}, {Bosco},
  {Bodendorf}, {Boenke}, {B{\"o}hringer}, {Bolzonella}, {Bonchi}, {Bonici},
  {Bonino}, {Bonino}, {Bonvin}, {Bon}, {Booth}, {Borgani}, {Borlaff},
  {Borsato}, {Bosco}, {Bose}, {Botticella}, {Boucaud}, {Bouche}, {Boucher},
  {Boutigny}, {Bouvard}, {Bouy}, {Bowler}, {Bozza}, {Bozzo}, {Branchini},
  {Brau-Nogue}, {Brekke}, {Bremer}, {Brescia}, {Breton}, {Brinchmann},
  {Brinckmann}, {Brockley-Blatt}, {Brodwin}, {Brouard}, {Brown}, {Bruton},
  {Bucko}, {Buddelmeijer}, {Buenadicha}, {Buitrago}, {Burger}, {Burigana},
  {Busillo}, {Busonero}, {Cabanac}, {Cabayol-Garcia}, {Cagliari}, {Caillat},
  {Caillat}, {Calabrese}, {Calabro}, {Calderone}, {Calura}, {Camacho Quevedo},
  {Camera}, {Campos}, {Canas-Herrera}, {Candini}, {Cantiello}, {Capobianco},
  {Cappellaro}, {Cappelluti}, {Cappi}, {Caputi}, {Cara}, {Carbone}, {Cardone},
  {Carella}, {Carlberg}, {Carle}, {Carminati}, {Caro}, {Carrasco}, {Carretero},
  {Carrilho}, {Carron Duque}, {Carry}, {Carvalho}, {Carvalho}, {Casas},
  {Casas}, {Casenove}, {Casey}, {Cassata}, {Castander}, {Castelao},
  {Castellano}, {Castiblanco}, {Castignani}, {Castro}, {Cavet}, {Cavuoti},
  {Chabaud}, {Chambers}, {Charles}, {Charlot}, {Chartab}, {Chary}, {Chaumeil},
  {Cho}, {Chon}, {Ciancetta}, {Ciliegi}, {Cimatti}, {Cimino}, {Cioni},
  {Claydon}, {Cleland}, {Cl{\'e}ment}, {Clements}, {Clerc}, {Clesse}, {Codis},
  {Cogato}, {Colbert}, {Cole}, {Coles}, {Collett}, {Collins}, {Colodro-Conde},
  {Colombo}, {Combes}, {Conforti}, {Congedo}, {Conseil}, {Conselice},
  {Contarini}, {Contini}, {Conversi}, {Cooray}, {Copin}, {Corasaniti},
  {Corcho-Caballero}, {Corcione}, {Cordes}, {Corpace}, {Correnti}, {Costanzi},
  {Costille}, {Courbin}, {Courcoult Mifsud}, {Courtois}, {Cousinou}, {Covone},
  {Cowell}, {Cragg}, {Cresci}, {Cristiani}, {Crocce}, {Cropper}, {E Crouzet},
  {Csizi}, {Cuby}, {Cucchetti}, {Cucciati}, {Cuillandre}, {Cunha}, {Cuozzo},
  {Daddi}, {D'Addona}, {Dafonte}, {Dagoneau}, {Dalessandro}, {Dalton},
  {D'Amico}, {Dannerbauer}, {Danto}, {Das}, {Da Silva}, {da Silva}, {Daste},
  {Davies}, {Davini}, {de Boer}, {Decarli}, {De Caro}, {Degaudenzi}, {Degni},
  {de Jong}, {de la Bella}, {de la Torre}, {Delhaise}, {Delley}, {Delucchi},
  {De Lucia}, {Denniston}, {De Paolis}, {De Petris}, {Derosa}, {Desai},
  {Desjacques}, {Despali}, {Desprez}, {De Vicente-Albendea}, {Deville}, {Dias},
  {D{\'\i}az-S{\'a}nchez}, {Diaz}, {Di Domizio}, {Diego}, {Di Ferdinando}, {Di
  Giorgio}, {Dimauro}, {Dinis}, {Dolag}, {Dolding}, {Dole}, {Dom{\'\i}nguez
  S{\'a}nchez}, {Dor{\'e}}, {Dournac}, {Douspis}, {Dreihahn}, {Droge}, {Dryer},
  {Dubath}, {Duc}, {Ducret}, {Duffy}, {Dufresne}, {Duncan}, {Dupac}, {Duret},
  {Durrer}, {Durret}, {Dusini}, {Ealet}, {Eggemeier}, {Eisenhardt}, {Elbaz},
  {Elkhashab}, {Ellien}, {Endicott}, {Enia}, {Erben}, {Escartin Vigo},
  {Escoffier}, {Escudero Sanz}, {Essert}, {Ettori}, {Ezziati}, {Fabbian},
  {Fabricius}, {Fang}, {Farina}, {Farina}, {Farinelli}, {Farrens}, {Faustini},
  {Feltre}, {Ferguson}, {Ferrando}, {Ferrari}, {Ferr{\'e}-Mateu}, {Ferreira},
  {Ferreras}, {Ferrero}, {Ferriol}, {Ferruit}, {Filleul}, {Finelli},
  {Finkelstein}, {Finoguenov}, {Fiorini}, {Flentge}, {Focardi}, {Fonseca},
  {Fontana}, {Fontanot}, {Fornari}, {Fosalba}, {Fossati}, {Fotopoulou},
  {Fouchez}, {Fourmanoit}, {Frailis}, {Fraix-Burnet}, {Franceschi}, {Franco},
  {Franzetti}, {Freihoefer}, {Frittoli}, {Frugier}, {Frusciante}, {Fumagalli},
  {Fumagalli}, {Fumana}, {Fu}, {Gabarra}, {Galeotta}, {Galluccio}, {Ganga},
  {Gao}, {Garc{\'\i}a-Bellido}, {Garcia}, {Gardner}, {Garilli},
  {Gaspar-Venancio}, {Gasparetto}, {Gautard}, {Gavazzi}, {Gaztanaga},
  {Genolet}, {Genova Santos}, {Gentile}, {George}, {Ghaffari}, {Giacomini},
  {Gianotti}, {Gibb}, {Gillard}, {Gillis}, {Ginolfi}, {Giocoli}, {Girardi},
  {Giri}, {Goh}, {G{\'o}mez-Alvarez}, {Gonzalez}, {Gonzalez}, {Gonzalez},
  {Gouyou Beauchamps}, {Gozaliasl}, {Gracia-Carpio}, {Grandis}, {Granett},
  {Granvik}, {Grazian}, {Gregorio}, {Grenet}, {Grillo}, {Grupp}, {Gruppioni},
  {Gruppuso}, {Guerbuez}, {Guerrini}, {Guidi}, {Guillard}, {Gutierrez},
  {Guttridge}, {Guzzo}, {Gwyn}, {Haapala}, {Haase}, {Haddow}, {Hailey}, {Hall},
  {Hall}, {Hamaus}, {Haridasu}, {Harnois-D{\'e}raps}, {Harper}, {Hartley},
  {Hasinger}, {Hassani}, {Hatch}, {Haugan}, {H{\"a}u{\ss}ler}, {Heavens},
  {Heisenberg}, {Helmi}, {Helou}, {Hemmati}, {Henares}, {Herent},
  {Hern{\'a}ndez-Monteagudo}, {Heuberger}, {Hewett}, {Heydenreich},
  {Hildebrandt}, {Hirschmann}, {Hjorth}, {Hoar}, {Hoekstra}, {Holland},
  {Holliman}, {Holmes}, {Hook}, {Horeau}, {Hormuth}, {Hornstrup}, {Hosseini},
  {Hu}, {Hudelot}, {Hudson}, {Huertas-Company}, {Huff}, {Hughes}, {Humphrey},
  {Hunt}, {Huynh}, {Ibata}, {Ichikawa}, {Iglesias-Groth}, {Ilbert}, {Ili{\'c}},
  {Ingoglia}, {Iodice}, {Israel}, {Israelsson}, {Izzo}, {Jablonka}, {Jackson},
  {Jacobson}, {Jafariyazani}, {Jahnke}, {Jansen}, {Jarvis}, {Jasche}, {Jauzac},
  {Jeffrey}, {Jhabvala}, {Jimenez-Teja}, {Jimenez Mu{\~n}oz}, {Joachimi},
  {Johansson}, {Joudaki}, {Jullo}, {Kajava}, {Kang}, {Kannawadi}, {Kansal},
  {Karagiannis}, {K{\"a}rcher}, {Kashlinsky}, {Kazandjian}, {Keck},
  {Keih{\"a}nen}, {Kerins}, {Kermiche}, {Khalil}, {Kiessling}, {Kiiveri},
  {Kilbinger}, {Kim}, {King}, {Kirkpatrick}, {Kitching}, {Kluge}, {Knabenhans},
  {Knapen}, {Knebe}, {Kneib}, {Kohley}, {Koopmans}, {Koskinen}, {Koulouridis},
  {Kou}, {Kov{\'a}cs}, {Kova\{{\v{c}}\}i{\'c}}, {Kowalczyk}, {Koyama},
  {Kraljic}, {Krause}, {Kruk}, {Kubik}, {Kuchner}, {Kuijken}, {K{\"u}mmel},
  {Kunz}, {Kurki-Suonio}, {Lacasa}, {Lacey}, {La Franca}, {Lagarde}, {Lahav},
  {Laigle}, {La Marca}, {La Marle}, {Lamine}, {Lam}, {Lan{\c{c}}on}, {Landt},
  {Langer}, {Lapi}, {Larcheveque}, {Larsen}, {Lattanzi}, {Laudisio}, {Laugier},
  {Laureijs}, {Lavaux}, {Lawrenson}, {Lazanu}, {Lazeyras}, {Le Boulc'h}, {Le
  Brun}, {Le Brun}, {Leclercq}, {Lee}, {Le Graet}, {Legrand}, {Leirvik}, {Le
  Jeune}, {Lembo}, {Le Mignant}, {Lepinzan}, {Lepori}, {Lesci}, {Lesgourgues},
  {Leuzzi}, {Levi}, {Liaudat}, {Libet}, {Liebing}, {Ligori}, {Lilje}, {Lin},
  {Linde}, {Linder}, {Lindholm}, {Linke}, {Li}, {Liu}, {Lloro}, {Lobo},
  {Lodieu}, {Lombardi}, {Lombriser}, {Lonare}, {Longo}, {L{\'o}pez-Caniego},
  {Lopez Lopez}, {Alvarez}, {Loureiro}, {Loveday}, {Lusso}, {Macias-Perez},
  {Maciaszek}, {Magliocchetti}, {Magnard}, {Magnier}, {Magro}, {Mahler},
  {Mainetti}, {Maino}, {Maiorano}, {Maiorano}, {Malavasi}, {Mamon}, {Mancini},
  {Mandelbaum}, {Manera}, {Manj{\'o}n-Garc{\'\i}a}, {Mannucci}, {Mansutti},
  {Manteiga Outeiro}, {Maoli}, {Maraston}, {Marcin}, {Marcos-Arenal},
  {Margalef-Bentabol}, {Marggraf}, {Marinucci}, {Marinucci}, {Markovic},
  {Marleau}, {Marpaud}, {Martignac}, {Mart{\'\i}n-Fleitas}, {Martin-Moruno},
  {Martin}, {Martinelli}, {Martinet}, {Martin}, {Martins}, {Marulli},
  {Massari}, {Massey}, {Masters}, {Matarrese}, {Matsuoka}, {Matthew},
  {Maughan}, {Mauri}, {Maurin}, {Maurogordato}, {McCarthy}, {McConnachie},
  {McCracken}, {McDonald}, {McEwen}, {McPartland}, {Medinaceli}, {Mehta},
  {Mei}, {Melchior}, {Melin}, {M{\'e}nard}, {Mendes}, {Mendez-Abreu},
  {Meneghetti}, {Mercurio}, {Merlin}, {Metcalf}, {Meylan}, {Migliaccio},
  {Mignoli}, {Miller}, {Miluzio}, {Milvang-Jensen}, {Mimoso}, {Miquel},
  {Miyatake}, {Mobasher}, {Mohr}, {Monaco}, {Mongui{\'o}}, {Montoro}, {Mora},
  {Moradinezhad Dizgah}, {Moresco}, {Moretti}, {Morgante}, {Morisset},
  {Moriya}, {Morris}, {Mortlock}, {Moscardini}, {Mota}, {Moustakas}, {Moutard},
  {M{\"u}ller}, {Munari}, {Murphree}, {Murray}, {Murray}, {Musi}, {Nadathur},
  {Nagam}, {Nagao}, {Naidoo}, {Nakajima}, {Nally}, {Natoli}, {Navarro-Alsina},
  {Navarro Girones}, {Neissner}, {Nersesian}, {Nesseris}, {Nguyen-Kim},
  {Nicastro}, {Nichol}, {Nielbock}, {Niemi}, {Nieto}, {Nilsson}, {Noller},
  {Norberg}, {Nourizonoz}, {Ntelis}, {Nucita}, {Nugent}, {Nunes}, {Nutma},
  {Ocampo}, {Odier}, {Oesch}, {Oguri}, {Magalhaes Oliveira}, {Onoue},
  {Oosterbroek}, {Oppizzi}, {Ordenovic}, {Osato}, {Pacaud}, {Pace}, {Padilla},
  {Paech}, {Pagano}, {Page}, {Palazzi}, {Paltani}, {Pamuk}, {Pandolfi},
  {Paoletti}, {Paolillo}, {Papaderos}, {Pardede}, {Parimbelli}, {Parmar},
  {Partmann}, {Pasian}, {Passalacqua}, {Paterson}, {Patrizii}, {Pattison},
  {Paulino-Afonso}, {Paviot}, {Peacock}, {Pearce}, {Pedersen}, {Peel},
  {Peletier}, {Pellejero Ibanez}, {Pello}, {Penny}, {Percival},
  {Perez-Garrido}, {Perotto}, {Pettorino}, {Pezzotta}, {Pezzuto}, {Philippon},
  {Piersanti}, {Pietroni}, {Piga}, {Pilo}, {Pires}, {Pisani}, {Pizzella},
  {Pizzuti}, {Plana}, {Polenta}, {Pollack}, {Poncet}, {P{\"o}ntinen}, {Pool},
  {Popa}, {Popa}, {Popp}, {Porciani}, {Porth}, {Potter}, {Poulain},
  {Pourtsidou}, {Pozzetti}, {Prandoni}, {Pratt}, {Prezelus}, {Prieto}, {Pugno},
  {Quai}, {Quilley}, {Racca}, {Raccanelli}, {R{\'a}cz}, {Radinovi{\'c}},
  {Radovich}, {Ragagnin}, {Ragnit}, {Raison}, {Ramos-Chernenko}, {Ranc},
  {Raylet}, {Rebolo}, {Refregier}, {Reimberg}, {Reiprich}, {Renk}, {Renzi},
  {Retre}, {Revaz}, {Reyl{\'e}}, {Reynolds}, {Rhodes}, {Ricci}, {Ricci},
  {Riccio}, {Ricken}, {Rissanen}, {Risso}, {Rix}, {Robin}, {Rocca-Volmerange},
  {Rocci}, {Rodenhuis}, {Rodighiero}, {Rodriguez Monroy}, {Rollins},
  {Romanello}, {Roman}, {Romelli}, {Romero-Gomez}, {Roncarelli}, {Rosati},
  {Rosset}, {Rossetti}, {Roster}, {Rottgering}, {Rozas-Fern{\'a}ndez}, {Ruane},
  {Rubino-Martin}, {Rudolph}, {Ruppin}, {Rusholme}, {Sacquegna},
  {S{\'a}ez-Casares}, {Saga}, {Saglia}, {Sahl{\'e}n}, {Saifollahi}, {Sakr},
  {Salvalaggio}, {Salvaterra}, {Salvati}, {Salvato}, {Salvignol},
  {S{\'a}nchez}, {Sanchez}, {Sanders}, {Sapone}, {Saponara}, {Sarpa}, {Sarron},
  {Sartori}, {Sassolas}, {Sauniere}, {Sauvage}, {Sawicki}, {Scaramella},
  {Scarlata}, {Scharr{\'e}}, {Schaye}, {Schewtschenko}, {Schindler},
  {Schinnerer}, {Schirmer}, {Schmidt}, {Schmidt}, {Schmidt}, {Schneider},
  {Schneider}, {Schneider}, {Sch{\"o}neberg}, {Schrabback}, {Schultheis},
  {Schulz}, {Schwartz}, {Sciotti}, {Scodeggio}, {Scognamiglio}, {Scott},
  {Scottez}, {Secroun}, {Sefusatti}, {Seidel}, {Seiffert}, {Sellentin},
  {Selwood}, {Semboloni}, {Sereno}, {Serjeant}, {Serrano}, {Shankar},
  {Sharples}, {Short}, {Shulevski}, {Shuntov}, {Sias}, {Sikkema}, {Silvestri},
  {Simon}, {Sirignano}, {Sirri}, {Skottfelt}, {Slezak}, {Sluse}, {Smith},
  {Smith}, {Smith}, {Smit}, {Soldano}, {Solheim}, {Sorce}, {Sorrenti},
  {Soubrie}, {Spinoglio}, {Spurio Mancini}, {Stadel}, {Stagnaro}, {Stanco},
  {Stanford}, {Starck}, {Stassi}, {Steinwagner}, {Stern}, {Stone}, {Strada},
  {Strafella}, {Stramaccioni}, {Surace}, {Sureau}, {Suyu}, {Swindells},
  {Szafraniec}, {Szapudi}, {Taamoli}, {Talia}, {Tallada-Cresp{\'\i}},
  {Tanidis}, {Tao}, {Tarr{\'\i}o}, {Tavagnacco}, {Taylor}, {Taylor}, {Taylor},
  {Teixeira}, {Tenti}, {Teodoro Idiago}, {Teplitz}, {Tereno}, {Tessore},
  {Testa}, {Testera}, {Tewes}, {Teyssier}, {Theret}, {Thizy}, {Thomas}, {Toba},
  {Toft}, {Toledo-Moreo}, {Tolstoy}, {Tommasi}, {Torbaniuk}, {Torradeflot},
  {Tortora}, {Tosi}, {Tosti}, {Trifoglio}, {Troja}, {Trombetti}, {Tronconi},
  {Tsedrik}, {Tsyganov}, {Tucci}, {Tutusaus}, {Uhlemann}, {Ulivi}, {Urbano},
  {Vacher}, {Vaillon}, {Valdes}, {Valentijn}, {Valenziano}, {Valieri},
  {Valiviita}, {Van den Broeck}, {Vassallo}, {Vavrek}, {Venemans}, {Venhola},
  {Ventura}, {Verdoes Kleijn}, {Vergani}, {Verma}, {Vernizzi}, {Veropalumbo},
  {Verza}, {Vescovi}, {Vibert}, {Viel}, {Vielzeuf}, {Viglione}, {Viitanen},
  {Villaescusa-Navarro}, {Vinciguerra}, {Visticot}, {Voggel}, {von
  Wietersheim-Kramsta}, {Vriend}, {Wachter}, {Walmsley}, {Walth}, {Walton},
  {Walton}, {Wander}, {Wang}, {Wang}, {Weaver}, {Weller}, {Whalen}, {Wiesmann},
  {Wilde}, {Williams}, {Winther}, {Wittje}, {Wong}, {Wright}, {Yankelevich},
  {Yeung}, {Youles}, {Yung}, {Zacchei}, {Zalesky}, {Zamorani}, {Zamorano
  Vitorelli}, {Zanoni Marc}, {Zennaro}, {Zerbi}, {Zinchenko}, {Zoubian},
  {Zucca}, \& {Zumalacarregui}}]{euclid_I_24}
{Euclid Collaboration: Mellier}, Y., {Abdurro'uf}, {Acevedo Barroso}, J.~A.,
  {et~al.} 2024,
  \href{https://ui.adsabs.harvard.edu/abs/2024arXiv240513491E}{\href{http://dx.doi.org/10.48550/arXiv.2405.13491}{\color{blue}arXiv
  e-prints}, arXiv:2405.13491}

\bibitem[{{Euclid Collaboration: Scaramella} {et~al.}(2022){Euclid
  Collaboration: Scaramella}, {Amiaux}, {Mellier}, {Burigana}, {Carvalho},
  {Cuillandre}, {Da Silva}, {Derosa}, {Dinis}, {Maiorano}, {Maris}, {Tereno},
  {Laureijs}, {Boenke}, {Buenadicha}, {Dupac}, {Gaspar Venancio},
  {G{\'o}mez-{\'A}lvarez}, {Hoar}, {Lorenzo Alvarez}, {Racca},
  {Saavedra-Criado}, {Schwartz}, {Vavrek}, {Schirmer}, {Aussel}, {Azzollini},
  {Cardone}, {Cropper}, {Ealet}, {Garilli}, {Gillard}, {Granett}, {Guzzo},
  {Hoekstra}, {Jahnke}, {Kitching}, {Maciaszek}, {Meneghetti}, {Miller},
  {Nakajima}, {Niemi}, {Pasian}, {Percival}, {Pottinger}, {Sauvage},
  {Scodeggio}, {Wachter}, {Zacchei}, {Aghanim}, {Amara}, {Auphan}, {Auricchio},
  {Awan}, {Balestra}, {Bender}, {Bodendorf}, {Bonino}, {Branchini},
  {Brau-Nogue}, {Brescia}, {Candini}, {Capobianco}, {Carbone}, {Carlberg},
  {Carretero}, {Casas}, {Castander}, {Castellano}, {Cavuoti}, {Cimatti},
  {Cledassou}, {Congedo}, {Conselice}, {Conversi}, {Copin}, {Corcione},
  {Costille}, {Courbin}, {Degaudenzi}, {Douspis}, {Dubath}, {Duncan}, {Dusini},
  {Farrens}, {Ferriol}, {Fosalba}, {Fourmanoit}, {Frailis}, {Franceschi},
  {Franzetti}, {Fumana}, {Gillis}, {Giocoli}, {Grazian}, {Grupp}, {Haugan},
  {Holmes}, {Hormuth}, {Hudelot}, {Kermiche}, {Kiessling}, {Kilbinger},
  {Kohley}, {Kubik}, {K{\"u}mmel}, {Kunz}, {Kurki-Suonio}, {Lahav}, {Ligori},
  {Lilje}, {Lloro}, {Mansutti}, {Marggraf}, {Markovic}, {Marulli}, {Massey},
  {Maurogordato}, {Melchior}, {Merlin}, {Meylan}, {Mohr}, {Moresco}, {Morin},
  {Moscardini}, {Munari}, {Nichol}, {Padilla}, {Paltani}, {Peacock},
  {Pedersen}, {Pettorino}, {Pires}, {Poncet}, {Popa}, {Pozzetti}, {Raison},
  {Rebolo}, {Rhodes}, {Rix}, {Roncarelli}, {Rossetti}, {Saglia}, {Schneider},
  {Schrabback}, {Secroun}, {Seidel}, {Serrano}, {Sirignano}, {Sirri},
  {Skottfelt}, {Stanco}, {Starck}, {Tallada-Cresp{\'\i}}, {Tavagnacco},
  {Taylor}, {Teplitz}, {Toledo-Moreo}, {Torradeflot}, {Trifoglio}, {Valentijn},
  {Valenziano}, {Verdoes Kleijn}, {Wang}, {Welikala}, {Weller}, {Wetzstein},
  {Zamorani}, {Zoubian}, {Andreon}, {Baldi}, {Bardelli}, {Boucaud}, {Camera},
  {Di Ferdinando}, {Fabbian}, {Farinelli}, {Galeotta}, {Graci{\'a}-Carpio},
  {Maino}, {Medinaceli}, {Mei}, {Neissner}, {Polenta}, {Renzi}, {Romelli},
  {Rosset}, {Sureau}, {Tenti}, {Vassallo}, {Zucca}, {Baccigalupi},
  {Balaguera-Antol{\'\i}nez}, {Battaglia}, {Biviano}, {Borgani}, {Bozzo},
  {Cabanac}, {Cappi}, {Casas}, {Castignani}, {Colodro-Conde}, {Coupon},
  {Courtois}, {Cuby}, {de la Torre}, {Desai}, {Dole}, {Fabricius}, {Farina},
  {Ferreira}, {Finelli}, {Flose-Reimberg}, {Fotopoulou}, {Ganga}, {Gozaliasl},
  {Hook}, {Keihanen}, {Kirkpatrick}, {Liebing}, {Lindholm}, {Mainetti},
  {Martinelli}, {Martinet}, {Maturi}, {McCracken}, {Metcalf}, {Morgante},
  {Nightingale}, {Nucita}, {Patrizii}, {Potter}, {Riccio}, {S{\'a}nchez},
  {Sapone}, {Schewtschenko}, {Schultheis}, {Scottez}, {Teyssier}, {Tutusaus},
  {Valiviita}, {Viel}, {Vriend}, \& {Whittaker}}]{euclid_pre_sca+al22}
{Euclid Collaboration: Scaramella}, R., {Amiaux}, J., {Mellier}, Y., {et~al.}
  2022, \href{http://dx.doi.org/10.1051/0004-6361/202141938}{\color{blue}\aap},
  \href{https://ui.adsabs.harvard.edu/abs/2022A&A...662A.112E}{662, A112}

\bibitem[{{Evrard} {et~al.}(2008){Evrard}, {Bialek}, {Busha}, {White}, {Habib},
  {Heitmann}, {Warren}, {Rasia}, {Tormen}, {Moscardini}, {Power}, {Jenkins},
  {Gao}, {Frenk}, {Springel}, {White}, \& {Diemand}}]{evr+al08}
{Evrard}, A.~E., {Bialek}, J., {Busha}, M., {et~al.} 2008,
  \href{http://dx.doi.org/10.1086/521616}{\color{blue}\apj},
  \href{http://adsabs.harvard.edu/abs/2008ApJ...672..122E}{672, 122}

\bibitem[{{Fadda} {et~al.}(1996){Fadda}, {Girardi}, {Giuricin}, {Mardirossian},
  \& {Mezzetti}}]{fad+al96}
{Fadda}, D., {Girardi}, M., {Giuricin}, G., {Mardirossian}, F., \& {Mezzetti},
  M. 1996, \href{http://dx.doi.org/10.1086/178180}{\color{blue}\apj},
  \href{http://ads.astro.puc.cl/abs/1996ApJ...473..670F}{473, 670}

\bibitem[{{Farahi} {et~al.}(2018){Farahi}, {Guglielmo}, {Evrard}, {Poggianti},
  {Adami}, {Ettori}, {Gastaldello}, {Giles}, {Maughan}, {Rapetti}, {Sereno},
  {Altieri}, {Baldry}, {Birkinshaw}, {Bolzonella}, {Bongiorno}, {Brown},
  {Chiappetti}, {Driver}, {Elyiv}, {Garilli}, {Guennou}, {Hopkins}, {Iovino},
  {Koulouridis}, {Liske}, {Maurogordato}, {Owers}, {Pacaud}, {Pierre},
  {Plionis}, {Ponman}, {Robotham}, {Sadibekova}, {Scodeggio}, {Tuffs}, \&
  {Valtchanov}}]{xxl_XXIII_far+al18}
{Farahi}, A., {Guglielmo}, V., {Evrard}, A.~E., {et~al.} 2018,
  \href{http://dx.doi.org/10.1051/0004-6361/201731321}{\color{blue}\aap},
  \href{https://ui.adsabs.harvard.edu/abs/2018A&A...620A...8F}{620, A8}

\bibitem[{{Ferragamo} {et~al.}(2021){Ferragamo}, {Barrena},
  {Rubi{\~n}o-Mart{\'\i}n}, {Aguado-Barahona}, {Streblyanska}, {Tramonte},
  {G{\'e}nova-Santos}, {Hempel}, \& {Lietzen}}]{fer+al21}
{Ferragamo}, A., {Barrena}, R., {Rubi{\~n}o-Mart{\'\i}n}, J.~A., {et~al.} 2021,
  \href{http://dx.doi.org/10.1051/0004-6361/202140382}{\color{blue}\aap},
  \href{https://ui.adsabs.harvard.edu/abs/2021A&A...655A.115F}{655, A115}

\bibitem[{{Ferragamo} {et~al.}(2020){Ferragamo}, {Rubi{\~n}o-Mart{\'\i}n},
  {Betancort-Rijo}, {Munari}, {Sartoris}, \& {Barrena}}]{fer+al20}
{Ferragamo}, A., {Rubi{\~n}o-Mart{\'\i}n}, J.~A., {Betancort-Rijo}, J.,
  {et~al.} 2020,
  \href{http://dx.doi.org/10.1051/0004-6361/201834837}{\color{blue}\aap},
  \href{https://ui.adsabs.harvard.edu/abs/2020A&A...641A..41F}{641, A41}

\bibitem[{{Fox} \& {Pen}(2002)}]{fo+pe02}
{Fox}, D.~C. \& {Pen}, U.-L. 2002,
  \href{http://dx.doi.org/10.1086/340897}{\color{blue}ApJ},
  \href{http://adsabs.harvard.edu/abs/2002ApJ...574...38F}{574, 38}

\bibitem[{{Gavazzi}(2005)}]{gav05}
{Gavazzi}, R. 2005,
  \href{http://dx.doi.org/10.1051/0004-6361:20053166}{\color{blue}\aap},
  \href{http://ads.astro.puc.cl/abs/2005A%26A...443..793G}{443, 793}

\bibitem[{{Ghirardini} {et~al.}(2024){Ghirardini}, {Bulbul}, {Artis}, {Clerc},
  {Garrel}, {Grandis}, {Kluge}, {Liu}, {Bahar}, {Balzer}, {Chiu}, {Comparat},
  {Gruen}, {Kleinebreil}, {Krippendorf}, {Merloni}, {Nandra}, {Okabe},
  {Pacaud}, {Predehl}, {Ramos-Ceja}, {Reiprich}, {Sanders}, {Schrabback},
  {Seppi}, {Zelmer}, {Zhang}, {Bornemann}, {Brunner}, {Burwitz}, {Coutinho},
  {Dennerl}, {Freyberg}, {Friedrich}, {Gaida}, {Gueguen}, {Haberl}, {Kink},
  {Lamer}, {Li}, {Liu}, {Maitra}, {Meidinger}, {Mueller}, {Miyatake},
  {Miyazaki}, {Robrade}, {Schwope}, \& {Stewart}}]{erosita_ghi+al24}
{Ghirardini}, V., {Bulbul}, E., {Artis}, E., {et~al.} 2024,
  \href{http://dx.doi.org/10.1051/0004-6361/202348852}{\color{blue}\aap},
  \href{https://ui.adsabs.harvard.edu/abs/2024A&A...689A.298G}{689, A298}

\bibitem[{{Gifford} \& {Miller}(2013)}]{gif+al13}
{Gifford}, D. \& {Miller}, C.~J. 2013,
  \href{http://dx.doi.org/10.1088/2041-8205/768/2/L32}{\color{blue}\apjl},
  \href{https://ui.adsabs.harvard.edu/abs/2013ApJ...768L..32G}{768, L32}

\bibitem[{{Girardi} {et~al.}(1993){Girardi}, {Biviano}, {Giuricin},
  {Mardirossian}, \& {Mezzetti}}]{gir+al93}
{Girardi}, M., {Biviano}, A., {Giuricin}, G., {Mardirossian}, F., \&
  {Mezzetti}, M. 1993,
  \href{http://dx.doi.org/10.1086/172256}{\color{blue}\apj},
  \href{https://ui.adsabs.harvard.edu/abs/1993ApJ...404...38G}{404, 38}

\bibitem[{{Girardi} {et~al.}(1998){Girardi}, {Giuricin}, {Mardirossian},
  {Mezzetti}, \& {Boschin}}]{gir+al98}
{Girardi}, M., {Giuricin}, G., {Mardirossian}, F., {Mezzetti}, M., \&
  {Boschin}, W. 1998,
  \href{http://dx.doi.org/10.1086/306157}{\color{blue}\apj},
  \href{http://ads.astro.puc.cl/abs/1998ApJ...505...74G}{505, 74}

\bibitem[{{Girardi} \& {Mezzetti}(2001)}]{gi+me01}
{Girardi}, M. \& {Mezzetti}, M. 2001,
  \href{http://dx.doi.org/10.1086/318665}{\color{blue}\apj},
  \href{http://adsabs.harvard.edu/abs/2001ApJ...548...79G}{548, 79}

\bibitem[{{Haines} {et~al.}(2015){Haines}, {Pereira}, {Smith}, {Egami},
  {Babul}, {Finoguenov}, {Ziparo}, {McGee}, {Rawle}, {Okabe}, \&
  {Moran}}]{haines+al15}
{Haines}, C.~P., {Pereira}, M.~J., {Smith}, G.~P., {et~al.} 2015,
  \href{http://dx.doi.org/10.1088/0004-637X/806/1/101}{\color{blue}\apj},
  \href{https://ui.adsabs.harvard.edu/abs/2015ApJ...806..101H}{806, 101}

\bibitem[{{Haines} {et~al.}(2013){Haines}, {Pereira}, {Smith}, {Egami},
  {Sanderson}, {Babul}, {Finoguenov}, {Merluzzi}, {Busarello}, {Rawle}, \&
  {Okabe}}]{haines+al13}
{Haines}, C.~P., {Pereira}, M.~J., {Smith}, G.~P., {et~al.} 2013,
  \href{http://dx.doi.org/10.1088/0004-637X/775/2/126}{\color{blue}\apj},
  \href{https://ui.adsabs.harvard.edu/abs/2013ApJ...775..126H}{775, 126}

\bibitem[{{Haines} {et~al.}(2009){Haines}, {Smith}, {Egami}, {Ellis}, {Moran},
  {Sanderson}, {Merluzzi}, {Busarello}, \& {Smith}}]{haines+al09}
{Haines}, C.~P., {Smith}, G.~P., {Egami}, E., {et~al.} 2009,
  \href{http://dx.doi.org/10.1088/0004-637X/704/1/126}{\color{blue}\apj},
  \href{https://ui.adsabs.harvard.edu/abs/2009ApJ...704..126H}{704, 126}

\bibitem[{{Hartigan} \& {Hartigan}(1985)}]{ha+ha85}
{Hartigan}, J.~A. \& {Hartigan}, P.~M. 1985,
  \href{http://dx.doi.org/10.1214/aos/1176346577}{\color{blue}Ann. Statist.},
  13, 13

\bibitem[{{Hilton} {et~al.}(2021){Hilton}, {Sif{\'o}n}, {Naess},
  {Madhavacheril}, {Oguri}, {Rozo}, {Rykoff}, {Abbott}, {Adhikari}, {Aguena},
  {Aiola}, {Allam}, {Amodeo}, {Amon}, {Annis}, {Ansarinejad}, {Aros-Bunster},
  {Austermann}, {Avila}, {Bacon}, {Battaglia}, {Beall}, {Becker}, {Bernstein},
  {Bertin}, {Bhandarkar}, {Bhargava}, {Bond}, {Brooks}, {Burke}, {Calabrese},
  {Carrasco Kind}, {Carretero}, {Choi}, {Choi}, {Conselice}, {da Costa},
  {Costanzi}, {Crichton}, {Crowley}, {D{\"u}nner}, {Denison}, {Devlin},
  {Dicker}, {Diehl}, {Dietrich}, {Doel}, {Duff}, {Duivenvoorden}, {Dunkley},
  {Everett}, {Ferraro}, {Ferrero}, {Fert{\'e}}, {Flaugher}, {Frieman},
  {Gallardo}, {Garc{\'\i}a-Bellido}, {Gaztanaga}, {Gerdes}, {Giles}, {Golec},
  {Gralla}, {Grandis}, {Gruen}, {Gruendl}, {Gschwend}, {Gutierrez}, {Han},
  {Hartley}, {Hasselfield}, {Hill}, {Hilton}, {Hincks}, {Hinton}, {Ho},
  {Honscheid}, {Hoyle}, {Hubmayr}, {Huffenberger}, {Hughes}, {Jaelani}, {Jain},
  {James}, {Jeltema}, {Kent}, {Knowles}, {Koopman}, {Kuehn}, {Lahav}, {Lima},
  {Lin}, {Lokken}, {Loubser}, {MacCrann}, {Maia}, {Marriage}, {Martin},
  {McMahon}, {Melchior}, {Menanteau}, {Miquel}, {Miyatake}, {Moodley},
  {Morgan}, {Mroczkowski}, {Nati}, {Newburgh}, {Niemack}, {Nishizawa},
  {Ogando}, {Orlowski-Scherer}, {Page}, {Palmese}, {Partridge},
  {Paz-Chinch{\'o}n}, {Phakathi}, {Plazas}, {Robertson}, {Romer}, {Carnero
  Rosell}, {Salatino}, {Sanchez}, {Schaan}, {Schillaci}, {Sehgal}, {Serrano},
  {Shin}, {Simon}, {Smith}, {Soares-Santos}, {Spergel}, {Staggs}, {Storer},
  {Suchyta}, {Swanson}, {Tarle}, {Thomas}, {To}, {Trac}, {Ullom}, {Vale}, {Van
  Lanen}, {Vavagiakis}, {De Vicente}, {Wilkinson}, {Wollack}, {Xu}, \&
  {Zhang}}]{act_hil+al21}
{Hilton}, M., {Sif{\'o}n}, C., {Naess}, S., {et~al.} 2021,
  \href{http://dx.doi.org/10.3847/1538-4365/abd023}{\color{blue}\apjs},
  \href{https://ui.adsabs.harvard.edu/abs/2021ApJS..253....3H}{253, 3}

\bibitem[{{Hoekstra} {et~al.}(2015){Hoekstra}, {Herbonnet}, {Muzzin}, {Babul},
  {Mahdavi}, {Viola}, \& {Cacciato}}]{hoe+al15}
{Hoekstra}, H., {Herbonnet}, R., {Muzzin}, A., {et~al.} 2015,
  \href{http://dx.doi.org/10.1093/mnras/stv275}{\color{blue}\mnras},
  \href{http://adsabs.harvard.edu/abs/2015MNRAS.449..685H}{449, 685}

\bibitem[{{Hoekstra} {et~al.}(2012){Hoekstra}, {Mahdavi}, {Babul}, \&
  {Bildfell}}]{hoe+al12}
{Hoekstra}, H., {Mahdavi}, A., {Babul}, A., \& {Bildfell}, C. 2012,
  \href{http://dx.doi.org/10.1111/j.1365-2966.2012.22072.x}{\color{blue}\mnras},
  \href{http://adsabs.harvard.edu/abs/2012MNRAS.427.1298H}{427, 1298}

\bibitem[{{Ivezi{\'c}} {et~al.}(2019){Ivezi{\'c}}, {Kahn}, {Tyson}, {Abel},
  {Acosta}, {Allsman}, {Alonso}, {AlSayyad}, {Anderson}, {Andrew}, {Angel},
  {Angeli}, {Ansari}, {Antilogus}, {Araujo}, {Armstrong}, {Arndt}, {Astier},
  {Aubourg}, {Auza}, {Axelrod}, {Bard}, {Barr}, {Barrau}, {Bartlett}, {Bauer},
  {Bauman}, {Baumont}, {Bechtol}, {Bechtol}, {Becker}, {Becla}, {Beldica},
  {Bellavia}, {Bianco}, {Biswas}, {Blanc}, {Blazek}, {Blandford}, {Bloom},
  {Bogart}, {Bond}, {Booth}, {Borgland}, {Borne}, {Bosch}, {Boutigny},
  {Brackett}, {Bradshaw}, {Brandt}, {Brown}, {Bullock}, {Burchat}, {Burke},
  {Cagnoli}, {Calabrese}, {Callahan}, {Callen}, {Carlin}, {Carlson},
  {Chandrasekharan}, {Charles-Emerson}, {Chesley}, {Cheu}, {Chiang}, {Chiang},
  {Chirino}, {Chow}, {Ciardi}, {Claver}, {Cohen-Tanugi}, {Cockrum}, {Coles},
  {Connolly}, {Cook}, {Cooray}, {Covey}, {Cribbs}, {Cui}, {Cutri}, {Daly},
  {Daniel}, {Daruich}, {Daubard}, {Daues}, {Dawson}, {Delgado}, {Dellapenna},
  {de Peyster}, {de Val-Borro}, {Digel}, {Doherty}, {Dubois},
  {Dubois-Felsmann}, {Durech}, {Economou}, {Eifler}, {Eracleous}, {Emmons},
  {Fausti Neto}, {Ferguson}, {Figueroa}, {Fisher-Levine}, {Focke}, {Foss},
  {Frank}, {Freemon}, {Gangler}, {Gawiser}, {Geary}, {Gee}, {Geha}, {Gessner},
  {Gibson}, {Gilmore}, {Glanzman}, {Glick}, {Goldina}, {Goldstein}, {Goodenow},
  {Graham}, {Gressler}, {Gris}, {Guy}, {Guyonnet}, {Haller}, {Harris},
  {Hascall}, {Haupt}, {Hernandez}, {Herrmann}, {Hileman}, {Hoblitt}, {Hodgson},
  {Hogan}, {Howard}, {Huang}, {Huffer}, {Ingraham}, {Innes}, {Jacoby}, {Jain},
  {Jammes}, {Jee}, {Jenness}, {Jernigan}, {Jevremovi{\'c}}, {Johns}, {Johnson},
  {Johnson}, {Jones}, {Juramy-Gilles}, {Juri{\'c}}, {Kalirai}, {Kallivayalil},
  {Kalmbach}, {Kantor}, {Karst}, {Kasliwal}, {Kelly}, {Kessler}, {Kinnison},
  {Kirkby}, {Knox}, {Kotov}, {Krabbendam}, {Krughoff}, {Kub{\'a}nek},
  {Kuczewski}, {Kulkarni}, {Ku}, {Kurita}, {Lage}, {Lambert}, {Lange},
  {Langton}, {Le Guillou}, {Levine}, {Liang}, {Lim}, {Lintott}, {Long},
  {Lopez}, {Lotz}, {Lupton}, {Lust}, {MacArthur}, {Mahabal}, {Mandelbaum},
  {Markiewicz}, {Marsh}, {Marshall}, {Marshall}, {May}, {McKercher}, {McQueen},
  {Meyers}, {Migliore}, {Miller}, {Mills}, {Miraval}, {Moeyens}, {Moolekamp},
  {Monet}, {Moniez}, {Monkewitz}, {Montgomery}, {Morrison}, {Mueller},
  {Muller}, {Mu{\~n}oz Arancibia}, {Neill}, {Newbry}, {Nief}, {Nomerotski},
  {Nordby}, {O'Connor}, {Oliver}, {Olivier}, {Olsen}, {O'Mullane}, {Ortiz},
  {Osier}, {Owen}, {Pain}, {Palecek}, {Parejko}, {Parsons}, {Pease},
  {Peterson}, {Peterson}, {Petravick}, {Libby Petrick}, {Petry},
  {Pierfederici}, {Pietrowicz}, {Pike}, {Pinto}, {Plante}, {Plate}, {Plutchak},
  {Price}, {Prouza}, {Radeka}, {Rajagopal}, {Rasmussen}, {Regnault}, {Reil},
  {Reiss}, {Reuter}, {Ridgway}, {Riot}, {Ritz}, {Robinson}, {Roby}, {Roodman},
  {Rosing}, {Roucelle}, {Rumore}, {Russo}, {Saha}, {Sassolas}, {Schalk},
  {Schellart}, {Schindler}, {Schmidt}, {Schneider}, {Schneider}, {Schoening},
  {Schumacher}, {Schwamb}, {Sebag}, {Selvy}, {Sembroski}, {Seppala}, {Serio},
  {Serrano}, {Shaw}, {Shipsey}, {Sick}, {Silvestri}, {Slater}, {Smith},
  {Smith}, {Sobhani}, {Soldahl}, {Storrie-Lombardi}, {Stover}, {Strauss},
  {Street}, {Stubbs}, {Sullivan}, {Sweeney}, {Swinbank}, {Szalay}, {Takacs},
  {Tether}, {Thaler}, {Thayer}, {Thomas}, {Thornton}, {Thukral}, {Tice},
  {Trilling}, {Turri}, {Van Berg}, {Vanden Berk}, {Vetter}, {Virieux},
  {Vucina}, {Wahl}, {Walkowicz}, {Walsh}, {Walter}, {Wang}, {Wang}, {Warner},
  {Wiecha}, {Willman}, {Winters}, {Wittman}, {Wolff}, {Wood-Vasey}, {Wu},
  {Xin}, {Yoachim}, \& {Zhan}}]{lsst_ive+al19}
{Ivezi{\'c}}, {\v{Z}}., {Kahn}, S.~M., {Tyson}, J.~A., {et~al.} 2019,
  \href{http://dx.doi.org/10.3847/1538-4357/ab042c}{\color{blue}\apj},
  \href{https://ui.adsabs.harvard.edu/abs/2019ApJ...873..111I}{873, 111}

\bibitem[{{Kaiser}(1984)}]{kai84}
{Kaiser}, N. 1984, \href{http://dx.doi.org/10.1086/184341}{\color{blue}\apjl},
  \href{http://adsabs.harvard.edu/abs/1984ApJ...284L...9K}{284, L9}

\bibitem[{{Kaiser}(1986)}]{kai86}
{Kaiser}, N. 1986, \mnras,
  \href{http://adsabs.harvard.edu/abs/1986MNRAS.222..323K}{222, 323}

\bibitem[{{Kay} {et~al.}(2012){Kay}, {Peel}, {Short}, {Thomas}, {Young},
  {Battye}, {Liddle}, \& {Pearce}}]{kay+al12}
{Kay}, S.~T., {Peel}, M.~W., {Short}, C.~J., {et~al.} 2012,
  \href{http://dx.doi.org/10.1111/j.1365-2966.2012.20623.x}{\color{blue}\mnras},
  \href{http://adsabs.harvard.edu/abs/2012MNRAS.422.1999K}{422, 1999}

\bibitem[{{Kim} {et~al.}(2024){Kim}, {Sayers}, {Sereno}, {Bartalucci},
  {Chappuis}, {De Grandi}, {De Luca}, {De Petris}, {Donahue}, {Eckert},
  {Ettori}, {Gaspari}, {Gastaldello}, {Gavazzi}, {Gavidia}, {Ghizzardi},
  {Iqbal}, {Kay}, {Lovisari}, {Maughan}, {Mazzotta}, {Okabe}, {Pointecouteau},
  {Pratt}, {Rossetti}, \& {Umetsu}}]{kim+al24}
{Kim}, J., {Sayers}, J., {Sereno}, M., {et~al.} 2024,
  \href{http://dx.doi.org/10.1051/0004-6361/202347399}{\color{blue}\aap},
  \href{https://ui.adsabs.harvard.edu/abs/2024A&A...686A..97K}{686, A97}

\bibitem[{{Kollmeier} {et~al.}(2019){Kollmeier}, {Anderson}, {Blanc},
  {Blanton}, {Covey}, {Crane}, {Drory}, {Frinchaboy}, {Froning}, {Johnson},
  {Kneib}, {Kreckel}, {Merloni}, {Pellegrini}, {Pogge}, {Ramirez}, {Rix},
  {Sayres}, {S{\'a}nchez-Gallego}, {Shen}, {Tkachenko}, {Trump}, {Tuttle},
  {Weijmans}, {Zasowski}, {Barbuy}, {Beaton}, {Bergemann}, {Bochanski},
  {Brandt}, {Casey}, {Cherinka}, {Eracleous}, {Fan}, {Garc{\'\i}a}, {Green},
  {Hekker}, {Lane}, {Longa-Pe{\~n}a}, {Mathur}, {Meza}, {Minchev}, {Myers},
  {Nidever}, {Nitschelm}, {O'Connell}, {Price-Whelan}, {Raddick}, {Rossi},
  {Sankrit}, {Simon}, {Stutz}, {Ting}, {Trakhtenbrot}, {Weaver}, {Willmer}, \&
  {Weinberg}}]{sdss_V_kol+19}
{Kollmeier}, J., {Anderson}, S.~F., {Blanc}, G.~A., {et~al.} 2019, in Bulletin
  of the American Astronomical Society, Vol.~51,
  \href{https://ui.adsabs.harvard.edu/abs/2019BAAS...51g.274K}{274}

\bibitem[{{Kopylov} \& {Kopylova}(2010)}]{ko+ko10}
{Kopylov}, A.~I. \& {Kopylova}, F.~G. 2010,
  \href{http://dx.doi.org/10.1134/S1990341310030016}{\color{blue}Astrophysical
  Bulletin}, \href{https://ui.adsabs.harvard.edu/abs/2010AstBu..65..205K}{65,
  205}

\bibitem[{{Kurtz} \& {Mink}(1998)}]{ku+mi98}
{Kurtz}, M.~J. \& {Mink}, D.~J. 1998,
  \href{http://dx.doi.org/10.1086/316207}{\color{blue}\pasp},
  \href{https://ui.adsabs.harvard.edu/abs/1998PASP..110..934K}{110, 934}

\bibitem[{{Kurtz} {et~al.}(1992){Kurtz}, {Mink}, {Wyatt}, {Fabricant},
  {Torres}, {Kriss}, \& {Tonry}}]{kur+al92}
{Kurtz}, M.~J., {Mink}, D.~J., {Wyatt}, W.~F., {et~al.} 1992, in Astronomical
  Society of the Pacific Conference Series, Vol.~25, Astronomical Data Analysis
  Software and Systems I, ed. D.~M. {Worrall}, C.~{Biemesderfer}, \&
  J.~{Barnes},
  \href{https://ui.adsabs.harvard.edu/abs/1992ASPC...25..432K}{432}

\bibitem[{{Lakhchaura} \& {Singh}(2014)}]{lak+al14}
{Lakhchaura}, K. \& {Singh}, K.~P. 2014,
  \href{http://dx.doi.org/10.1088/0004-6256/147/6/156}{\color{blue}\aj},
  \href{https://ui.adsabs.harvard.edu/abs/2014AJ....147..156L}{147, 156}

\bibitem[{{Lesci} {et~al.}(2022){Lesci}, {Marulli}, {Moscardini}, {Sereno},
  {Veropalumbo}, {Maturi}, {Giocoli}, {Radovich}, {Bellagamba}, {Roncarelli},
  {Bardelli}, {Contarini}, {Covone}, {Ingoglia}, {Nanni}, \&
  {Puddu}}]{les+al22}
{Lesci}, G.~F., {Marulli}, F., {Moscardini}, L., {et~al.} 2022,
  \href{http://dx.doi.org/10.1051/0004-6361/202040194}{\color{blue}\aap},
  \href{https://ui.adsabs.harvard.edu/abs/2022A&A...659A..88L}{659, A88}

\bibitem[{{Lesci} {et~al.}(2023){Lesci}, {Veropalumbo}, {Sereno}, {Marulli},
  {Moscardini}, \& {Giocoli}}]{les+al23}
{Lesci}, G.~F., {Veropalumbo}, A., {Sereno}, M., {et~al.} 2023,
  \href{http://dx.doi.org/10.1051/0004-6361/202346261}{\color{blue}\aap},
  \href{https://ui.adsabs.harvard.edu/abs/2023A&A...674A..80L}{674, A80}

\bibitem[{{Limousin} {et~al.}(2013){Limousin}, {Morandi}, {Sereno},
  {Meneghetti}, {Ettori}, {Bartelmann}, \& {Verdugo}}]{lim+al13}
{Limousin}, M., {Morandi}, A., {Sereno}, M., {et~al.} 2013,
  \href{http://dx.doi.org/10.1007/s11214-013-9980-y}{\color{blue}\ssr},
  \href{http://adsabs.harvard.edu/abs/2013SSRv..177..155L}{177, 155}

\bibitem[{{Logan} {et~al.}(2022){Logan}, {Maughan}, {Diaferio}, {Duffy},
  {Geller}, {Rines}, \& {Sohn}}]{log+al22}
{Logan}, C. H.~A., {Maughan}, B.~J., {Diaferio}, A., {et~al.} 2022,
  \href{http://dx.doi.org/10.1051/0004-6361/202243347}{\color{blue}\aap},
  \href{https://ui.adsabs.harvard.edu/abs/2022A&A...665A.124L}{665, A124}

\bibitem[{{{\L}okas} \& {Mamon}(2001)}]{lo+ma01}
{{\L}okas}, E.~L. \& {Mamon}, G.~A. 2001,
  \href{http://dx.doi.org/10.1046/j.1365-8711.2001.04007.x}{\color{blue}\mnras},
  \href{https://ui.adsabs.harvard.edu/abs/2001MNRAS.321..155L}{321, 155}

\bibitem[{{Lovisari} {et~al.}(2020){Lovisari}, {Schellenberger}, {Sereno},
  {Ettori}, {Pratt}, {Forman}, {Jones}, {Andrade-Santos}, {Randall}, \&
  {Kraft}}]{lov+al20}
{Lovisari}, L., {Schellenberger}, G., {Sereno}, M., {et~al.} 2020,
  \href{http://dx.doi.org/10.3847/1538-4357/ab7997}{\color{blue}\apj},
  \href{https://ui.adsabs.harvard.edu/abs/2020ApJ...892..102L}{892, 102}

\bibitem[{{Mamon} {et~al.}(2013){Mamon}, {Biviano}, \& {Bou{\'e}}}]{mam+al13}
{Mamon}, G.~A., {Biviano}, A., \& {Bou{\'e}}, G. 2013,
  \href{http://dx.doi.org/10.1093/mnras/sts565}{\color{blue}\mnras},
  \href{https://ui.adsabs.harvard.edu/abs/2013MNRAS.429.3079M}{429, 3079}

\bibitem[{{Mamon} {et~al.}(2010){Mamon}, {Biviano}, \& {Murante}}]{mam+al10}
{Mamon}, G.~A., {Biviano}, A., \& {Murante}, G. 2010,
  \href{http://dx.doi.org/10.1051/0004-6361/200913948}{\color{blue}\aap},
  \href{https://ui.adsabs.harvard.edu/abs/2010A&A...520A..30M}{520, A30}

\bibitem[{{Mamon} \& {{\L}okas}(2005)}]{ma+lo05}
{Mamon}, G.~A. \& {{\L}okas}, E.~L. 2005,
  \href{http://dx.doi.org/10.1111/j.1365-2966.2005.09400.x}{\color{blue}\mnras},
  \href{https://ui.adsabs.harvard.edu/abs/2005MNRAS.363..705M}{363, 705}

\bibitem[{{Maurogordato} {et~al.}(1997){Maurogordato}, {Proust}, {Cappi},
  {Slezak}, \& {Martin}}]{mau+al97}
{Maurogordato}, S., {Proust}, D., {Cappi}, A., {Slezak}, E., \& {Martin}, J.~M.
  1997, \href{http://dx.doi.org/10.1051/aas:1997165}{\color{blue}\aaps},
  \href{https://ui.adsabs.harvard.edu/abs/1997A&AS..123..411M}{123, 411}

\bibitem[{{Medezinski} {et~al.}(2018){Medezinski}, {Oguri}, {Nishizawa},
  {Speagle}, {Miyatake}, {Umetsu}, {Leauthaud}, {Murata}, {Mandelbaum},
  {Sif{\'o}n}, {Strauss}, {Huang}, {Simet}, {Okabe}, {Tanaka}, \&
  {Komiyama}}]{hsc_med+al18b}
{Medezinski}, E., {Oguri}, M., {Nishizawa}, A.~J., {et~al.} 2018,
  \href{http://dx.doi.org/10.1093/pasj/psy009}{\color{blue}\pasj},
  \href{https://ui.adsabs.harvard.edu/abs/2018PASJ...70...30M}{70, 30}

\bibitem[{{Melchior} {et~al.}(2015){Melchior}, {Suchyta}, {Huff}, {Hirsch},
  {Kacprzak}, {Rykoff}, {Gruen}, {Armstrong}, {Bacon}, {Bechtol}, {Bernstein},
  {Bridle}, {Clampitt}, {Honscheid}, {Jain}, {Jouvel}, {Krause}, {Lin},
  {MacCrann}, {Patton}, {Plazas}, {Rowe}, {Vikram}, {Wilcox}, {Young}, {Zuntz},
  {Abbott}, {Abdalla}, {Allam}, {Banerji}, {Bernstein}, {Bernstein}, {Bertin},
  {Buckley-Geer}, {Burke}, {Castander}, {da Costa}, {Cunha}, {Depoy}, {Desai},
  {Diehl}, {Doel}, {Estrada}, {Evrard}, {Neto}, {Fernandez}, {Finley},
  {Flaugher}, {Frieman}, {Gaztanaga}, {Gerdes}, {Gruendl}, {Gutierrez},
  {Jarvis}, {Karliner}, {Kent}, {Kuehn}, {Kuropatkin}, {Lahav}, {Maia},
  {Makler}, {Marriner}, {Marshall}, {Merritt}, {Miller}, {Miquel}, {Mohr},
  {Neilsen}, {Nichol}, {Nord}, {Reil}, {Roe}, {Roodman}, {Sako}, {Sanchez},
  {Santiago}, {Schindler}, {Schubnell}, {Sevilla-Noarbe}, {Sheldon}, {Smith},
  {Soares-Santos}, {Swanson}, {Sypniewski}, {Tarle}, {Thaler}, {Thomas},
  {Tucker}, {Walker}, {Wechsler}, {Weller}, \& {Wester}}]{mel+al15}
{Melchior}, P., {Suchyta}, E., {Huff}, E., {et~al.} 2015,
  \href{http://dx.doi.org/10.1093/mnras/stv398}{\color{blue}\mnras},
  \href{https://ui.adsabs.harvard.edu/abs/2015MNRAS.449.2219M}{449, 2219}

\bibitem[{{Monteiro-Oliveira} {et~al.}(2017){Monteiro-Oliveira}, {Cypriano},
  {Machado}, {Lima Neto}, {Ribeiro}, {Sodr{\'e}}, \& {Dupke}}]{mon+al17b}
{Monteiro-Oliveira}, R., {Cypriano}, E.~S., {Machado}, R.~E.~G., {et~al.} 2017,
  \href{http://dx.doi.org/10.1093/mnras/stw3238}{\color{blue}\mnras},
  \href{https://ui.adsabs.harvard.edu/abs/2017MNRAS.466.2614M}{466, 2614}

\bibitem[{{Munari} {et~al.}(2013){Munari}, {Biviano}, {Borgani}, {Murante}, \&
  {Fabjan}}]{mun+al13}
{Munari}, E., {Biviano}, A., {Borgani}, S., {Murante}, G., \& {Fabjan}, D.
  2013, \href{http://dx.doi.org/10.1093/mnras/stt049}{\color{blue}\mnras},
  \href{http://adsabs.harvard.edu/abs/2013MNRAS.430.2638M}{430, 2638}

\bibitem[{{Navarro} {et~al.}(1996){Navarro}, {Frenk}, \& {White}}]{nfw96}
{Navarro}, J.~F., {Frenk}, C.~S., \& {White}, S.~D.~M. 1996, \apj,
  \href{http://adsabs.harvard.edu/abs/1996ApJ...462..563N}{462, 563}

\bibitem[{{Noordeh} {et~al.}(2020){Noordeh}, {Canning}, {King}, {Allen},
  {Mantz}, {Morris}, {Ehlert}, {von der Linden}, {Brandt}, {Luo}, {Xue}, \&
  {Kelly}}]{noo+al20}
{Noordeh}, E., {Canning}, R.~E.~A., {King}, A., {et~al.} 2020,
  \href{http://dx.doi.org/10.1093/mnras/staa2682}{\color{blue}\mnras},
  \href{https://ui.adsabs.harvard.edu/abs/2020MNRAS.498.4095N}{498, 4095}

\bibitem[{{Okabe} \& {Smith}(2016)}]{ok+sm16}
{Okabe}, N. \& {Smith}, G.~P. 2016,
  \href{http://dx.doi.org/10.1093/mnras/stw1539}{\color{blue}\mnras},
  \href{http://adsabs.harvard.edu/abs/2016MNRAS.461.3794O}{461, 3794}

\bibitem[{{Old} {et~al.}(2014){Old}, {Skibba}, {Pearce}, {Croton}, {Muldrew},
  {Mu{\~n}oz-Cuartas}, {Gifford}, {Gray}, {von der Linden}, {Mamon},
  {Merrifield}, {M{\"u}ller}, {Pearson}, {Ponman}, {Saro}, {Sepp}, {Sif{\'o}n},
  {Tempel}, {Tundo}, {Wang}, \& {Wojtak}}]{old+al14}
{Old}, L., {Skibba}, R.~A., {Pearce}, F.~R., {et~al.} 2014,
  \href{http://dx.doi.org/10.1093/mnras/stu545}{\color{blue}\mnras},
  \href{https://ui.adsabs.harvard.edu/abs/2014MNRAS.441.1513O}{441, 1513}

\bibitem[{{Planck Collaboration: Ade} {et~al.}(2014{\natexlab{a}}){Planck
  Collaboration: Ade}, {Aghanim}, {Armitage-Caplan}, {Arnaud}, {Ashdown},
  {Atrio-Barandela}, {Aumont}, {Aussel}, {Baccigalupi}, \&
  et~al.}]{planck_2013_XXIX}
{Planck Collaboration: Ade}, P.~A.~R., {Aghanim}, N., {Armitage-Caplan}, C.,
  {et~al.} 2014{\natexlab{a}},
  \href{http://dx.doi.org/10.1051/0004-6361/201321523}{\color{blue}\aap},
  \href{http://adsabs.harvard.edu/abs/2014A%26A...571A..29P}{571, A29}

\bibitem[{{Planck Collaboration: Ade} {et~al.}(2014{\natexlab{b}}){Planck
  Collaboration: Ade}, {Aghanim}, {Armitage-Caplan}, {Arnaud}, {Ashdown},
  {Atrio-Barandela}, {Aumont}, {Baccigalupi}, {Banday}, \&
  et~al.}]{planck_2013_XX}
{Planck Collaboration: Ade}, P.~A.~R., {Aghanim}, N., {Armitage-Caplan}, C.,
  {et~al.} 2014{\natexlab{b}},
  \href{http://dx.doi.org/10.1051/0004-6361/201321521}{\color{blue}\aap},
  \href{http://adsabs.harvard.edu/abs/2014A%26A...571A..20P}{571, A20}

\bibitem[{{Planck Collaboration: Ade} {et~al.}(2016{\natexlab{a}}){Planck
  Collaboration: Ade}, {Aghanim}, {Arnaud}, {Ashdown}, {Aumont}, {Baccigalupi},
  {Banday}, {Barreiro}, {Barrena}, \& et~al.}]{planck_2015_XXVII}
{Planck Collaboration: Ade}, P.~A.~R., {Aghanim}, N., {Arnaud}, M., {et~al.}
  2016{\natexlab{a}},
  \href{http://dx.doi.org/10.1051/0004-6361/201525823}{\color{blue}\aap},
  \href{http://adsabs.harvard.edu/abs/2016A%26A...594A..27P}{594, A27}

\bibitem[{{Planck Collaboration: Ade} {et~al.}(2016{\natexlab{b}}){Planck
  Collaboration: Ade}, {Aghanim}, {Arnaud}, {Ashdown}, {Aumont}, {Baccigalupi},
  {Banday}, {Barreiro}, {Bartlett}, \& et~al.}]{planck_2015_XXIV}
{Planck Collaboration: Ade}, P.~A.~R., {Aghanim}, N., {Arnaud}, M., {et~al.}
  2016{\natexlab{b}},
  \href{http://dx.doi.org/10.1051/0004-6361/201525833}{\color{blue}\aap},
  \href{http://adsabs.harvard.edu/abs/2016A%26A...594A..24P}{594, A24}

\bibitem[{{Power} {et~al.}(2012){Power}, {Knebe}, \& {Knollmann}}]{pow+al12}
{Power}, C., {Knebe}, A., \& {Knollmann}, S.~R. 2012,
  \href{http://dx.doi.org/10.1111/j.1365-2966.2011.19820.x}{\color{blue}\mnras},
  \href{https://ui.adsabs.harvard.edu/abs/2012MNRAS.419.1576P}{419, 1576}

\bibitem[{{Pratt} {et~al.}(2019){Pratt}, {Arnaud}, {Biviano}, {Eckert},
  {Ettori}, {Nagai}, {Okabe}, \& {Reiprich}}]{pra+al19}
{Pratt}, G.~W., {Arnaud}, M., {Biviano}, A., {et~al.} 2019,
  \href{http://dx.doi.org/10.1007/s11214-019-0591-0}{\color{blue}\ssr},
  \href{https://ui.adsabs.harvard.edu/abs/2019SSRv..215...25P}{215, 25}

\bibitem[{{Rines} {et~al.}(2016){Rines}, {Geller}, {Diaferio}, \&
  {Hwang}}]{rin+al16}
{Rines}, K.~J., {Geller}, M.~J., {Diaferio}, A., \& {Hwang}, H.~S. 2016,
  \href{http://dx.doi.org/10.3847/0004-637X/819/1/63}{\color{blue}\apj},
  \href{https://ui.adsabs.harvard.edu/abs/2016ApJ...819...63R}{819, 63}

\bibitem[{{Saro} {et~al.}(2013){Saro}, {Mohr}, {Bazin}, \& {Dolag}}]{sar+al13}
{Saro}, A., {Mohr}, J.~J., {Bazin}, G., \& {Dolag}, K. 2013,
  \href{http://dx.doi.org/10.1088/0004-637X/772/1/47}{\color{blue}\apj},
  \href{http://adsabs.harvard.edu/abs/2013ApJ...772...47S}{772, 47}

\bibitem[{{Sartoris} {et~al.}(2016){Sartoris}, {Biviano}, {Fedeli}, {Bartlett},
  {Borgani}, {Costanzi}, {Giocoli}, {Moscardini}, {Weller}, {Ascaso},
  {Bardelli}, {Maurogordato}, \& {Viana}}]{sar+al16}
{Sartoris}, B., {Biviano}, A., {Fedeli}, C., {et~al.} 2016,
  \href{http://dx.doi.org/10.1093/mnras/stw630}{\color{blue}\mnras},
  \href{https://ui.adsabs.harvard.edu/abs/2016MNRAS.459.1764S}{459, 1764}

\bibitem[{{Sereno}(2015)}]{ser15_comalit_III}
{Sereno}, M. 2015,
  \href{http://dx.doi.org/10.1093/mnras/stu2505}{\color{blue}\mnras},
  \href{http://adsabs.harvard.edu/abs/2015MNRAS.450.3665S}{450, 3665}

\bibitem[{{Sereno}(2016)}]{ser16_lira}
{Sereno}, M. 2016,
  \href{http://dx.doi.org/10.1093/mnras/stv2374}{\color{blue}\mnras},
  \href{http://adsabs.harvard.edu/abs/2016MNRAS.455.2149S}{455, 2149}

\bibitem[{{Sereno} {et~al.}(2017{\natexlab{a}}){Sereno}, {Covone}, {Izzo},
  {Ettori}, {Coupon}, \& {Lieu}}]{ser+al17_psz2lens}
{Sereno}, M., {Covone}, G., {Izzo}, L., {et~al.} 2017{\natexlab{a}},
  \href{http://dx.doi.org/10.1093/mnras/stx2085}{\color{blue}\mnras},
  \href{http://adsabs.harvard.edu/abs/2017MNRAS.472.1946S}{472, 1946}

\bibitem[{{Sereno} \& {Ettori}(2015{\natexlab{a}})}]{se+et15_comalit_IV}
{Sereno}, M. \& {Ettori}, S. 2015{\natexlab{a}},
  \href{http://dx.doi.org/10.1093/mnras/stv814}{\color{blue}\mnras},
  \href{http://adsabs.harvard.edu/abs/2015MNRAS.450.3675S}{450, 3675}

\bibitem[{{Sereno} \& {Ettori}(2015{\natexlab{b}})}]{se+et15_comalit_I}
{Sereno}, M. \& {Ettori}, S. 2015{\natexlab{b}},
  \href{http://dx.doi.org/10.1093/mnras/stv810}{\color{blue}\mnras},
  \href{http://adsabs.harvard.edu/abs/2015MNRAS.450.3633S}{450, 3633}

\bibitem[{{Sereno} \& {Ettori}(2017)}]{se+et17_comalit_V}
{Sereno}, M. \& {Ettori}, S. 2017,
  \href{http://dx.doi.org/10.1093/mnras/stx576}{\color{blue}\mnras},
  \href{http://adsabs.harvard.edu/abs/2017MNRAS.468.3322S}{468, 3322}

\bibitem[{{Sereno} {et~al.}(2017{\natexlab{b}}){Sereno}, {Ettori},
  {Meneghetti}, {Sayers}, {Umetsu}, {Merten}, {Chiu}, \&
  {Zitrin}}]{ser+al17_CLUMP_M1206}
{Sereno}, M., {Ettori}, S., {Meneghetti}, M., {et~al.} 2017{\natexlab{b}},
  \href{http://dx.doi.org/10.1093/mnras/stx326}{\color{blue}\mnras},
  \href{http://adsabs.harvard.edu/abs/2017MNRAS.467.3801S}{467, 3801}

\bibitem[{{Sereno} {et~al.}(2015){Sereno}, {Ettori}, \&
  {Moscardini}}]{ser+al15_comalit_II}
{Sereno}, M., {Ettori}, S., \& {Moscardini}, L. 2015,
  \href{http://dx.doi.org/10.1093/mnras/stv809}{\color{blue}\mnras},
  \href{http://adsabs.harvard.edu/abs/2015MNRAS.450.3649S}{450, 3649}

\bibitem[{{Sereno} {et~al.}(2018{\natexlab{a}}){Sereno}, {Giocoli}, {Izzo},
  {Marulli}, {Veropalumbo}, {Ettori}, {Moscardini}, {Covone}, {Ferragamo},
  {Barrena}, \& {Streblyanska}}]{ser+al18_psz2lens}
{Sereno}, M., {Giocoli}, C., {Izzo}, L., {et~al.} 2018{\natexlab{a}},
  \href{http://dx.doi.org/10.1038/s41550-018-0508-y}{\color{blue}Nature
  Astronomy}, \href{https://ui.adsabs.harvard.edu/abs/2018NatAs...2..744S}{2,
  744}

\bibitem[{{Sereno} {et~al.}(2021){Sereno}, {Lovisari}, {Cui}, \&
  {Schellenberger}}]{ser+al21}
{Sereno}, M., {Lovisari}, L., {Cui}, W., \& {Schellenberger}, G. 2021,
  \href{http://dx.doi.org/10.1093/mnras/stab2435}{\color{blue}\mnras},
  \href{https://ui.adsabs.harvard.edu/abs/2021MNRAS.507.5214S}{507, 5214}

\bibitem[{{Sereno} {et~al.}(2018{\natexlab{b}}){Sereno}, {Umetsu}, {Ettori},
  {Sayers}, {Chiu}, {Meneghetti}, {Vega-Ferrero}, \&
  {Zitrin}}]{ser+al18_CLUMP_I}
{Sereno}, M., {Umetsu}, K., {Ettori}, S., {et~al.} 2018{\natexlab{b}},
  \href{http://dx.doi.org/10.3847/2041-8213/aac6d9}{\color{blue}\apj},
  \href{https://ui.adsabs.harvard.edu/abs/2018ApJ...860L...4S}{860, L4}

\bibitem[{{Sif{\'o}n} {et~al.}(2016){Sif{\'o}n}, {Battaglia}, {Hasselfield},
  {Menanteau}, {Barrientos}, {Bond}, {Crichton}, {Devlin}, {D{\"u}nner},
  {Hilton}, {Hincks}, {Hlozek}, {Huffenberger}, {Hughes}, {Infante},
  {Kosowsky}, {Marsden}, {Marriage}, {Moodley}, {Niemack}, {Page}, {Spergel},
  {Staggs}, {Trac}, \& {Wollack}}]{sif+al16}
{Sif{\'o}n}, C., {Battaglia}, N., {Hasselfield}, M., {et~al.} 2016,
  \href{http://dx.doi.org/10.1093/mnras/stw1284}{\color{blue}\mnras},
  \href{http://adsabs.harvard.edu/abs/2016MNRAS.461..248S}{461, 248}

\bibitem[{{Sif{\'o}n} {et~al.}(2015){Sif{\'o}n}, {Hoekstra}, {Cacciato},
  {Viola}, {K{\"o}hlinger}, {van der Burg}, {Sand}, \& {Graham}}]{sif+al15}
{Sif{\'o}n}, C., {Hoekstra}, H., {Cacciato}, M., {et~al.} 2015,
  \href{http://dx.doi.org/10.1051/0004-6361/201424435}{\color{blue}\aap},
  \href{https://ui.adsabs.harvard.edu/abs/2015A&A...575A..48S}{575, A48}

\bibitem[{{Sif{\'o}n} {et~al.}(2014){Sif{\'o}n}, {Menanteau}, {Hughes},
  {Carrasco}, \& {Barrientos}}]{sif+al14}
{Sif{\'o}n}, C., {Menanteau}, F., {Hughes}, J.~P., {Carrasco}, M., \&
  {Barrientos}, L.~F. 2014,
  \href{http://dx.doi.org/10.1051/0004-6361/201321638}{\color{blue}\aap},
  \href{https://ui.adsabs.harvard.edu/abs/2014A&A...562A..43S}{562, A43}

\bibitem[{{Smith} {et~al.}(2016){Smith}, {Mazzotta}, {Okabe}, {Ziparo},
  {Mulroy}, {Babul}, {Finoguenov}, {McCarthy}, {Lieu}, {Bah{\'e}}, {Bourdin},
  {Evrard}, {Futamase}, {Haines}, {Jauzac}, {Marrone}, {Martino}, {May},
  {Taylor}, \& {Umetsu}}]{smi+al16}
{Smith}, G.~P., {Mazzotta}, P., {Okabe}, N., {et~al.} 2016,
  \href{http://dx.doi.org/10.1093/mnrasl/slv175}{\color{blue}\mnras},
  \href{http://adsabs.harvard.edu/abs/2016MNRAS.456L..74S}{456, L74}

\bibitem[{{Sohn} {et~al.}(2020){Sohn}, {Geller}, {Diaferio}, \&
  {Rines}}]{soh+al20}
{Sohn}, J., {Geller}, M.~J., {Diaferio}, A., \& {Rines}, K.~J. 2020,
  \href{http://dx.doi.org/10.3847/1538-4357/ab6e6a}{\color{blue}\apj},
  \href{https://ui.adsabs.harvard.edu/abs/2020ApJ...891..129S}{891, 129}

\bibitem[{{Tiret} {et~al.}(2007){Tiret}, {Combes}, {Angus}, {Famaey}, \&
  {Zhao}}]{tir+al07}
{Tiret}, O., {Combes}, F., {Angus}, G.~W., {Famaey}, B., \& {Zhao}, H.~S. 2007,
  \href{http://dx.doi.org/10.1051/0004-6361:20078569}{\color{blue}\aap},
  \href{https://ui.adsabs.harvard.edu/abs/2007A&A...476L...1T}{476, L1}

\bibitem[{{Tonry} \& {Davis}(1979)}]{to+da79}
{Tonry}, J. \& {Davis}, M. 1979,
  \href{http://dx.doi.org/10.1086/112569}{\color{blue}\aj},
  \href{https://ui.adsabs.harvard.edu/abs/1979AJ.....84.1511T}{84, 1511}

\bibitem[{{Umetsu}(2020)}]{ume20}
{Umetsu}, K. 2020,
  \href{http://dx.doi.org/10.1007/s00159-020-00129-w}{\color{blue}\aapr},
  \href{https://ui.adsabs.harvard.edu/abs/2020A&ARv..28....7U}{28, 7}

\bibitem[{{Umetsu} {et~al.}(2014){Umetsu}, {Medezinski}, {Nonino}, {Merten},
  {Postman}, {Meneghetti}, {Donahue}, {Czakon}, {Molino}, {Seitz}, {Gruen},
  {Lemze}, {Balestra}, {Ben{\'{\i}}tez}, {Biviano}, {Broadhurst}, {Ford},
  {Grillo}, {Koekemoer}, {Melchior}, {Mercurio}, {Moustakas}, {Rosati}, \&
  {Zitrin}}]{ume+al14}
{Umetsu}, K., {Medezinski}, E., {Nonino}, M., {et~al.} 2014,
  \href{http://dx.doi.org/10.1088/0004-637X/795/2/163}{\color{blue}\apj},
  \href{http://adsabs.harvard.edu/abs/2014ApJ...795..163U}{795, 163}

\bibitem[{{Umetsu} {et~al.}(2020){Umetsu}, {Sereno}, {Lieu}, {Miyatake},
  {Medezinski}, {Nishizawa}, {Giles}, {Gastaldello}, {McCarthy}, {Kilbinger},
  {Birkinshaw}, {Ettori}, {Okabe}, {Chiu}, {Coupon}, {Eckert}, {Fujita},
  {Higuchi}, {Koulouridis}, {Maughan}, {Miyazaki}, {Oguri}, {Pacaud}, {Pierre},
  {Rapetti}, \& {Smith}}]{ume+al20}
{Umetsu}, K., {Sereno}, M., {Lieu}, M., {et~al.} 2020,
  \href{http://dx.doi.org/10.3847/1538-4357/ab6bca}{\color{blue}\apj},
  \href{https://ui.adsabs.harvard.edu/abs/2020ApJ...890..148U}{890, 148}

\bibitem[{{Umetsu} {et~al.}(2016){Umetsu}, {Zitrin}, {Gruen}, {Merten},
  {Donahue}, \& {Postman}}]{ume+al16b}
{Umetsu}, K., {Zitrin}, A., {Gruen}, D., {et~al.} 2016,
  \href{http://dx.doi.org/10.3847/0004-637X/821/2/116}{\color{blue}\apj},
  \href{http://adsabs.harvard.edu/abs/2016ApJ...821..116U}{821, 116}

\bibitem[{{Voit}(2005)}]{voi05}
{Voit}, G.~M. 2005,
  \href{http://dx.doi.org/10.1103/RevModPhys.77.207}{\color{blue}Reviews of
  Modern Physics}, \href{http://adsabs.harvard.edu/abs/2005RvMP...77..207V}{77,
  207}

\bibitem[{{von der Linden} {et~al.}(2014{\natexlab{a}}){von der Linden},
  {Allen}, {Applegate}, {Kelly}, {Allen}, {Ebeling}, {Burchat}, {Burke},
  {Donovan}, {Morris}, {Blandford}, {Erben}, \& {Mantz}}]{wtg_I_14}
{von der Linden}, A., {Allen}, M.~T., {Applegate}, D.~E., {et~al.}
  2014{\natexlab{a}},
  \href{http://dx.doi.org/10.1093/mnras/stt1945}{\color{blue}\mnras},
  \href{http://adsabs.harvard.edu/abs/2014MNRAS.439....2V}{439, 2}

\bibitem[{{von der Linden} {et~al.}(2014{\natexlab{b}}){von der Linden},
  {Mantz}, {Allen}, {Applegate}, {Kelly}, {Morris}, {Wright}, {Allen},
  {Burchat}, {Burke}, {Donovan}, \& {Ebeling}}]{lin+al14}
{von der Linden}, A., {Mantz}, A., {Allen}, S.~W., {et~al.} 2014{\natexlab{b}},
  \href{http://dx.doi.org/10.1093/mnras/stu1423}{\color{blue}\mnras},
  \href{http://adsabs.harvard.edu/abs/2014MNRAS.443.1973V}{443, 1973}

\bibitem[{{Wainer} \& {Thissen}(1976)}]{wa+th76}
{Wainer}, H. \& {Thissen}, D. 1976,
  \href{http://dx.doi.org/10.1007/BF02291695}{\color{blue}Ann. Statist}, 41, 41

\bibitem[{{Wojtak} {et~al.}(2007){Wojtak}, {{\L}okas}, {Mamon},
  {Gottl{\"o}ber}, {Prada}, \& {Moles}}]{woj+al07}
{Wojtak}, R., {{\L}okas}, E.~L., {Mamon}, G.~A., {et~al.} 2007,
  \href{http://dx.doi.org/10.1051/0004-6361:20066813}{\color{blue}\aap},
  \href{http://ads.astro.puc.cl/abs/2007A%26A...466..437W}{466, 437}

\bibitem[{{Yahil} \& {Vidal}(1977)}]{ya+vi77}
{Yahil}, A. \& {Vidal}, N.~V. 1977,
  \href{http://dx.doi.org/10.1086/155257}{\color{blue}\apj},
  \href{https://ui.adsabs.harvard.edu/abs/1977ApJ...214..347Y}{214, 347}

\end{thebibliography}

\begin{appendix}

\section{Analytical model}
\label{sec_nfw}

In this work, we describe the halo matter profile with a NFW model \citep{nfw96},
\begin{equation}
\label{eq_nfw_1}
M(<r) = M_\Delta \frac{g_\text{NFW}(c_\Delta)}{g_\text{NFW}(x)} \, ,
\end{equation}
where $x = r / r_\text{s}$ is the dimensionless radius, the scale radius is $r_\text{s} = r_\Delta / c_\Delta$, and
\begin{equation}
\label{eq_nfw_2}
g_\text{NFW} (x) = \frac{1}{\ln(1+x) - \frac{x}{1+x}} \, .
\end{equation}

Anisotropy in the velocity distribution in a spherical system can be described as
\begin{equation}
\label{eq_vel_ani_1}
\beta_\sigma = 1 - \frac{\sigma^2_\theta}{\sigma^2_\text{r}} \, ,
\end{equation}
where $\sigma_\text{r}$ and $\sigma_\theta$ stand for the radial and tangential component of the velocity dispersion, respectively. \citet[ML,][]{ma+lo05} proposed a realistic model for orbits that are nearly isotropic in the inner regions and anisotropic at large radii,
\begin{equation}
\label{eq_vel_ani_2}
\beta_\text{ML} = \frac{1}{2} \frac{r}{r + r_\beta} \, ,
\end{equation}
where $r_\beta$ is the anisotropy radius. In this work, we considered $r_\beta$ identical to the scale radius $r_\text{s}$ \citep{mam+al10,mam+al13,tir+al07}.

The radial  velocity dispersion $\sigma_\text{r}$ can be obtained as a solution of the spherical Jeans equation (see, for example, Eq. (A1) in \citealt{mam+al13}). We assume that the tracers follow the mass distribution. Solving the integral for the ML anisotropy model with $r_\beta = r_\text{s}$, we obtain
\begin{equation}
\label{eq_vel_ani_3}
\sigma_\text{r}^2 = V_\Delta^2 c_\Delta g_\text{NFW}(c_\Delta) \tilde{\sigma}_\text{r}^2 (x)\, ,
\end{equation}
where $V_\Delta^2 = G M_\Delta /r_\Delta$ is the squared circular velocity at $r_\Delta$ and
\begin{multline}
\label{eq_vel_ani_4}
\tilde{\sigma}_\text{r}^2 (x) = - \left[
     1 + 2x + 2x (1+x) \ln\left(\frac{x}{1+x} \right) 
\right] \\
+ \frac{1}{6} x (1+x)  \\
\times \left(
        \pi^2 + \frac{3}{x} + 9 \ln(x) 
        - 3 (1+x) (3 x - 1)  \frac{\ln (1+x)}{x^2}  \right. \\
        \left. + 3 \ln^2(1+x) - 6\, \text{Li}_2(-x) \right) \, ,
\end{multline}
where $\text{Li}_2$ is the dilogarithm.

The kinetic energy for arbitrary $\beta_\sigma$ can be expressed as an integral, see, for example, Eq. (23) in \citet{lo+ma01}, whose solution for the ML model can be written as
\begin{equation}
\label{eq_vel_ani_5}
T(<r) = \frac{| W_\infty|}{2}  \tilde{T} (x) \, ,
\end{equation}
where $W_\infty$ is the potential energy at infinity \citep[Eq. 22 in][]{lo+ma01},
\begin{equation}
\label{eq_vel_ani_6}
W_\infty = - \frac{1}{2} c_\Delta g_\text{NFW}^2(c_\Delta) V_\Delta^2 M_\Delta\, ,
\end{equation}
and
\begin{multline}
\label{eq_vel_ani_7}
\tilde{T} (x) = 
- 2  \ln(1 + x)  \\
+ x  \left(
    \frac{\pi^2}{3} - \frac{1}{1 + x} + \ln^2(1 + x)  + 2\, \text{Li}_2(-x) 
    \right) \, .
\end{multline}
The 1D velocity dispersion is obtained as
\begin{equation}
\label{eq_vel_ani_8}
3\, \sigma_\text{1D}(r) = 2 \frac{T(<r)} {M(<r)} \, .
\end{equation}

Finally, the line-of-sight velocity dispersion at projected radius $R$, $\sigma_\text{los}(R)$ \citep[Eq. 44 in][]{lo+ma01}, and the velocity dispersion within a projected radius $R$, $\sigma_\text{ap}(R)$ \citep[Eqs. 46-48 in][]{lo+ma01} can be obtained by numerical integration.



\section{Dynamical masses}
\label{sec_tab_masses}

\begin{sidewaystable*}
\caption{Dynamical masses.}
\label{tab_masses}
\resizebox{\hsize}{!}{
\begin{tabular}{l rrrrrr r@{$\,\pm\,$}lr@{$\,\pm\,$}l cc r@{$\,\pm\,$}l r@{$\,\pm\,$}l cc r@{$\,\pm\,$}l r@{$\,\pm\,$}l}
\hline \hline
PSZ2 & RA & DEC & $N_\text{can}$ & $N_\text{members}$ & $N_{3r_\text{MMF3}}$ & $N_\text{200c}$ & \multicolumn{2}{c}{$z$} &  \multicolumn{2}{c}{$\sigma_\text{ap}$} & $R_\text{ap}$ & $R_\text{max}$ &  \multicolumn{2}{c}{$M_{\sigma, \text{200c}}$} & \multicolumn{2}{c}{$M_{\sigma, \text{500c}}$} & $c_\text{200c}$ & $r_\text{200c}$ &  \multicolumn{2}{c}{$\sigma_\text{ap, 200c}$} &  \multicolumn{2}{c}{$\sigma_\text{1D,200c}$} \\
\hline
G$033.81+77.18$ & $207.2206$ & $26.5908$ & 289 & 154 & 226 & 154 & $0.0629$ & $0.0002$ & $796.5$ & $49.7$ & $1.5$ & $1.6$ & $5.8$ & $1.1$ & $4.0$ & $0.7$ & $3.8$ & $1.7$ & $786.2$ & $49.0$ & $809.1$ & $50.5$ \\
G$041.45+29.10$ & $259.4366$ & $19.6766$ & 113 & 110 & 110 & 110 & $0.1771$ & $0.0005$ & $1331.5$ & $91.9$ & $1.0$ & $1.7$ & $21.2$ & $4.4$ & $14.5$ & $3.0$ & $3.9$ & $2.5$ & $1233.2$ & $85.1$ & $1269.2$ & $87.6$ \\
G$042.81+56.61$ & $230.6228$ & $27.7078$ & 318 & 211 & 256 & 210 & $0.0721$ & $0.0003$ & $1162.1$ & $47.9$ & $2.2$ & $2.5$ & $18.0$ & $2.2$ & $12.4$ & $1.5$ & $3.8$ & $2.5$ & $1148.9$ & $47.3$ & $1182.3$ & $48.7$ \\
G$048.10+57.16$ & $230.2848$ & $30.6354$ & 335 & 165 & 242 & 165 & $0.0781$ & $0.0003$ & $822.7$ & $52.7$ & $1.5$ & $1.7$ & $6.3$ & $1.2$ & $4.3$ & $0.8$ & $3.8$ & $1.7$ & $811.7$ & $52.0$ & $835.3$ & $53.5$ \\
G$049.22+30.87$ & $260.0411$ & $26.6253$ & 664 & 367 & 470 & 368 & $0.1603$ & $0.0002$ & $1048.7$ & $39.9$ & $2.0$ & $2.1$ & $12.9$ & $1.5$ & $8.8$ & $1.0$ & $3.8$ & $2.1$ & $1042.9$ & $39.7$ & $1073.1$ & $40.8$ \\
G$053.53+59.52$ & $227.5525$ & $33.5083$ & 321 & 222 & 270 & 222 & $0.1138$ & $0.0003$ & $1069.0$ & $47.6$ & $1.7$ & $2.1$ & $13.2$ & $1.8$ & $9.0$ & $1.2$ & $3.8$ & $2.2$ & $1041.7$ & $46.4$ & $1071.9$ & $47.7$ \\
G$056.77+36.32$ & $255.6774$ & $34.0601$ & 186 & 96 & 133 & 96 & $0.0989$ & $0.0005$ & $1123.5$ & $77.5$ & $1.9$ & $2.3$ & $15.7$ & $3.3$ & $10.8$ & $2.2$ & $3.8$ & $2.3$ & $1102.8$ & $76.1$ & $1134.8$ & $78.3$ \\
G$080.41-33.24$ & $336.5245$ & $17.3643$ & 173 & 137 & 151 & 137 & $0.11$ & $0.0003$ & $830.4$ & $51.7$ & $1.1$ & $1.7$ & $5.9$ & $1.1$ & $4.1$ & $0.8$ & $3.8$ & $1.7$ & $797.5$ & $49.6$ & $820.7$ & $51.0$ \\
G$226.18+76.79$ & $178.8243$ & $23.4047$ & 238 & 131 & 162 & 131 & $0.1413$ & $0.0004$ & $1064.0$ & $77.9$ & $1.5$ & $2.1$ & $12.4$ & $2.7$ & $8.5$ & $1.9$ & $3.8$ & $2.1$ & $1027.1$ & $75.2$ & $1056.9$ & $77.4$ \\
G$238.69+63.26$ & $168.2267$ & $13.4346$ & 509 & 229 & 330 & 229 & $0.1671$ & $0.0003$ & $869.4$ & $46.6$ & $1.8$ & $1.8$ & $7.5$ & $1.2$ & $5.1$ & $0.8$ & $3.8$ & $1.8$ & $871.6$ & $46.7$ & $896.9$ & $48.0$ \\
\hline
\end{tabular}
}
\tablefoot{The full table is available in electronic form. Velocity dispersions in units of [$\text{km~s}^{-1}$]. Lengths and radii in units of [$\text{Mpc}$]. Masses in units of [$10^{14}M_\odot$].
RA and DEC (cols.~2 and 3) are the J2000 coordinates of the X-ray peak. $N_\text{can}$ (col.~4) is the number of initial candidates. $N_\text{members}$ (col.~5) is the number of members used to compute the velocity dispersion. $N_{3r_\text{MMF3}}$ (col.~6) is the number of members within an aperture of $3~r_\text{MMF3}$. $N_\text{200c}$ (col.~7) is the number of members within $r_\text{200c}$.
$z$ (col.~8) is the estimated redshift.
$\sigma_\text{ap}$ (col.~9) is the aperture velocity dispersion within the effective aperture $R_\text{ap}$ (col.~10). $R_\text{max}$ (col.~11) is the maximum radius of the member galaxies. $M_{\sigma, \text{200c}}$ (col.~12) is the dynamical mass within $r_\text{200c}$. $M_{\sigma, \text{500c}}$ (col.~13) is the dynamical mass within $r_\text{500c}$. $c_\text{200c}$ (col.~14) is the estimated concentration for $M_{\sigma, \text{200c}}$, computed with a mass-concentration relation for the $C_\text{BI}$ value of the marginalised posterior mass distribution.
$r_\text{200c}$ (col.~15) is the estimated overdensity radius for the central value of $M_{\sigma, \text{200c}}$.
$\sigma_\text{ap, 200c}$ (col.~16) is the aperture velocity dispersion within $r_\text{200c}$. $\sigma_\text{1D,200c}$ (col.~17) is the 1D velocity dispersion within $r_\text{200c}$.
}
\end{sidewaystable*}

The estimates of redshifts, velocity dispersions, and dynamical masses, and the results of our analysis are summarised in Table~\ref{tab_masses}, whose full version is only available in electronic form.
Catalogues and updates can be found at \url{http://xmm-heritage.oas.inaf.it/}
 or \url{http://pico.oabo.inaf.it/\textasciitilde sereno/CHEX-MATE/sigma/}. 

\section{Measured redshifts}

\label{sec_efosc2}

This work is based in part on the first wave of spectroscopic redshifts in the cluster fields of view measured with observations collected at ESO under programmes 0110.A-4192 and 0111.1-0186. Redshift reliability is estimated with the ${\cal R}$ statistics \citep{to+da79}, that is, the reliability factor of the cross correlation peak, larger than or equal to 3 \citep{ku+mi98}. Redshifts are available in electronic form. As an example of the data format, in Table~\ref{tab_her033} we list an extract of the redshifts collected in the field of view of the cluster PSZ2~G263.68-22.55.

\begin{table}[htbp]
\caption{Extract of the galaxies with measured spectroscopic redshift in the field of PSZ2 G263.68-22.55.}
\label{tab_her033}
\centering
\begin{tabular}{llr@{$\,\pm\,$}lr}
\hline
\multicolumn{1}{c}{RA} & \multicolumn{1}{c}{DEC} & \multicolumn{2}{c}{$cz$}  & ${\cal R}$ \\
   &     & \multicolumn{2}{c}{[$\text{km~s}^{-1}$]} &    \\   
\hline
$101.32300$ & $-54.24084$ & $50625$ & $37$ & $10.6$ \\
$101.32897$ & $-54.22427$ & $49309$ & $46$ & $8.8$ \\
$101.34888$ & $-54.24609$ & $48816$ & $42$ & $8.9$ \\
$101.35916$ & $-54.24773$ & $48840$ & $36$ & $9.3$ \\
$101.36487$ & $-54.21105$ & $47999$ & $47$ & $7.9$ \\
$101.37290$ & $-54.22694$ & $49105$ & $52$ & $7.3$ \\
$101.37981$ & $-54.20316$ & $48364$ & $45$ & $8.8$ \\
$101.38369$ & $-54.23370$ & $50055$ & $87$ & $4.2$ \\
$101.38995$ & $-54.23248$ & $48577$ & $34$ & $9.5$ \\
$101.39023$ & $-54.20097$ & $51662$ & $56$ & $6.9$ \\
\hline
\end{tabular}
\tablefoot{Coordinates RA and DEC are in J2000; redshift are given in units of $\text{km~s}^{-1}$; ${\cal R}$ is the \texttt{xcsao} reliability of the correlation. The full tables are available in electronic format.}
\end{table}








\end{appendix}

\end{document}